\documentclass[twoside,11pt,preprint]{article}

\usepackage[abbrvbib, preprint]{jmlr2e}
\usepackage{amsmath,amssymb}
\usepackage{ulem}
\usepackage{xcolor}
\usepackage{soul}

\usepackage{lastpage}

\usepackage{multirow}
\usepackage{relsize}
\usepackage{comment}
\usepackage{graphicx}
\usepackage{listings}
\usepackage{xcolor}
\usepackage{booktabs}
\usepackage{fancyref}
\usepackage[parfill]{parskip}
\usepackage{todonotes}
\usepackage{import}
\usepackage{dsfont}
\usepackage{wrapfig} 
\usepackage{lineno}
\usepackage{float}
\usepackage{caption}
\usepackage[skip=10pt]{subcaption}
\usepackage{bm}
\usepackage{bbm}
\usepackage{commath}
\usepackage[section]{placeins}
\usepackage{fancyhdr}
\usepackage{natbib}
\usepackage{datetime}

\usepackage{algorithm}
\usepackage{algorithmic}
\mathchardef\mhyphen="2D % Define a "math hyphen"

\newtheorem{assumption}{Assumption}

\newcommand{\indep}{\perp \!\!\! \perp} % independence symbol
\newcommand{\notindep}{\not\!\perp \!\!\! \perp} % not independence symbol

\numberwithin{equation}{section}

\usepackage{amsmath}
\usepackage{tikz}
\usetikzlibrary{calc}
\usetikzlibrary{matrix}
\usetikzlibrary{chains}
\usetikzlibrary{positioning}
\usetikzlibrary{decorations.pathreplacing}
\usetikzlibrary{arrows}
\usetikzlibrary{bayesnet}

\usepackage{graphicx}

\usepackage{url}

\newcommand{\myref}[2]{\hyperref[#2]{#1 \ref*{#2}}}

\ShortHeadings{Causal Discovery in Financial Markets}{Sadeghi, Gopal, and Fesanghary}
\firstpageno{1}

\begin{document}

\title{Causal Discovery in Financial Markets: \\
A Framework for Nonstationary Time-Series Data
}

\author{\name Agathe Sadeghi\thanks{work performed while interning on the Bloomberg Quant Research team} \email asadeghi@stevens.edu \\
       \addr Stevens Institute of Technology \\
       Hoboken, NJ, USA
       \AND
       \name Achintya Gopal \email agopal6@bloomberg.net \\
       \addr Bloomberg \\
       New York, NY, USA
       \AND
       \name Mohammad Fesanghary \email mfesanghary1@bloomberg.net \\
       \addr Bloomberg \\
       New York, NY, USA
       }

\editor{My editor}

\maketitle

\begin{abstract}%   <- trailing '%' for backward compatibility of .sty file
This paper introduces a new causal structure learning method for nonstationary time series data, a common data type found in fields such as finance, economics, healthcare, and environmental science.
Our work builds upon the constraint-based causal discovery from nonstationary data algorithm (CD-NOD). We introduce a refined version (CDNOTS)\footnote{Open-source implementation: \texttt{causal-ts}, \url{https://github.com/bloomberg/causal-ts}.} which is designed specifically to account for lagged dependencies in time series data. We compare the performance of different algorithmic choices, such as the type of conditional independence test and the significance level, to help select the best hyperparameters given various scenarios of sample size, problem dimensionality, and availability of computational resources. 
Using the results from the simulated data, we apply CDNOTS to a broad range of real-world financial applications in order to identify causal connections among nonstationary time series data, 
thereby illustrating applications in factor-based investing, portfolio diversification, and comprehension of market dynamics. 

\end{abstract}


%%%%%%%%%%%%%%%%%%%%%%%%%%%%%%%%%
\section{Introduction}

Actions have consequences; that explains causality in simple words.
Going beyond correlation or association, it is a guide which helps us gain a deeper understanding of the relationships between events, allowing to perceive the impact of one phenomenon on another. An empirical causal framework would consist of three phases: (a) causal discovery; (b) causal inference; (c) causal explainability. In the first phase, the causal network is estimated from the data based on independence tests. When the network is discovered, then different scenarios can be tested on the structure. Afterwards, we can dive deep into the why and explanations of the scenarios implied by the discovered network.

In this paper, our focus is on causal discovery in time series, which aims to uncover the causal dependency structure from the observed data (\myref{Figure}{fig:final_arch}). Time series is the fundamental type of data used in many domains, from finance to climate science and engineering, requiring customized algorithms to take into account characteristics such as temporality, lead-lag dependency, and nonstationarity. 

Most causal discovery methods can be categorized into three types of algorithms: constraint-based (e.g., \cite{spirtes00,huang20}), score-based (e.g., \cite{Chickering03,Silander2012,Lam22}), and functional causal model-based (FCM) (e.g., \cite{shimizu6,shimizu11,hyvarinen10}). The constraint-based causal discovery algorithm infers causal dependency between nodes by running statistical independence tests among conditional and marginal probabilities to establish the network structure \citep{borboudakis18, nogueira22}. Two widely used algorithms of this type are Peters and Clark (PC), and Fast Causal Inference (FCI) \citep{spirtes00}. The score-based algorithm defines a score criterion (e.g., likelihood scores \citep{Schwarz78,Buntine1991}) and optimizes for the causal structure that best fits the data. On the other hand, FCMs intend to model the functional mechanisms of the variables. This calls for additional assumptions on the distribution of the data or domain-specific knowledge.

Causal discovery algorithms require assumptions about the data generating process; thus, each algorithm has limitations depending on the realism of the assumptions.
For example, some only work on linear systems \citep{granger69}, some assume a Gaussian distribution for the variables \citep{ramsey18}, and many assume there are no latent (unobserved) confounders\footnote{Some research has been conducted to first eliminate the impact of hidden variables and then apply the structural learning model (e.g., \cite{nandy19}).}.\citep{kun9}. Further, implementing causal discovery algorithms on time series data adds additional challenges, such as nonstationarity, to the existing issues. There are ways to reduce the nonstationarity of time series data such as using stock returns as opposed to prices. However, empirically, we find this is insufficient to make the process stationary. Unlike evaluation of predictive performance, in general, causal discovery algorithms, when applied in practice, do not have ground truth data or ways to test the accuracy of the discovered network via randomized controlled trials. This leads to the importance of ensuring that the assumptions of the causal discovery algorithm pertain to the situations in which they are used. In our case studies (\myref{Section}{sec:case_studies}), we show that, in the financial examples we experiment with, the data is neither linear, Gaussian, nor stationary.

There is a line of approaches addressing the time series causal discovery problem such as Granger causality \citep{granger69}, Dynamic Bayesian Networks and Hidden Markov models \citep{moraffah21}, Mutual Information \citep{palus1}, Transfer Entropy \citep{schreiber0}, time series Fast Causal Inference \citep{entner10}, and PCMCI \citep{runge19}. Since most of the literature mentioned are focused on stationary time series and most current approaches do not handle heterogeneous variable types their approaches might lead to spurious causal relations if the time series contains distributional shifts. Some work has been developed to address this gap (e.g., \cite{peters15,zhang17}).

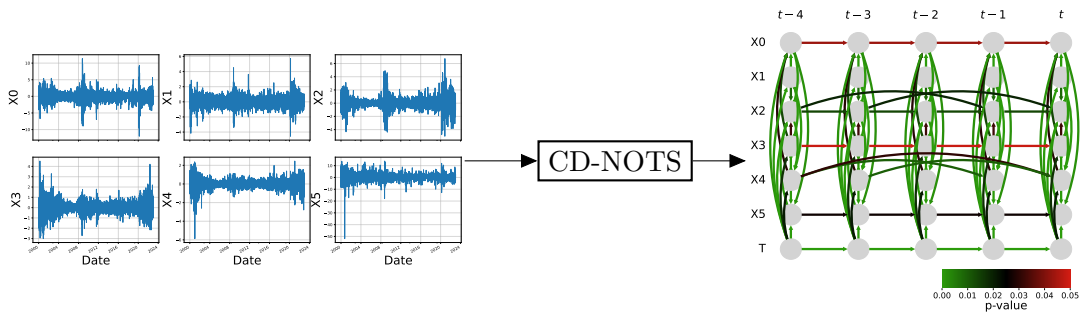
\begin{figure*}[!bt]
  \begin{center}
  \begin{tikzpicture}[
    every neuron/.style={
      circle,
      % draw,
      minimum size=0.3cm,
      very thick
    },
    every data/.style={
      rectangle,
      % draw,
      minimum size=0.4cm,
      thick
    },
  ]
      % \draw[gray, thick] (0,0.5) -- (10,0.5);
  
    \node[inner sep=0pt] (ff) at ($ (2.5,0) $) {\includegraphics[width=.4\textwidth]{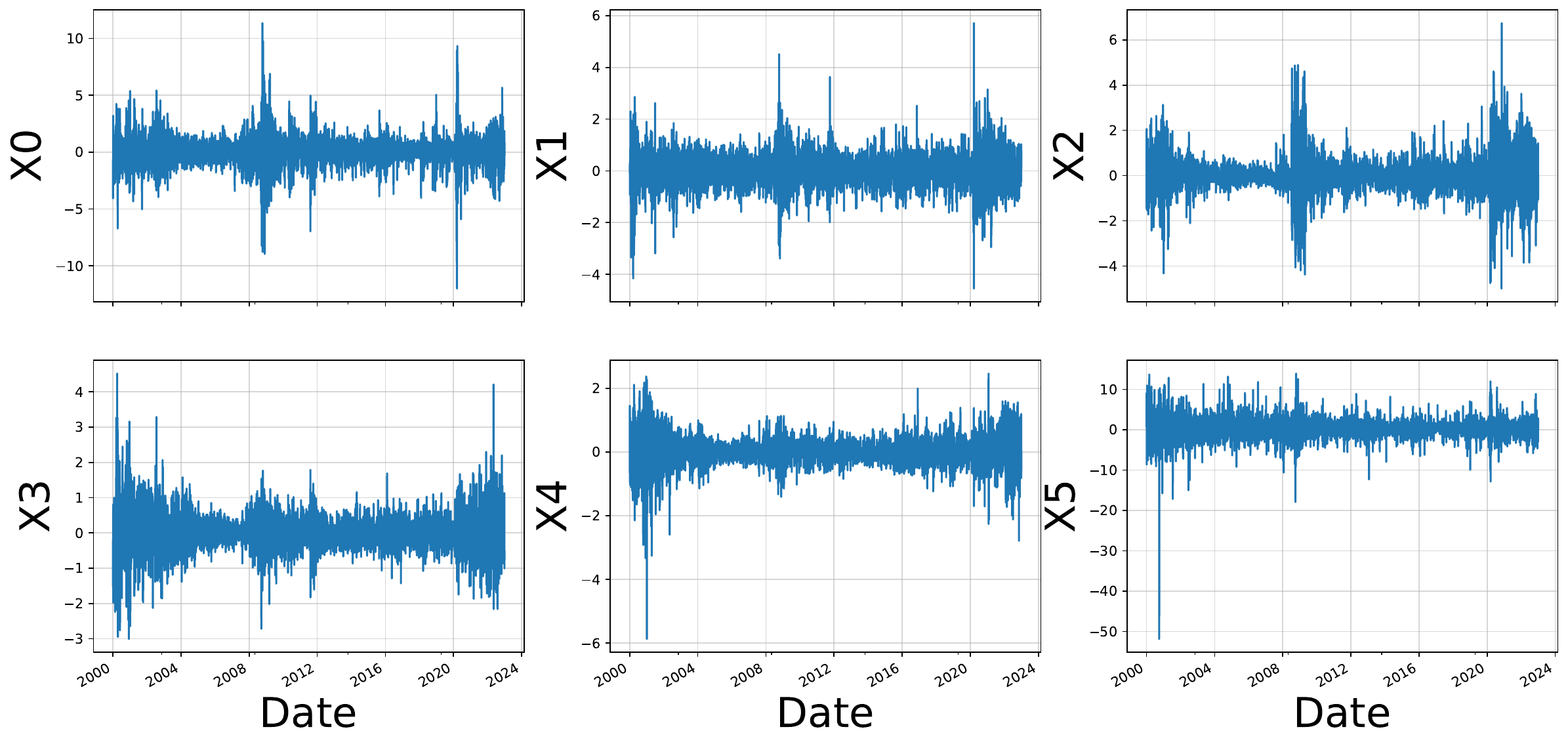}};
    \node [align=center,every data/.try, data 1/.try, minimum width=0.3cm, draw] (cdnots) at ($ (ff.east) + (2.0, 0) $) {CDNOTS};
    \node[inner sep=0pt] (graph) at ($ (cdnots.east) + (3.0, 0) $) {\includegraphics[width=.3\textwidth]{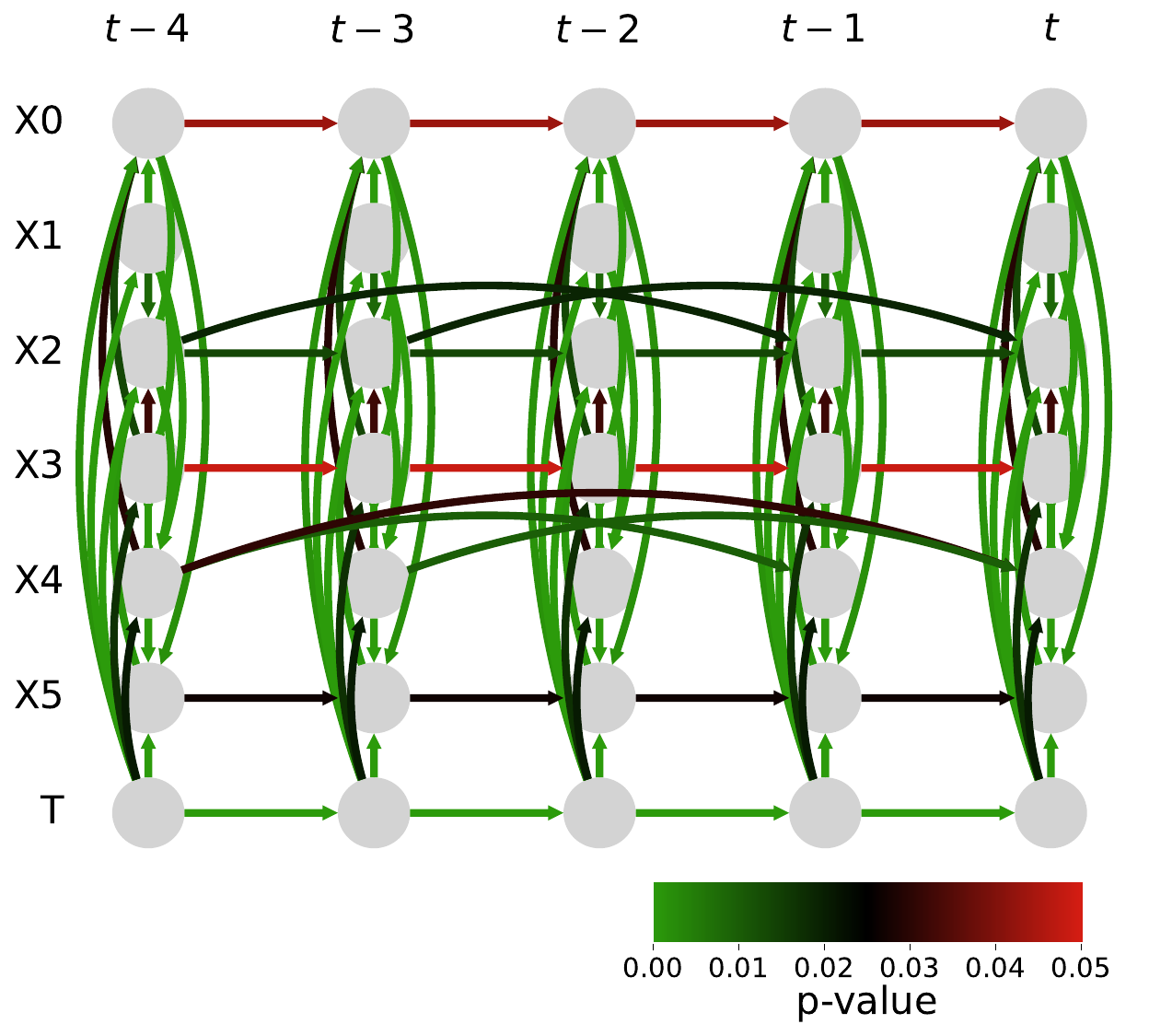}};

        \draw [black,solid,->] ($(ff.east)$) -- ($(cdnots.west)$);
        \draw [black,solid,->] ($(cdnots.east)$) -- ($(graph.west)$);
  
  \end{tikzpicture}
  \end{center}
  \caption{High-level diagram of the problem; we insert some time series data into our causal discovery algorithm, from which we get a causal diagram of how the variables are related.}\label{fig:final_arch}
  
  \end{figure*}

We design an algorithm for causal discovery from nonstationary time series, referred to as the “Constraint-based Causal Discovery from Nonstationary Time Series (CDNOTS)" hereafter.  This algorithm is an extension of the CD-NOD algorithm \citep{huang20} to handle time series data\footnote{\citet{huang20} discusses in their work a way in which their algorithm can be extended to lagged relationships; however, in \myref{Section}{sec:cdnod_algo_compare}, we show a simple example in which their extension would discover an incorrect graph.}. CDNOTS handles nonstationarity as well as pseudo-confounders (defined in \myref{Section}{sec:methodology}) while considering lagged relationships and enforces consistency of the graph through time. It is a nonparametric approach, meaning it can capture both linear and non-linear causal relations and also work with non-Gaussian distributions (e.g., \cite{wang23}). In addition, it is able to detect both lagged and contemporaneous causal effects. The algorithm starts from a full graph, considering nodes for lagged time series, and designates a node to the time itself in order to model nonstationarity. Then, it uses Conditional Independence (CI) tests to uncover the causal skeleton. Leveraging background knowledge (e.g., \cite{wongchokprasitti23}), discoverable structures, and causal change independence, it orients the edges. We experiment CDNOTS with four different tests: Kernel-based Conditional Independence Test (KCIT) \citep{zhang12}, Randomized Conditional Correlation Test (RCoT) \citep{strobl17}, Conditional Mutual Information K-nearest neighbor (CMIknn) Test \citep{runge17}, and Partial Correlation (ParCorr) \citep{Baba2004}.

The structure of this paper is as follows. In \myref{Section}{sec:methodology}, we explain our methodology in detail. In \myref{Section}{sec:evaluations}, we conduct an investigation of simulated and real financial data to assess the framework’s validity and performance. Finally, in \myref{Section}{sec:case_studies}, we summarize the key findings of the paper and propose potential avenues for future research.

%%%%%%%%%%%%%%%%%%%%%%%%%%%%%%%%%
\section{Methodology}\label{sec:methodology}

Following the CD-NOD approach, our algorithm, CDNOTS, consists of four steps: adding the time-indexed node (which helps in handling nonstationarity), discovering the causal skeleton (undirected causal graph) using conditional independence tests, orienting the edges via prior knowledge (e.g., the arrow of time) and identifiable structures (e.g., V-structures) and, finally, orienting remaining edges based on causal change independence.

With suitable assumptions, the causal network using only observational data, can be estimated by conditional independence tests \citep{scholkopf21}. We follow the same set of assumptions commonly used in the causal network field. 
\begin{assumption}[Pseudo Causal Sufficiency]\label{assu:suff}
We assume that any potential confounders can be expressed as a smooth function of time. Hence, the time-indexed node is the representative of all the confounders.
\end{assumption}
\begin{assumption}[Causal Faithfulness]\label{assu:faith}
We assume that the conditional independence between the observed data is a true reflection of the absence of a direct relation in the variables given the conditional set. In other words, if $X$ is conditionally independent of $Y$ given $Z$ ($X \indep Y | Z$), then this is true in the underlying causal network.
\end{assumption}
\begin{assumption}[Randomness]\label{assu:rand}
The data points are randomly selected from the population implied by the causal model (i.e., there is no selection bias).
\end{assumption}
\begin{assumption}[Causal Consistency]\label{assu:cons}
We assume that the causal relations between the variables, including the time-indexed node, are consistent through time. This means that when lagged relations are present, the contemporaneous causal relations are the same in each lag and the cross lag causal relations repeat themselves.
\end{assumption}

We note that most prior work assumes sufficiency, i.e., cannot handle any unobserved confounders.

%%%%%%
\subsection{CDNOTS Algorithm}

Our developed approach, CDNOTS, identifies causal relationships between time series. We show a high-level diagram of our algorithm in \myref{Figure}{fig:high_level_diagram}.

\paragraph{Stage 1} Assume we have $N$ time series. Within our network setup, each time series is treated as a distinct node, denoted as $V_{i,t}$, representing the time series $i$ at time $t$. To handle nonstationarity, we introduce a time-indexed node, $U_t$. Specifically, for any nonstationary time series $i$, we anticipate a connection between $V_{i,t}$ and $U_t$ across all time points. We start from a full graph including the time-indexed node, where all the edges are drawn out.

\paragraph{Stage 2} With \myref{Assumptions}{assu:suff}, \ref{assu:faith} and \ref{assu:rand} in place, we reconstruct the causal skeleton. Employing CI tests, we investigate the independence between a node and the time-indexed node given a subset of nodes $S$. This process figures out the edges between nodes $U_t$ and $V_{i,t}$. The null hypothesis is that the pair are conditionally independent, implying that their conditional joint probability equals the product of their conditional marginals. If the obtained p-value falls below the user specified confidence level \(\alpha\), signifying dependence, the edge between the pair remains. The hypothesis test is:
\begin{align*}
    H_0: P(V_{i,t},U_{t}|S)=P(V_{i,t}|S)\, P(U_{t}|S)\,,\\
    H_1: P(V_{i,t},U_{t}|S)\neq P(V_{i,t}|S)\,P(U_{t}|S)\,,
\end{align*}
which we test for all subsets $S \subseteq \{V_{k,t-l}\ |\ k=1,...,N;l=0,...,L \} \setminus \{V_{i,t}\}$ where $L$ denotes the maximum considered lag. Note that this hypothesis examines the relationship between the time-indexed node and other nodes at the identical time point $t$, maintaining consistency across time $t-l$. 

Afterwards, utilizing the same test, we evaluate the independence of node pairs $V_{i,t-l}$ and $V_{j,t}$ excluding $U_t$. The hypothesis test is:
\begin{align*}
    H_0: P(V_{i,t-l},V_{j,t}|S)=P(V_{i,t-l}|S)\,P(V_{j,t}|S)\,,\\
    H_1: P(V_{i,t-l},V_{j,t}|S)\neq P(V_{i,t-l}|S)\,P(V_{j,t}|S)\,,
\end{align*}
where $S \subseteq \{U_t\} \cup \{V_{k,t-m}\ |\ k=1,...,N; m=0,...,L\} \setminus \{V_{i,t-l},V_{j,t}\}$. It is worth noting that when $l=0$, indicating no lag, we capture the contemporaneous relationships between nodes. On the other hand, when $l>0$, we discover the lagged dependencies among the time series. Under \myref{Assumption}{assu:cons}, we maintain the structure consistent over time. This means that the discovered dependencies between $V_{i,t-l},V_{j,t}$ remain the same for $V_{i,t-l-m},V_{j,t-m}; m=1,...,L-l$. The discovered network is undirected and has the edge minimality condition. This condition signifies that the network has the smallest number of causal changing mechanisms. This is due to the faithfulness across the entire network \citep{ghassemi18}.

\paragraph{Stage 3} Once we have derived the network skeleton, our next step involves orienting the edges based on prior knowledge and identifiable structures.
If there are any nonstationary nodes, meaning there exists an edge between the node and the time-indexed node, then the orientation is from the time-indexed node to the nonstationary node, as time is what causes the node's corresponding time series to be nonstationary. Moreover, all the edges where $l<m$ between pairs $V_{i,t-l}$ and $V_{j,t-m}$, the orientation is from $V_{i,t-m}$ to $V_{j,t-l}$, as logically the effect would happen after the cause. These directed edges along with the conditional independence tests can be used to orient other edges using the unique properties of V-structures (colliders) \citep{spirtes00,Meek1995}.

\paragraph{Stage 4} There might still remain some edges that are not yet directed. In the fourth stage, we rely on the independent changes of the causal modules \citep{Pearl09}. A dependency measure is defined as an extension of the Hilbert Schmidt Independence Criterion \citep{hilbert8}, which shows the level of dependency between causal modules. This is done by developing a kernel embedding of nonstationary conditional distributions and using their Gram matrix to create a test statistic \citep{huang17}. We apply the Meek orientation rule \citep{Meek1995} on top of this to orient other edges. 
Note that this stage is only applicable to the edges between two nonstationary nodes. There may remain edges that are undirected even after the third and fourth stages; in this scenario, no comments can be given on the causal direction of the corresponding pair of nodes.

\input{cdnots_algo}
%%%%%%
\subsection{Testing Assumptions}

In practice, an important step when using algorithms that learn from data is to evaluate (or validate) the results of the algorithm. Running interventional studies is one approach; however, this is often infeasible, if not impossible, in fields such as finance. Instead, as mentioned in the introduction, evaluation of causal discovery algorithms in the practice rests upon evaluating the realism of the assumptions underlying the algorithm.

We categorize assumptions into three types:
\begin{enumerate}
    \item \textit{Objective Assumptions:} 
    These are assumptions which are statistically testable such as stationarity and linearity. In \myref{Section}{sec:objective}, we discuss specific ways we can test these two properties. Since these assumptions are testable, we argue that any causal discovery algorithms reliant on these assumptions should ensure that the data aligns with them.
    \item \textit{Algorithmic Assumptions (or Algorithmic Hyperparameters):} These assumptions refer to decisions in the causal discovery algorithm which can neither be directly evaluated nor justified purely by domain expertise. Examples include the significance threshold used in the conditional independence test and the type of conditional independence test. While there is some theories and studies that can be used to justify one threshold or test over another (e.g., our simulated experiments in \myref{Section}{sec:evaluations}), we recommend testing the robustness of the causal discovery algorithm's results to these hyperparameters.
    \item \textit{Subjective Assumptions:} These are assumptions which cannot be statistically tested and require domain expertise to evaluate. Pseudo-sufficiency, consistency, faithfulness, and randomness (assumptions made by CDNOTS) fall in this category.
    In our case studies, while we maintain confidence in the validity of these assumptions, the credibility of our algorithm's results comes from the reader's belief in these assumptions. We discuss the intuition behind these assumptions in \myref{Section}{sec:subjective}.
\end{enumerate}

\subsubsection{Objective Assumptions}\label{sec:objective}

One of the main assumptions in prior causal discovery algorithms is stationarity, which can result in serious inaccuracies when dealing with inherently nonstationary data, such as those commonly encountered in fiance and economics. Since our algorithm does not rely on this assumption, we can analyze the graph identified by our algorithm to see if nonstationary relationships are found.

While many recent algorithms do not assume linearity, for comprehensive analysis, we test this assumption post-hoc in our case studies (\myref{Section}{sec:case_studies}) using a modified version of a previously employed algorithm \citep{peters14}. Typically, prior work assuming linear relationships also assume additive noise. Accordingly, when we test for linearity, we also test for additive noise. The algorithm used in \citet{peters14} involves performing a regression followed by an independence test on the noise. For instance, if the null hypothesis we are testing is that $X$ and $Y$ have a linear relationship given $Z$ (where we allow for $Z$ to have a non-linear relationship with $Y$ and the noise to be heteroscedastic with respect to $Z$), we perform a Gaussian process regression $\hat{f}$ using a linear kernel for $X$ and a non-linear kernel (e.g., RBF) for $Z$. Subsequently, we perform a CI test on $Y - \hat{f}(X, Y) \indep X | Z$. Note that our test is more generic than testing if \textit{all} relationships are linear.

\subsubsection{Subjective Assumptions}\label{sec:subjective}
Almost every causal discovery algorithm is built upon subjective assumptions. It is crucial to justify these assumptions due to their subjective nature. Here, we provide an intuitive explanation of these assumptions in order to allow readers to judge the realism of these assumptions.

\paragraph{Pseudo-sufficiency} Any potential confounder described as a smooth function of time can be captured by the time-indexed node; while, if there exists a confounder which is stationary and not changing through time, the algorithm will not capture it.

\paragraph{Consistency} The consistency assumption states that the causal links remain unchanged through time.
We assert any deviations observed in the causal graph over time in \myref{Section}{sec:case_studies} can be attributed to sampling variations and data availability. 

\paragraph{Faithfulness} This assumption can be violated in many ways, including xor connections (or gating mechanisms), deterministic functions, or cancelling paths \citep{marx2021}. In our case studies, we do not have situations of deterministic functions which can be directly tested from the data. 

\paragraph{Randomness} The randomness assumption comes from the CI tests which enables the derivation of the null distribution (statistical validation). We note that this assumption does not hold for time series\footnote{This is a limitation for all constraint-based causal discovery algorithms that use p-values of CI tests to define a threshold, and this is a gap to be filled in future work.}; this implies that the p-value threshold used does not accurately control the false positive rate. We argue that testing the robustness of our algorithm to different p-value thresholds mitigates this concern.

%%%%%%
\subsection{Importance of Modeling Lagged Relationships}\label{sec:lagged_rel}

We show with a toy example how, if one were to not model lagged relationships but the data contains them, the graph would find nonstationary variables. Say we have one variable $X_t$ where:
\begin{align*}
    dX_t &= -X_t dt  + \sqrt{2}\ dW_t\,,
\\  X_0 &= 0\,.
\end{align*}
where $W_t$ is a Wiener process.

To discretize, we set $t$ to be integers. Since this is simply an Ornstein-Uhlenbeck (OU) process, we have:
\begin{align*}
    P(X_t | X_u) &= P(X_t |\  \mathcal{N}(X_u e^{-\abs{t-u}}, e^{-\abs{t-u}})) \, ,\, u < t\,.
\end{align*}
From this, we can see that $P(X_t | X_0)$ is a time dependent distribution whereas $P(X_t | X_{t-1})$ is not. This implies if we ran CDNOTS assuming no lagged relationships (i.e., CD-NOD), $X_t$ would be considered nonstationary; whereas, if we ran CDNOTS assuming a single lag, then $X_{t-1} \rightarrow X_t$ and $X_t$ would not be considered nonstationary. 
%%%%%%
\subsection{Comparison to Algorithm in CD-NOD}\label{sec:cdnod_algo_compare}
\citet{huang20} presents an algorithm to extend CD-NOD to time series. However, we find that in the first step in the algorithm (`Detection of changing modules') to remove connections from the time varying node $U$ and observed variables $V_{i, t}$, the tests only use a subset of $\{V_{j, t}\}_{j=1}^{N} \setminus U$ and do not include lags in the conditioning set. We can see in \myref{Figure}{fig:cdnod_algo_compare} that $Y_t \notindep U$ and $Y_t \notindep U | X_t$; however, $Y_t \indep U | X_t, X_{t-1}$, which implies that $Y_t$ is stationary. As CD-NOD does not condition on $X_{t-1}$, it would keep the edge between $U$ and $Y_t$, resulting in incorrectly assuming time series $Y$ is nonstationary.
\begin{figure}[!bt]
\begin{minipage}{\linewidth}
\begin{center}
\begin{tikzpicture}[
  every neuron/.style={
    circle,
    % draw,
    minimum size=0.3cm,
    very thick
  },
  every data/.style={
    rectangle,
    % draw,
    minimum size=0.4cm,
    thick
  },
]

    \node[latent,minimum size=0.9cm]                   (Xt1)      {$X_{t-1}$} ; %
    \node[latent, minimum size=0.9cm, right=of Xt1]                   (Xt)      {$X_{t}$} ; %
    \node[latent,minimum size=0.9cm, below=of Xt1]                   (Yt1)      {$Y_{t-1}$} ; %
    \node[latent,minimum size=0.9cm, right=of Yt1]                   (Yt)      {$Y_{t}$} ; %

    \node[latent,minimum size=0.9cm, above=of $(Xt1)!0.5!(Xt)$]                   (C)      {$U$} ; %
  \edge {C} {Xt};
  \edge {C} {Xt1};
  \edge {Xt1} {Yt1};
  \edge {Xt} {Yt};
  \edge {Yt1} {Yt};

\end{tikzpicture}

\end{center}
\caption{Diagram of example causal graph for which CD-NOD's time series algorithm would make a mistake.}\label{fig:cdnod_algo_compare}
\end{minipage}
\end{figure}
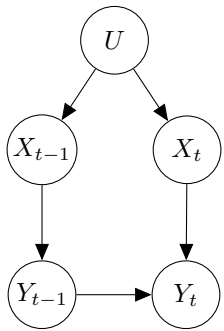
%

%%%%%%%%%%%%%%%%%%%%%%%%%%%%%%%%%
\section{Evaluation}\label{sec:evaluations}
One major design decision in any constrained-based algorithms, including ours, is the CI test. In order to ascertain which test might be most effective, we test our algorithm on different simulated datasets. In total, we tested graphs with 3, 4, 5, 6, 8, 10, and 15 nodes, for a total of 350 randomly generated graphs (50 graphs per number of nodes). For each graph, we tested having 50, 150, 300, 500, and 1,000 data points. The relationship between any connected nodes can be either linear or non-linear (quadratic, exponential, or sine). The max lag we allowed was five.

For the conditional independence tests, we specifically test using partial correlations (ParCorr, a conditional independence test that assumes linearity), KCIT, RCoT , and CMIknn. For both KCIT and RCoT, there are a few variations in how the p-value can be computed, specifically using the Satterthwaite–Welch (SW) method \citep{Welch38,Satterthwaite46} or Hall–Buckley–Eagleson (HBE) method \citep{Hall83,Buckley1988}. The goal of these algorithms is to approximate the CDF of a weighted sum of squared normals, which is the limiting behavior of the test statistic for KCIT and RCoT. The hyperparameters used for each test can be found in \myref{Appendix}{sec:hparams}. 

In \myref{Figure}{fig:ci_fscore_3d}, we compare the average F-score\footnote{F-score is the harmonic mean of precision and recall metrics i.e., F-score$=2\,\frac{precision\times recall}{precision+recall}\,$.} for each CI test across a varying number of nodes in the graph and a varying number of data points per graph. As expected, the F-score improves as the number of data points increase. Possibly most interesting is that ParCorr performs the best for low data regimes. Further, we can see, as found by previous work \citep{Bodenham2015}, that the SW variant of KCIT tends to be weaker than HBE (except when there is a large number of nodes and only 50 data points). However, there seems to be little difference for RCoT. Hence, relying on results in \citet{Bodenham2015}, we recommend the HBE variant. 
\begin{figure}[!bt]
\begin{minipage}{\linewidth}
    \centerline{\includegraphics[width=\linewidth]{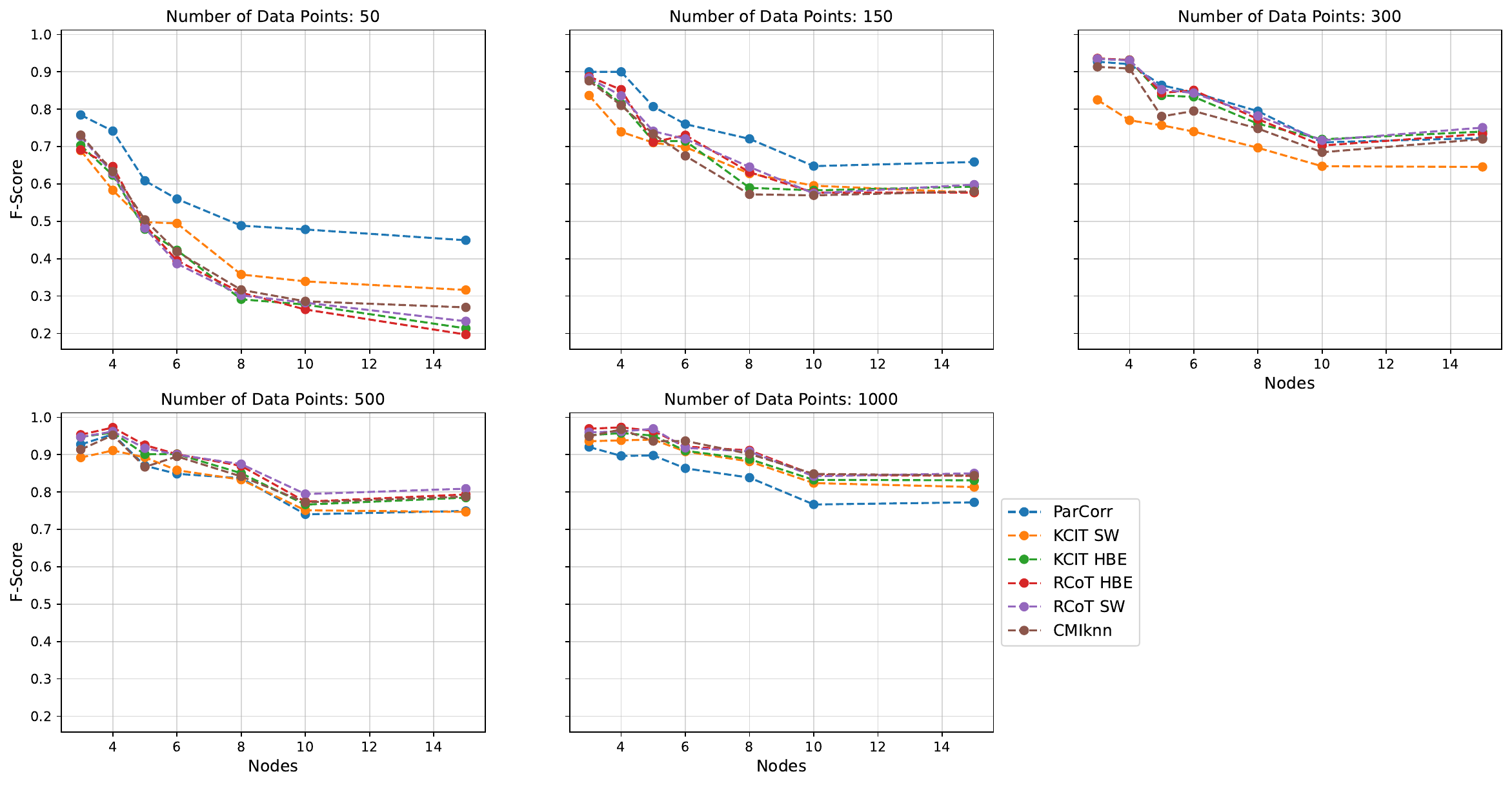}}
    \caption{F-score evaluation for CDNOTS with different CI tests, tested out on different simulated datasets.}\label{fig:ci_fscore_3d}
\end{minipage}
\end{figure}
From our simulation studies, we recommend using RCoT for large datasets since it is much faster than KCIT on a CPU. However, we do note that the difference in speed is much less when using a GPU implementation of KCIT and RCoT; thus, if a GPU is available, we recommend KCIT since it does not have a randomized component. Finally, while CMIknn is quite competitive, we do not recommend its usage on large problems due to its lengthy runtime.

Another important metric to consider while choosing a CI test is the computational cost of it. \myref{Figure}{fig:ci_runtime} shows the average runtime of different tests for different sample sizes. In general, methods like CMIKnn are several orders of magnitude slower than simple linear tests like ParCorr which limits their applicability in very large samples. In terms of computations, KCIT is positioned in the middle and its cost grows in $\mathcal{O}(n^3)$ with sample size. RCoT has time complexity of $\mathcal{O}(m^2n)$ where $m$ is the number of Fourier terms (typically in order of 10-100) set by the user which gives it a considerable computation saving when compared to KCIT. In general, as expected the number of conditions that needs to be tested grows rapidly with the number of nodes as shown in \myref{Figure}{fig:ci_tests}. Note that the number of tests needed for a given network size will not be exactly the same for different CI tests due to fact that any edge removal in early states of skeleton discovery (which depends on the p-value generated by the given test) will result in a substantial decrease in condition sets that need to be tested and hence the total number of tests.
\begin{figure}[!bt]
\begin{minipage}{\linewidth}
    \centerline{\includegraphics[width=0.45\linewidth]{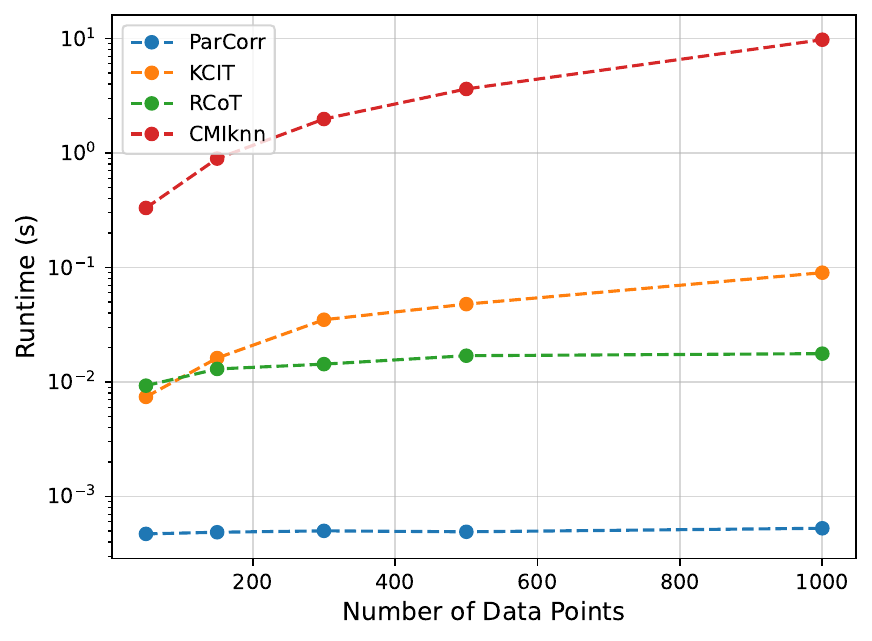}}
    \caption{Runtime evaluation for CDNOTS for different CI tests. Note the y-axis is in log scale.}\label{fig:ci_runtime}
\end{minipage}
\end{figure}
\begin{figure}[!bt]
\begin{minipage}{\linewidth}
    \centerline{\includegraphics[width=\linewidth]{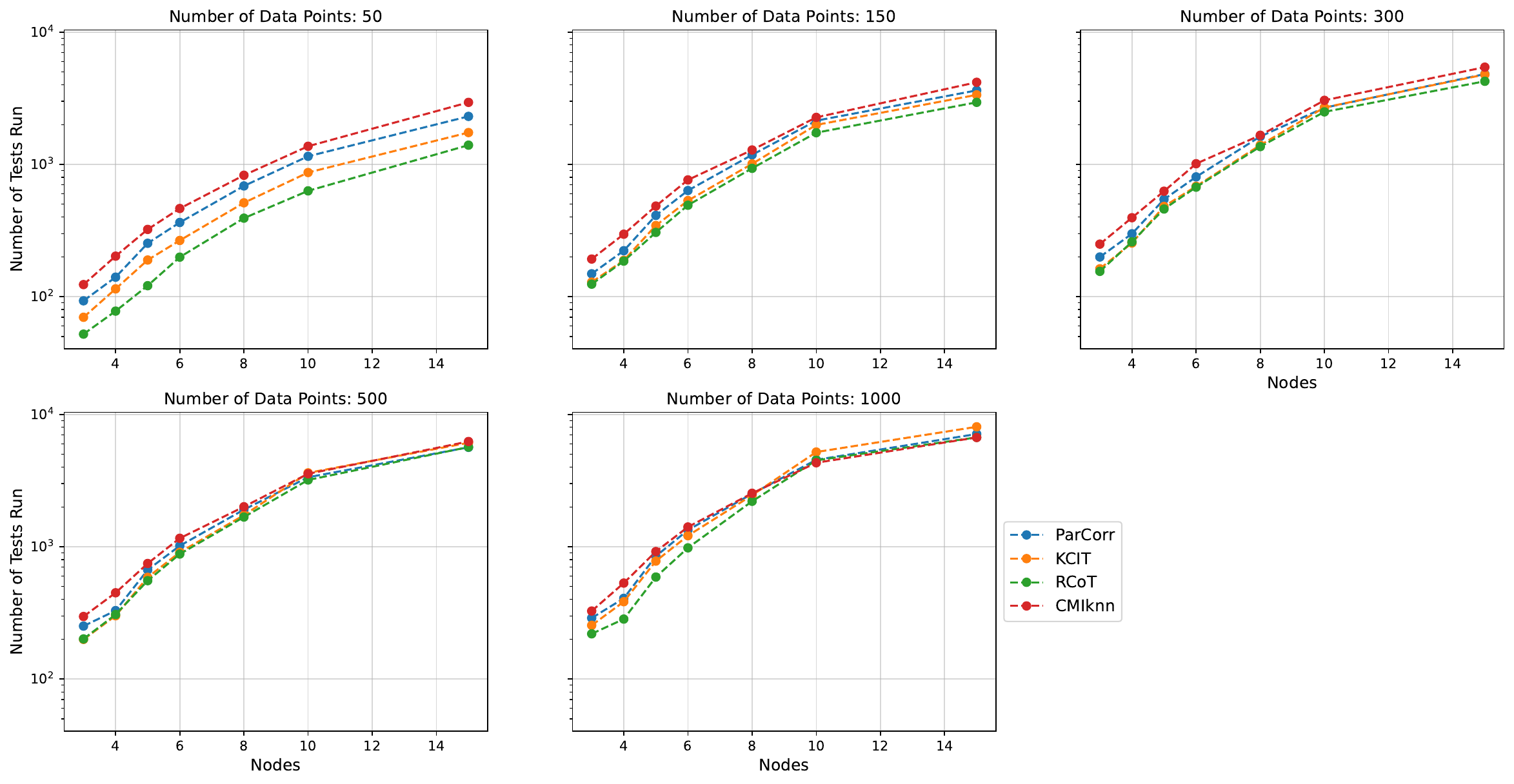}}
    \caption{Number of tests run for CDNOTS for different CI tests. Note the y-axis is in log scale.}\label{fig:ci_tests}
\end{minipage}
\end{figure}

Another important parameter of the algorithm is the significance level \(\alpha\). We ran a series of experiments with \(\alpha\) levels [0.01, 0.05, 0.1] to study the sensitivity of the algorithm to the choice of \(\alpha\). The results for F-score have been shown in \myref{Figure}{fig:cdnots_alpha_sensitivity_f_score3d}. Please see \myref{Appendix}{sec:alpha_cdnod} for other metrics. As expected, a lower \(\alpha\) results in higher precision across all data ranges and CI tests. The only exception to this is RCoT in low data regimes, but this is expected because RCoT, by design, assumes there are large number of data points and its results are not reliable for small datasets. In terms of F-score, larger \(\alpha\) values perform better in low data regimes; however, in large data regimes, small alphas perform better.
\begin{figure}[!bt]
\begin{minipage}{\linewidth}

\begin{subfigure}{\linewidth}
    \centerline{\includegraphics[width=\linewidth]{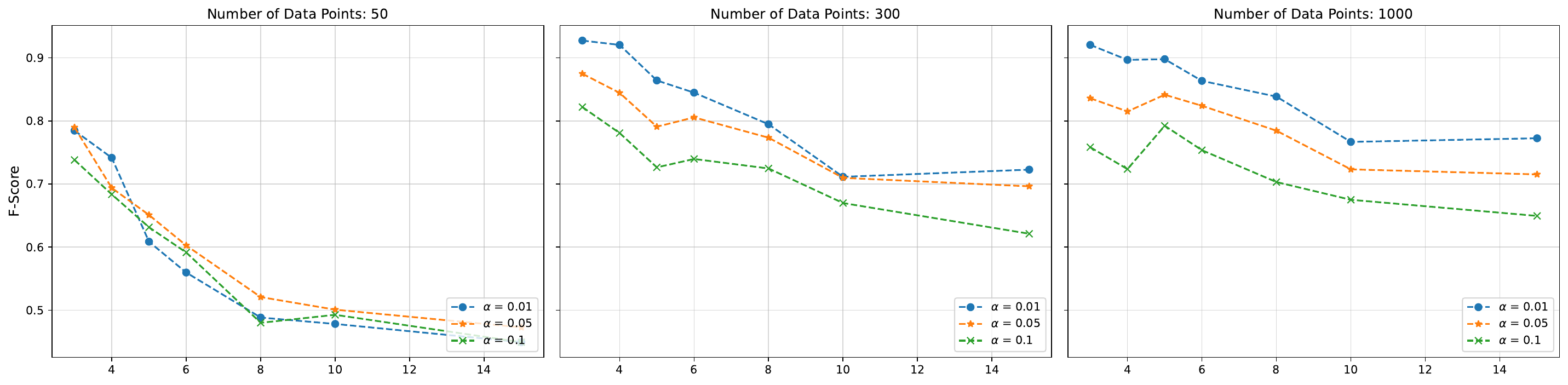}}
    \caption{ParCorr}
    \end{subfigure}
    
    \begin{subfigure}{\linewidth}
    \centerline{\includegraphics[width=\linewidth]{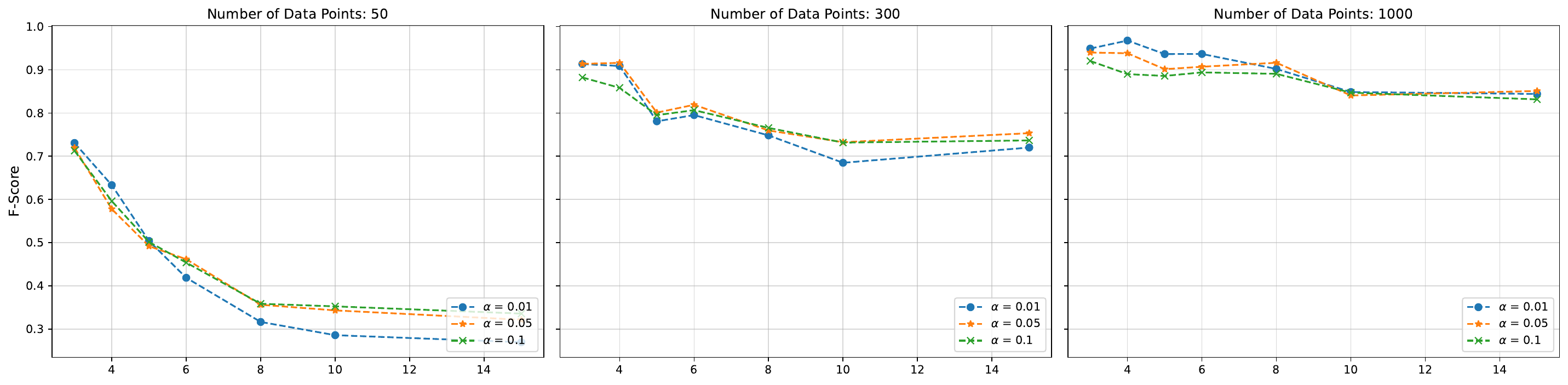}}
    \caption{CMIKnn}
    \end{subfigure}

    \begin{subfigure}{\linewidth}
    \centerline{\includegraphics[width=\linewidth]{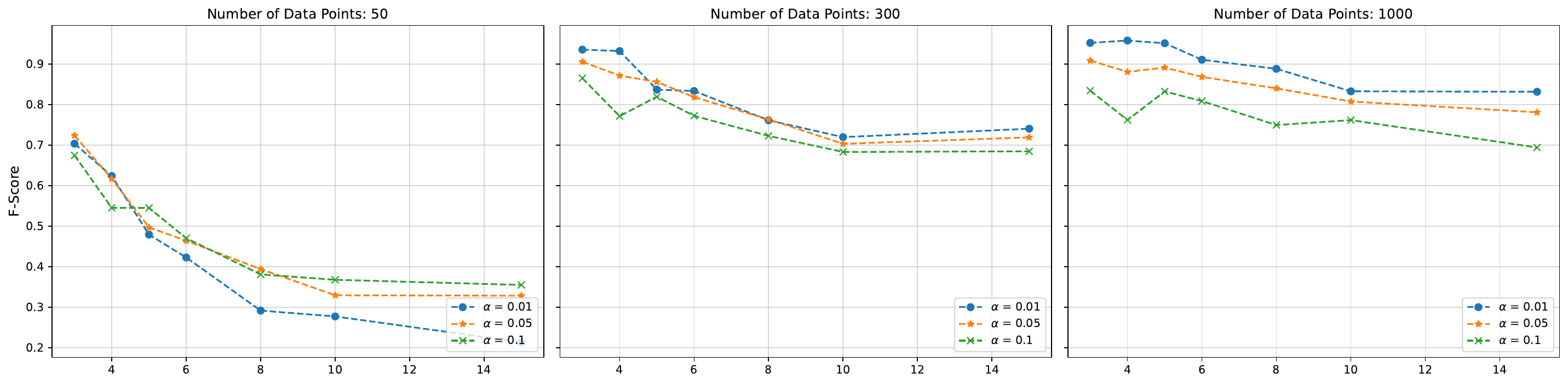}}
    \caption{KCIT}
    \end{subfigure}

    \begin{subfigure}{\linewidth}
    \centerline{\includegraphics[width=\linewidth]{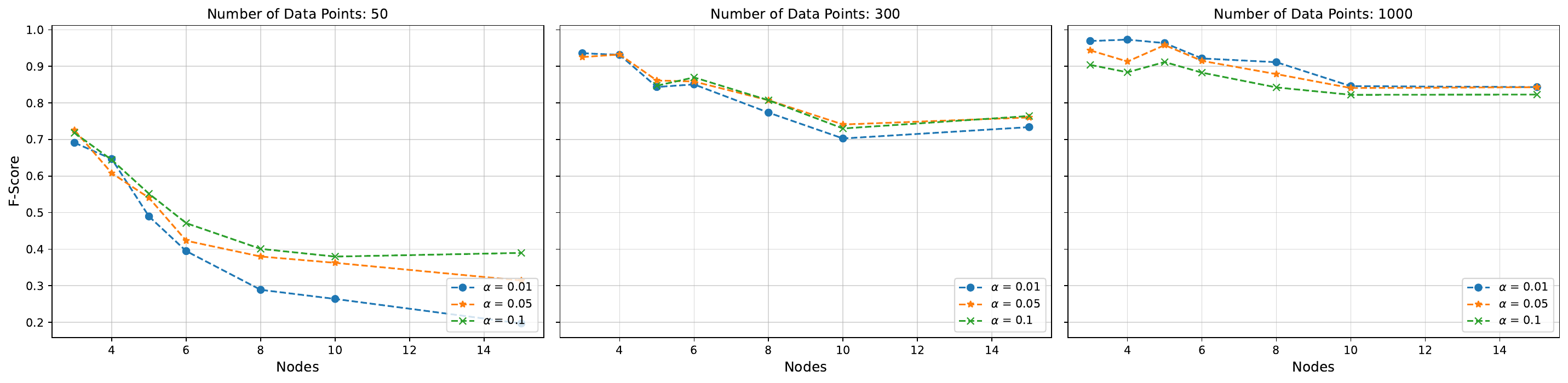}}
    \caption{RCoT}
    \end{subfigure}

    \caption{F-score evaluation, comparing different confidence levels.}\label{fig:cdnots_alpha_sensitivity_f_score3d}

\end{minipage}
\end{figure}

Comparing against the best tests (KCIT HBE, RCoT HBE, ParCorr, CMIknn),  we consider PCMCI \citep{runge19} as a benchmark since it is a comparable PC-based algorithm and is widely used in the causal discovery literature. The main difference in PCMCI over prior PC algorithms is its pruning of the set of variables which to condition on in the CI test. In \myref{Figure}{fig:pcmci_fscore_3d}, we see that CDNOTS consistently outperforms PCMCI.
\begin{figure}[!bt]
\begin{minipage}{\linewidth}
    \centerline{\includegraphics[width=\linewidth]{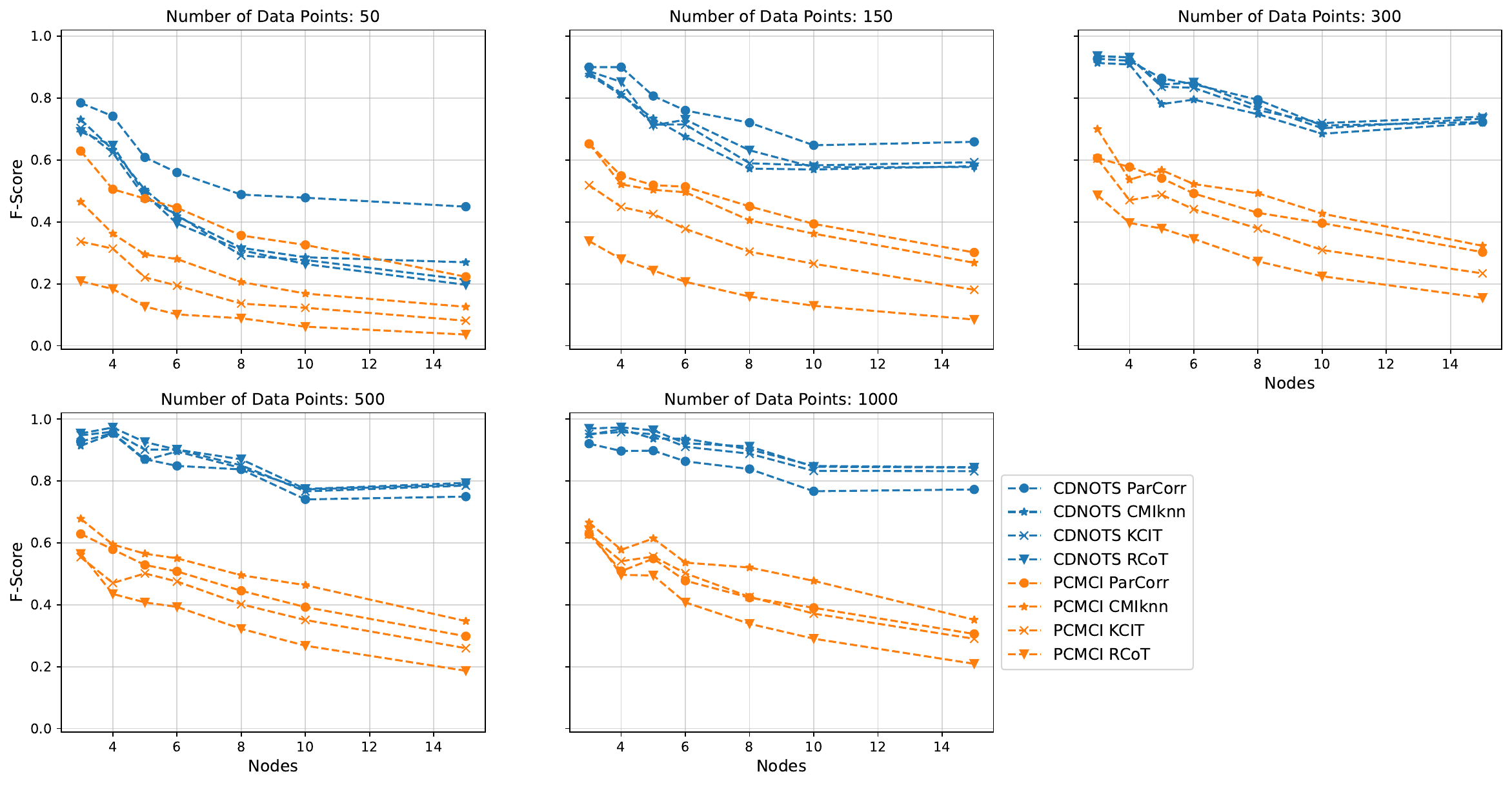}}
    \caption{F-score evaluation, comparing CDNOTS and PCMCI, tested out on many different simulated datasets.}\label{fig:pcmci_fscore_3d}
\end{minipage}
\end{figure}
% 
%%%%%%%%%%%%%%%%%%%%%%%%%%%%%%%%%
\section{Case Studies}\label{sec:case_studies}
Most of the work done in factor investing and economic indices' analysis rely on the correlation between the variables and do not consider the fundamental question of ``why'' \citep{lopez23}. We apply CDNOTS to three financial datasets: 1) Fama-French factors \citep{Fama1993} and Apple's returns, 2) unemployment, CPI, and PPI of different countries, and 3) Price-to-Book ratio and stock returns of financial companies in the S\&P 500. In each case study, we explain the data and properties we see in the data or that has been found in prior work, discuss the implications of the causal networks found by CDNOTS and, finally, show that the assumptions of stationarity, linearity, and no lagged relationships would be incorrect to assume and thus, any algorithm making said assumptions would be inaccurate to use on these datasets. 

For all the graphs in this section, we color by the maximum p-value for that edge across all tests ran during the causal skeleton step (Stage 2). While we recognize that p-values are not a metric to measure the strength of the relationship, we can think of the vicinity of the p-value to 0.05 as a measure of the sensitivity of the graph to that hyperparameter (i.e., the threshold). While in our experiments in \myref{Section}{sec:evaluations} we found that using a p-value of 0.01 to be better for larger datasets, we run our experiments with a p-value of 0.05, both to be consistent with prior work \citep{strobl17} as well as to show the edges that would be removed by changing from 0.05 to 0.01.

%%%%%%
\subsection{Fama-French Factors}\label{sec:kcit_ff}

\paragraph{Data} For our first case study, we analyze the relationship between the Fama-French factors and daily Apple's returns from the beginning of 2000 to the end of 2022. In the data in \myref{Figure}{fig:factors_data}, we show that there is some correlation between the variables (e.g., the spike in volatility around 2008 and during COVID Pandemic can be seen in all six subplots) and, accordingly, all variables show volatility clustering.
\begin{figure}[!bt]
\begin{center}
    \begin{subfigure}{0.9\linewidth}
        \centerline{\includegraphics[width=\linewidth]{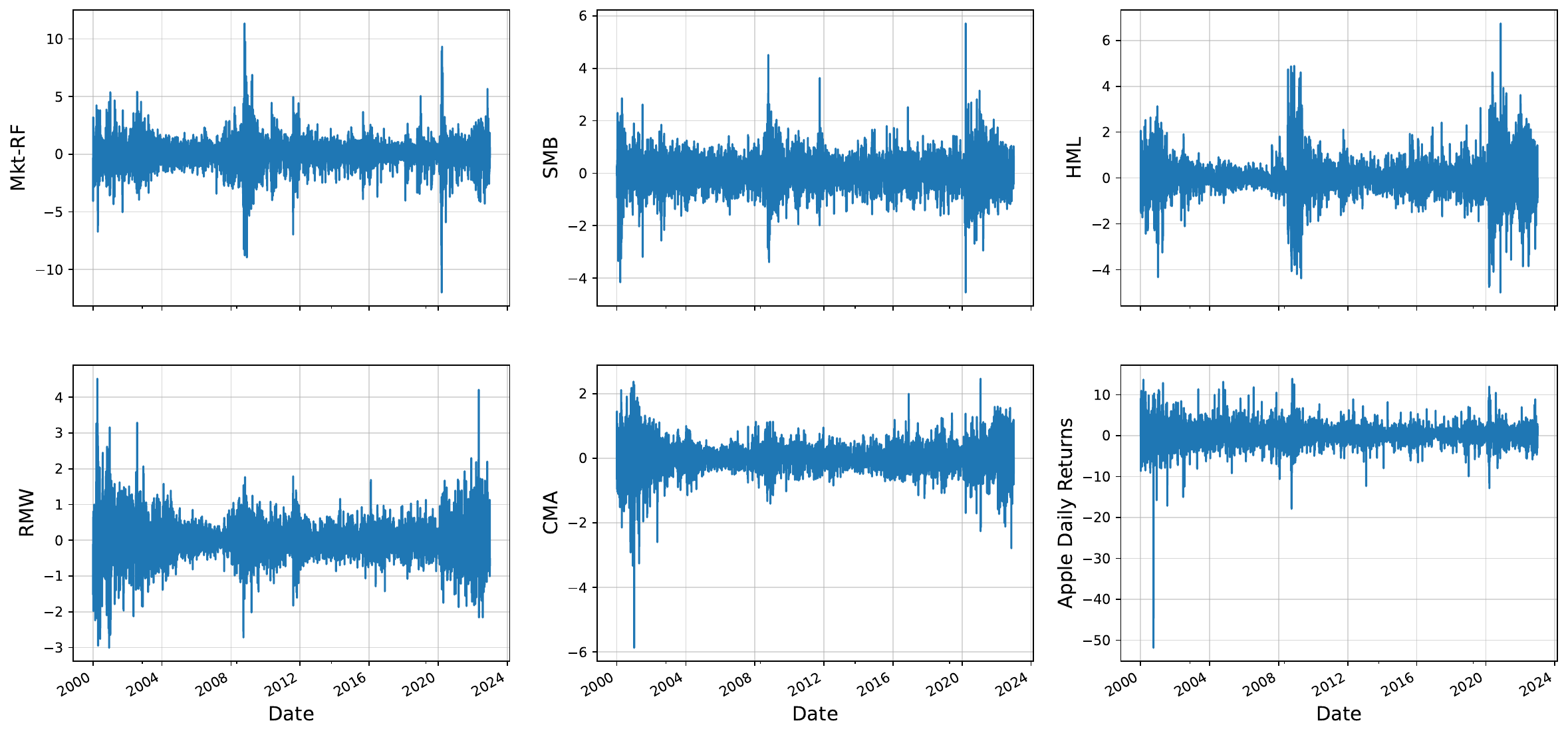}}
    \end{subfigure}
    \caption{Time series of Fama-French factors and Apple's returns.}\label{fig:factors_data}
\end{center}
\end{figure}

\paragraph{Results and Discussion} 

In \myref{Figure}{fig:factors_full} and  \myref{Figure}{fig:factors_full_no_t}, we run CDNOTS with KCIT on the whole time range with and without the $T$ node, respectively, or, in other words, test the impact of modeling nonstationarity. While the number of edges in the contemporaneous graph is about the same (one edge difference, specifically between SMB and CMA), there are fewer lagged relationships in \myref{Figure}{fig:factors_full}. This implies the dependence we see through time is better explained as a nonstationary relationship.

It should be noted that there is a connection from $T$ to Apple's returns (AAPL\_RET). This is to be expected for two reasons: 1) over the course of the last 22 years, Apple's business model has changed, leading to exposures to different factors over time, and 2) the changing relationship between the factors and Apple's returns.
\begin{figure}[!bt]
\begin{minipage}{\linewidth}
    \begin{subfigure}{0.45\linewidth}
        \centerline{\includegraphics[width=\linewidth]{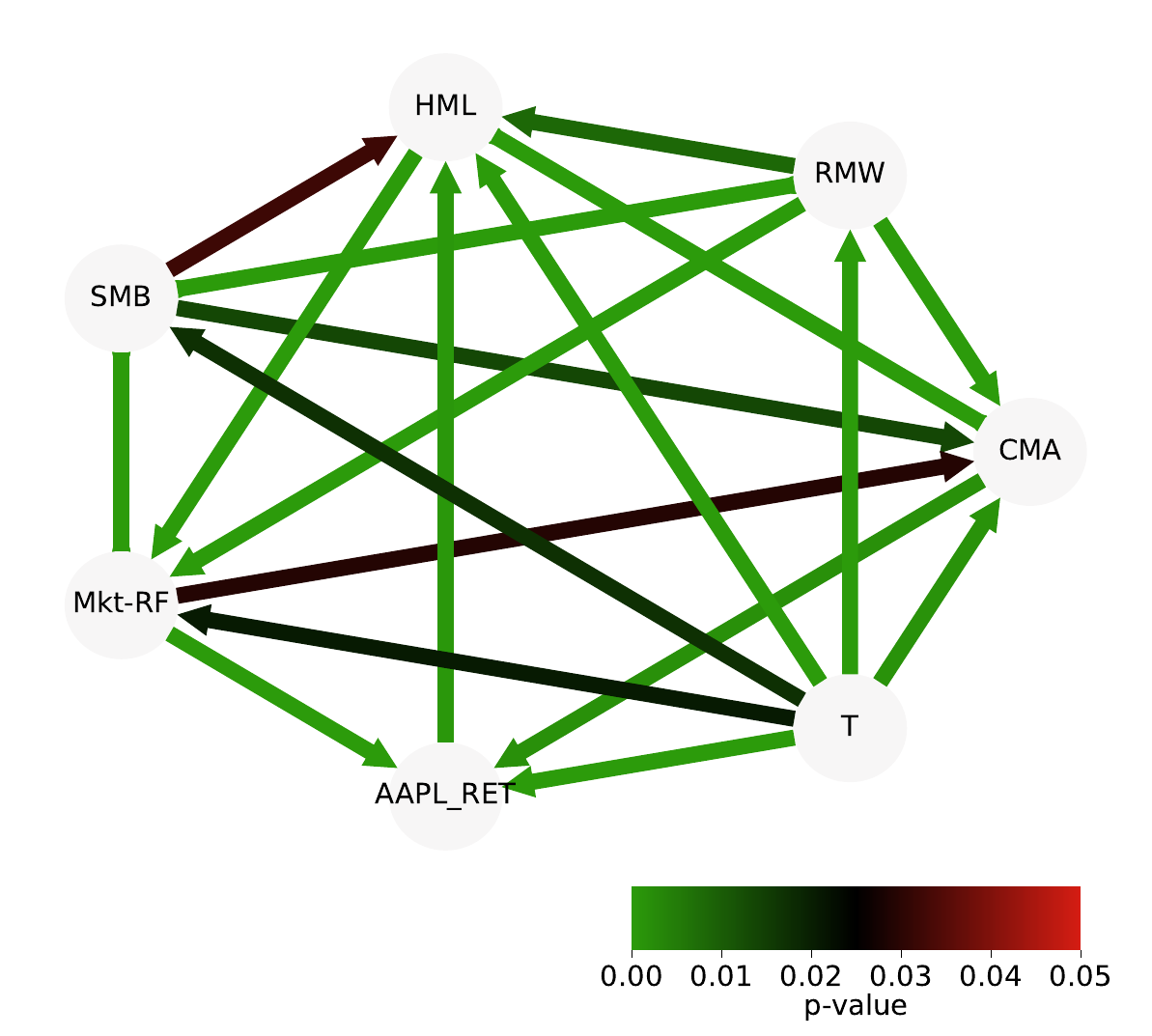}}
    \end{subfigure}\hfill
    \begin{subfigure}{0.45\linewidth}
    \centerline{\includegraphics[width=\linewidth]{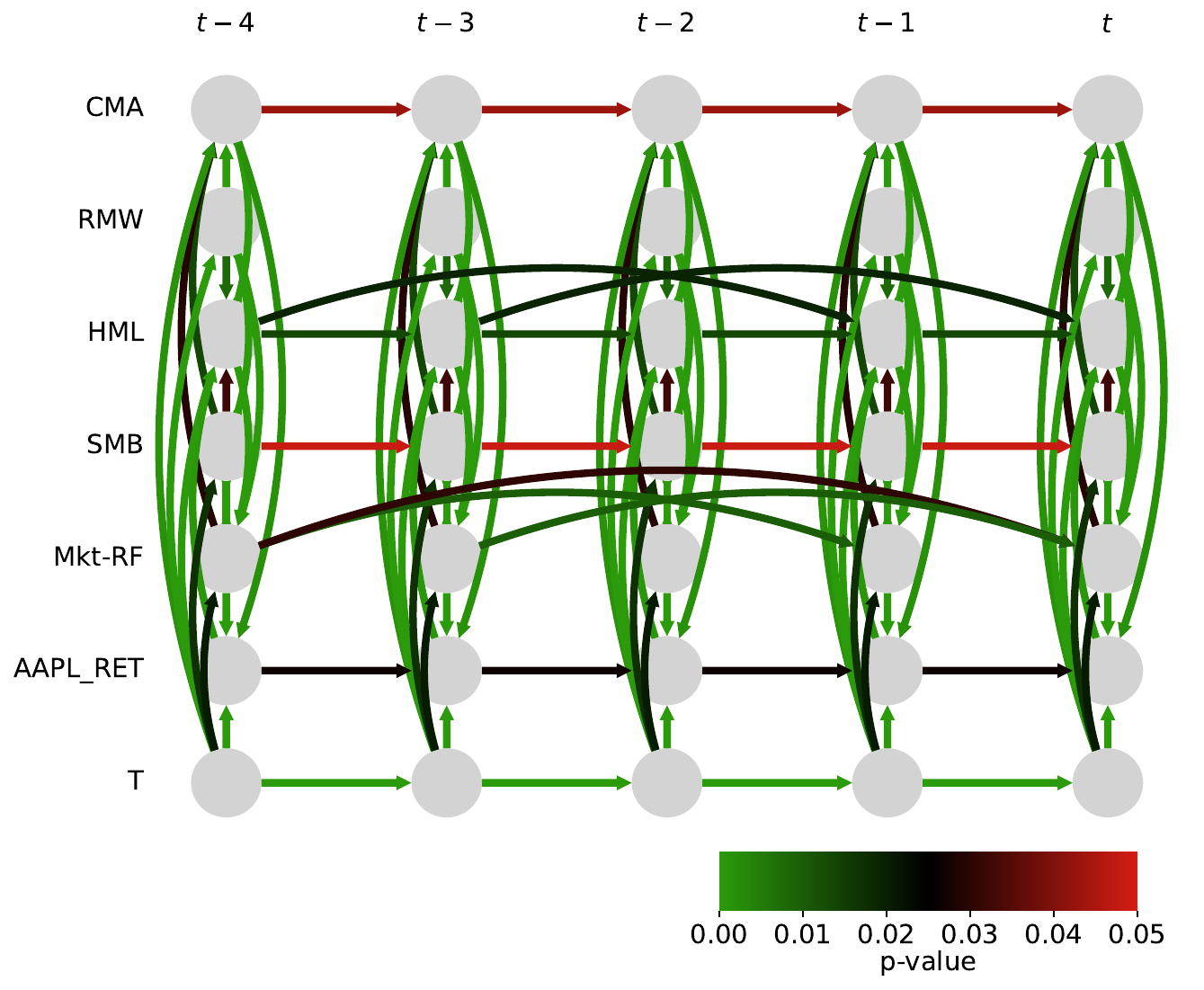}}
    \end{subfigure}
    \caption{Results of CDNOTS KCIT run from beginning of 2000 to end of 2022. \textbf{Left} Contemporaneous graph \textbf{Right} Full time series graph}\label{fig:factors_full}
\end{minipage}
\end{figure}
\begin{figure}[!bt]
\begin{minipage}{\linewidth}
    \begin{subfigure}{0.45\linewidth}
        \centerline{\includegraphics[width=\linewidth]{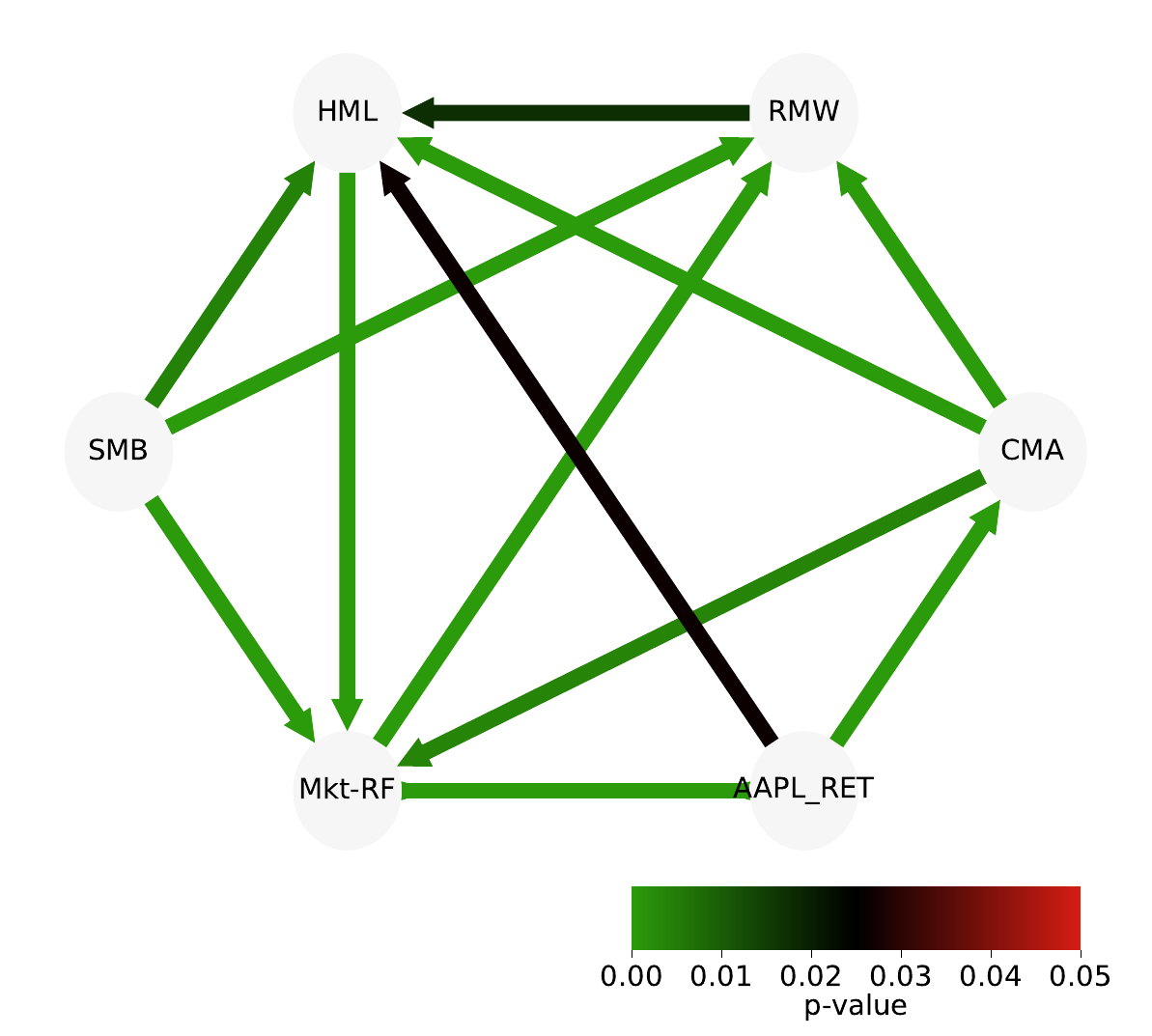}}
    \end{subfigure}\hfill
    \begin{subfigure}{0.45\linewidth}
    \centerline{\includegraphics[width=\linewidth]{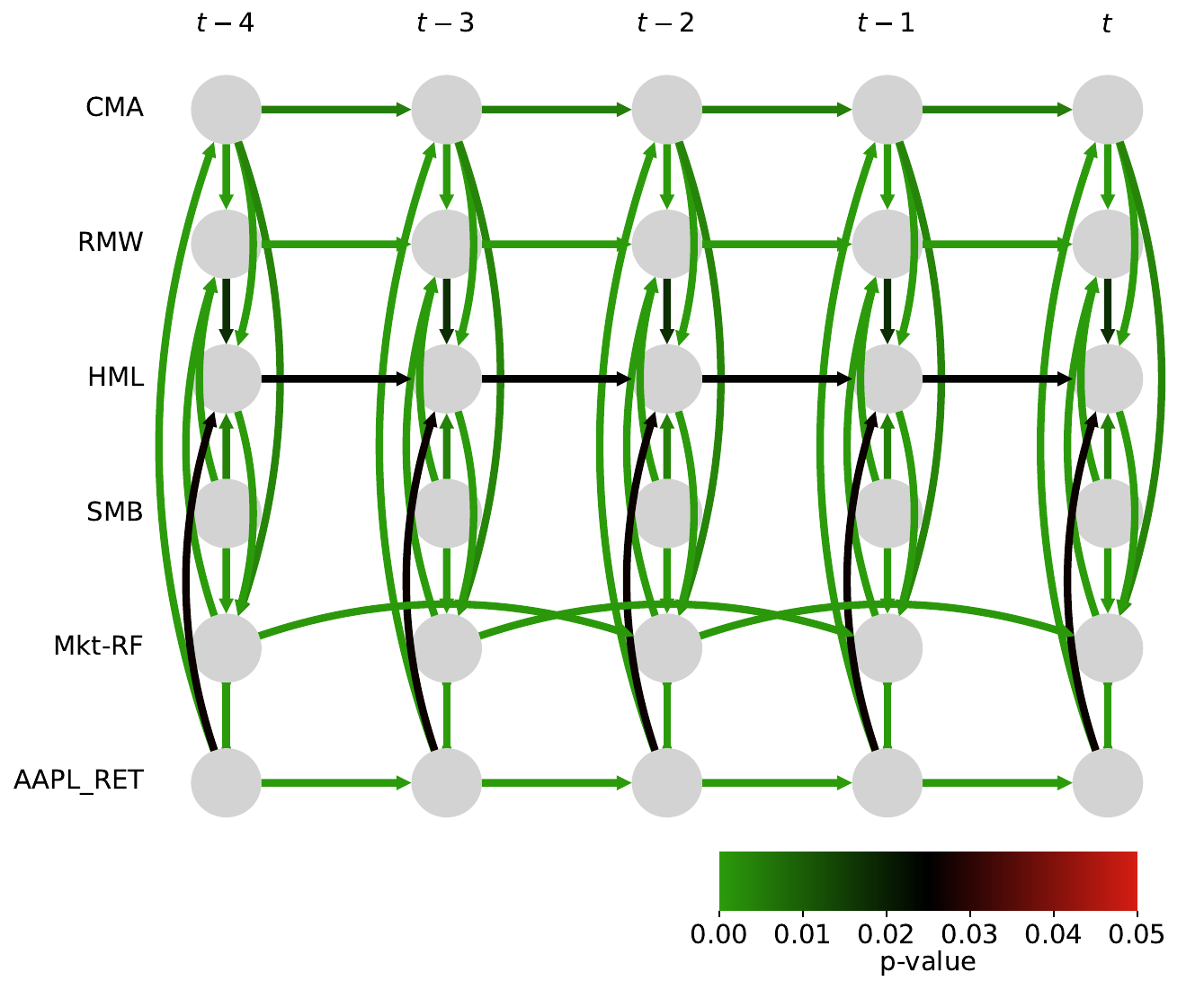}}
    \end{subfigure}
    \caption{Results of CDNOTS KCIT run from beginning of 2000 to end of 2022 without the time-indexed node. \textbf{Left} Contemporaneous graph \textbf{Right} Full time series graph}\label{fig:factors_full_no_t}
\end{minipage}
\end{figure}
\paragraph{Evaluating Assumptions} 

We note that the fact that there is a relationship between $T$ and all the Fama-French factors (nonstationarity) implies that any causal discovery algorithm assuming stationarity would give invalid results. Similarly, we find statistically significant lagged relationships. Further, for our linearity test, we test the linearity of Apple's returns and each parent factor (CMA and Mkt-RF) and find the linearity hypothesis can be rejected for both factors; in other words, assuming linearity in the relationships between variables would be incorrect and thus could lead to incorrectly identified causal relationships.
%%%%%%
\subsection{Economic Data}

\paragraph{Data} For our second case study, we analyze the relationship between the month-over-month percentage change in Consumer Price Index (CPI), the month-over-month percentage change in Producer Price Index (PPI), and unemployment for the U.S., Japan, Canada, India, Italy, the U.K., and France. We use monthly data from the beginning of 2000 to the end of 2023.
\begin{figure}[!b]
\begin{minipage}{\linewidth}
    \begin{subfigure}{0.45\linewidth}
    \centerline{\includegraphics[width=\linewidth]{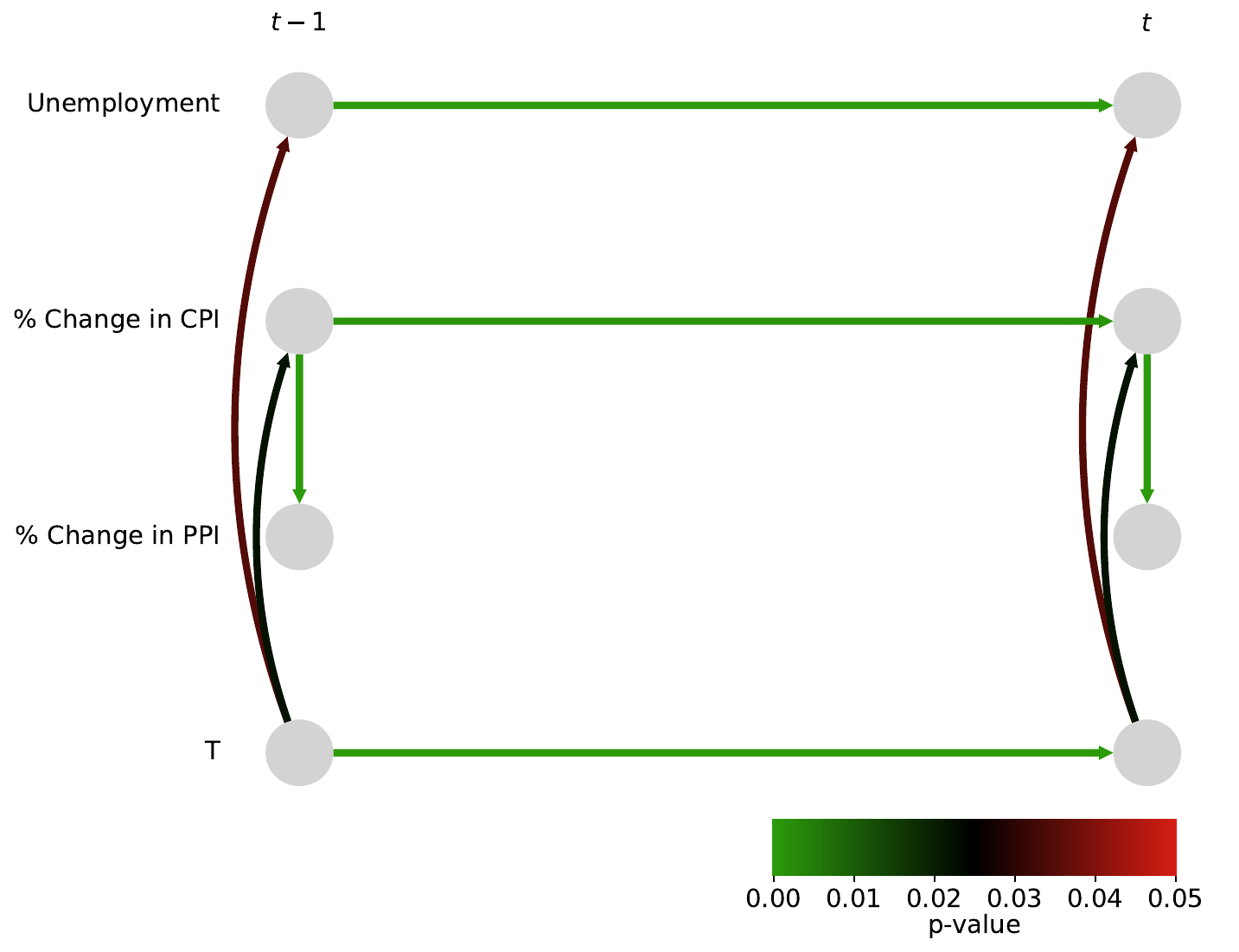}}
    \caption{U.S.}
    \end{subfigure}\hfill
    \begin{subfigure}{0.45\linewidth}
        \centerline{\includegraphics[width=\linewidth]{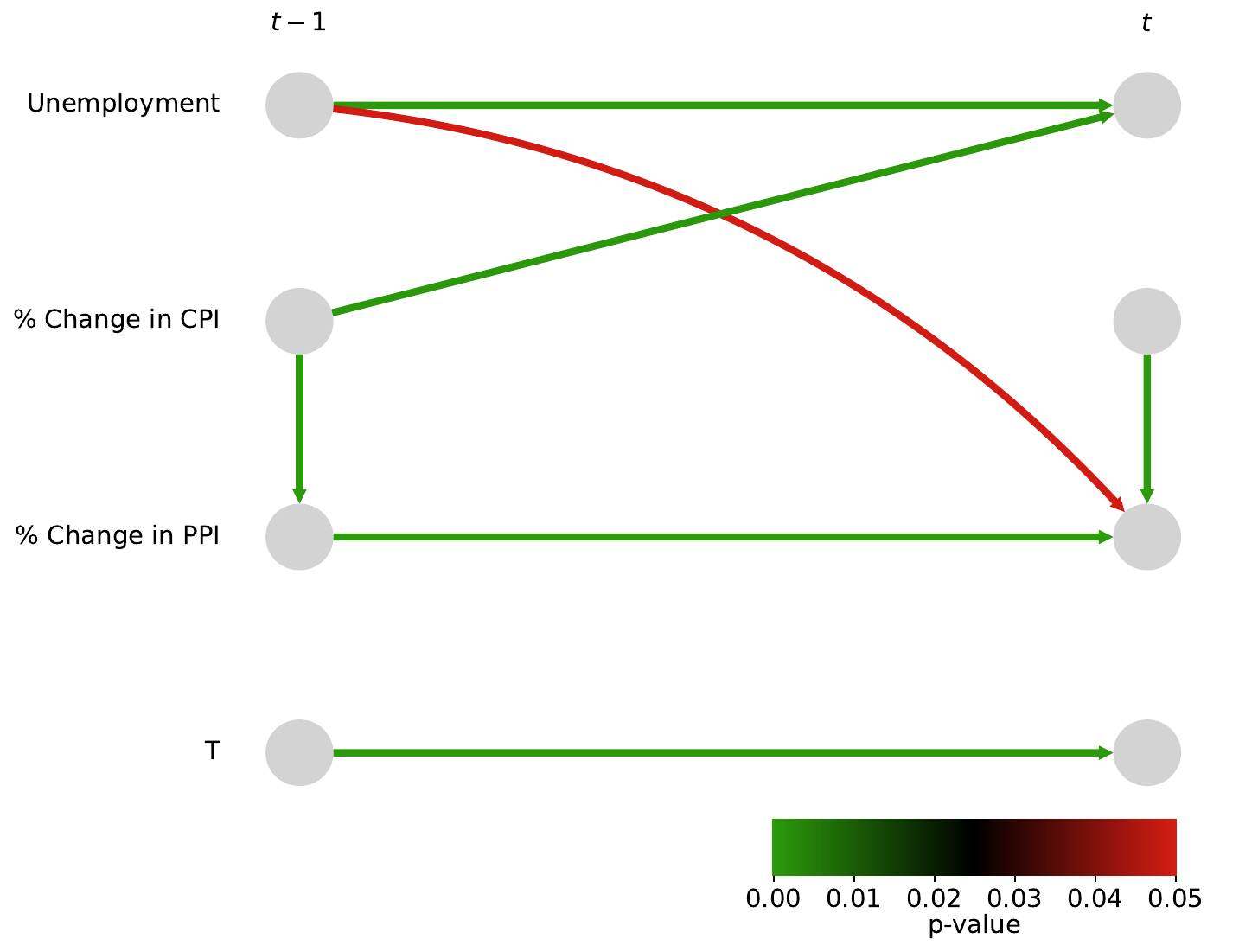}}
    \caption{Canada}
    \end{subfigure}
    
    \begin{subfigure}{0.45\linewidth}
        \centerline{\includegraphics[width=\linewidth]{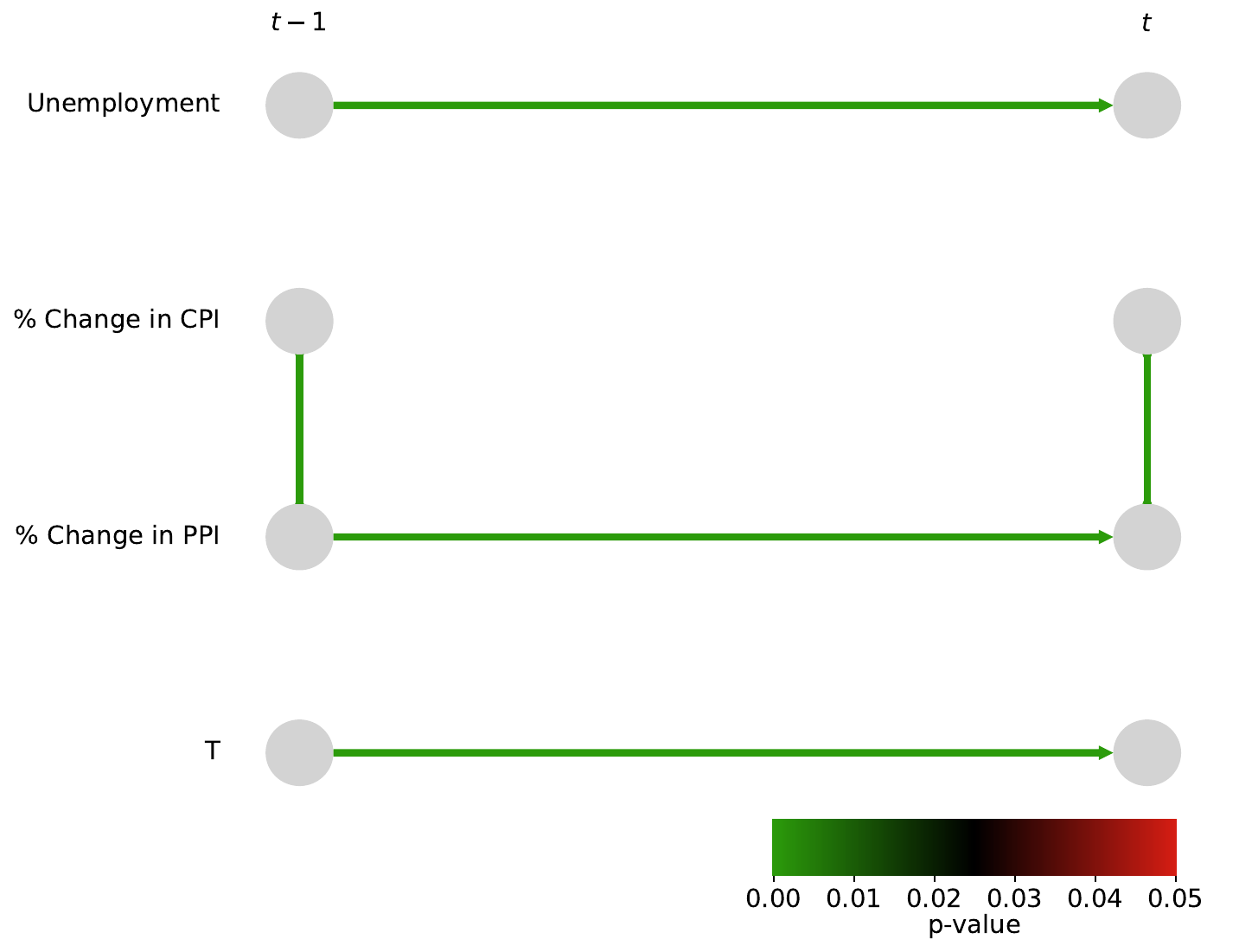}}
    \caption{Japan}
    \end{subfigure}\hfill
    \begin{subfigure}{0.45\linewidth}
    \centerline{\includegraphics[width=\linewidth]{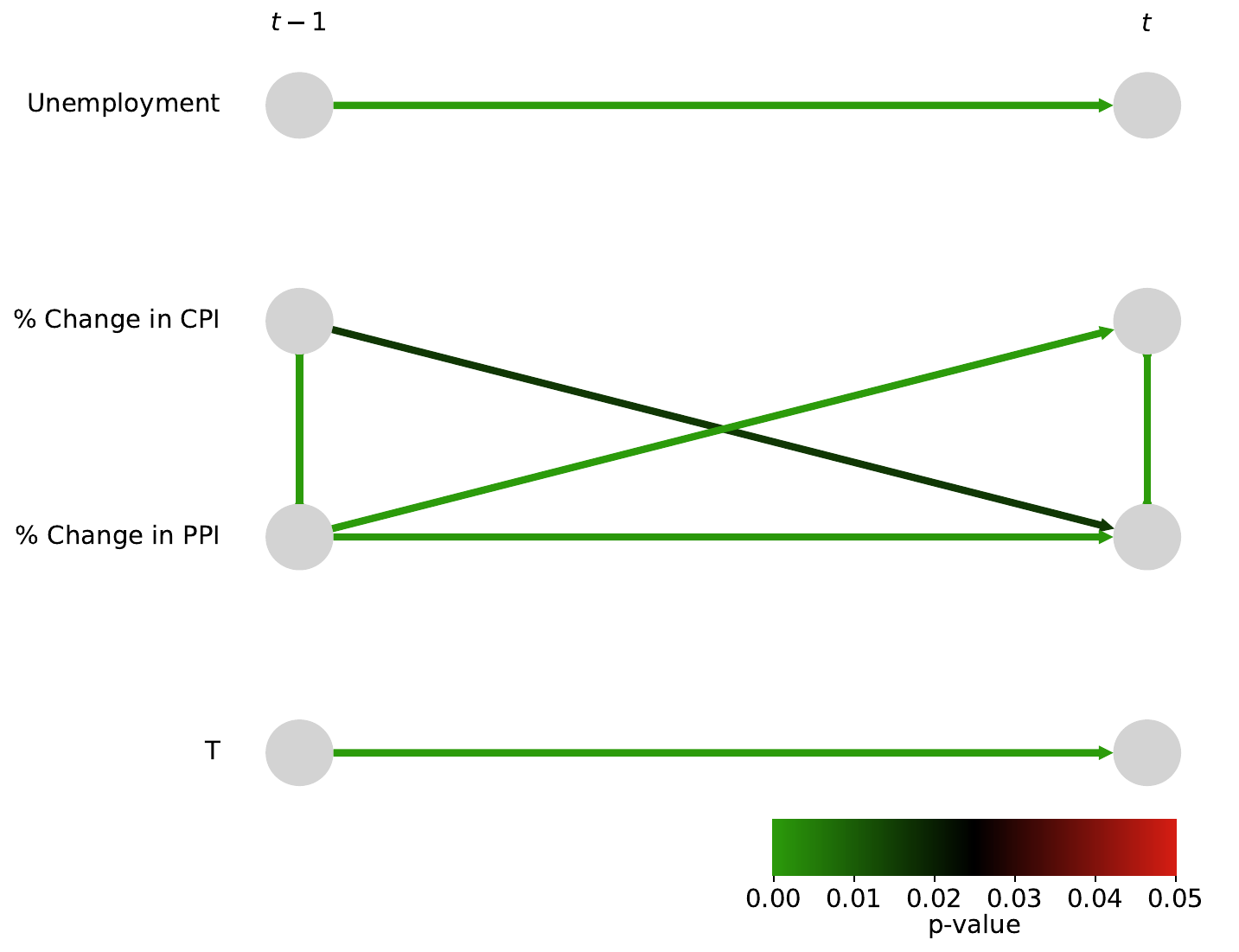}}
    \caption{France}
    \end{subfigure}
    
    \begin{subfigure}{0.45\linewidth}
        \centerline{\includegraphics[width=\linewidth]{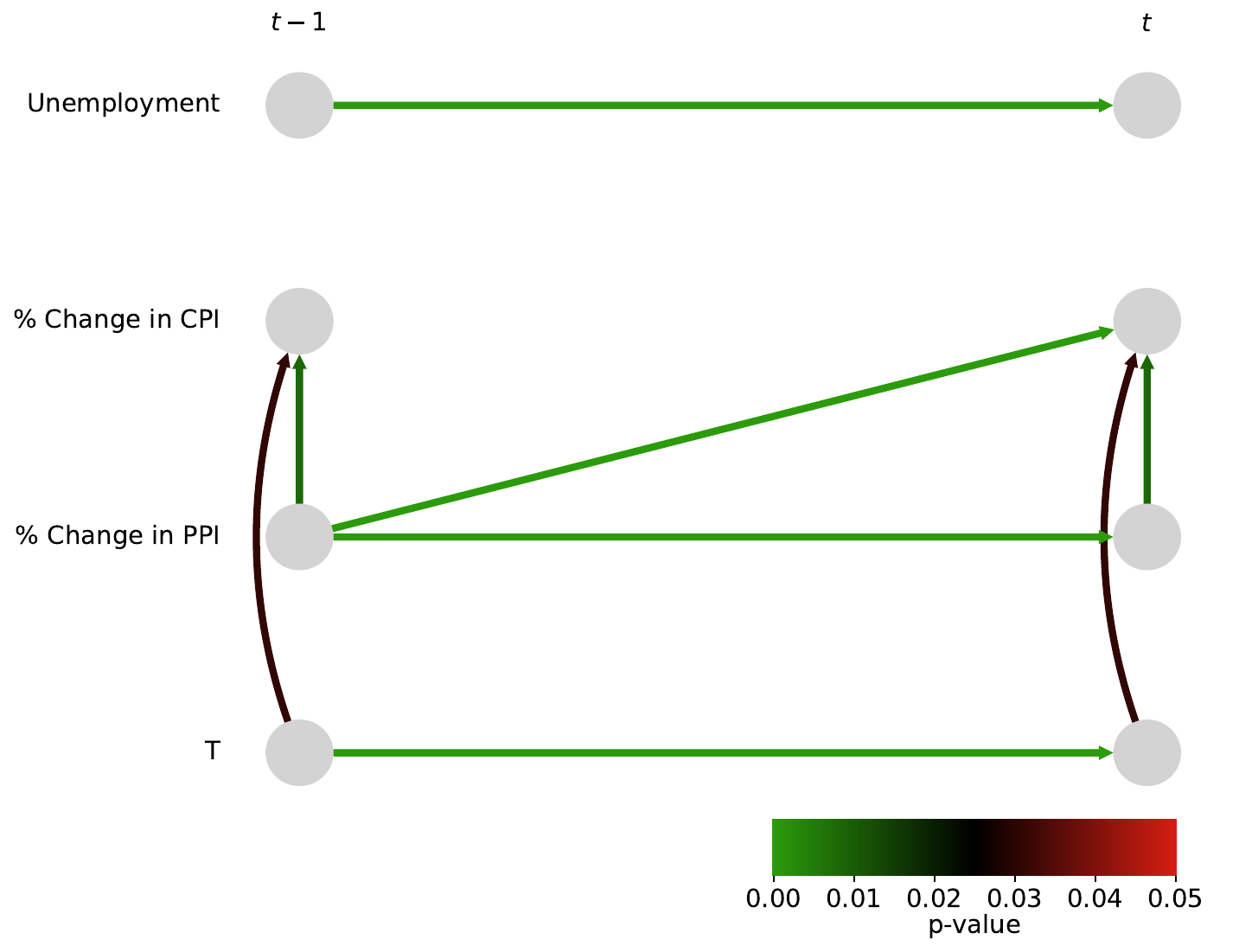}}
    \caption{United Kingdom}
    \end{subfigure}\hfill
    \begin{subfigure}{0.45\linewidth}
        \centerline{\includegraphics[width=\linewidth]{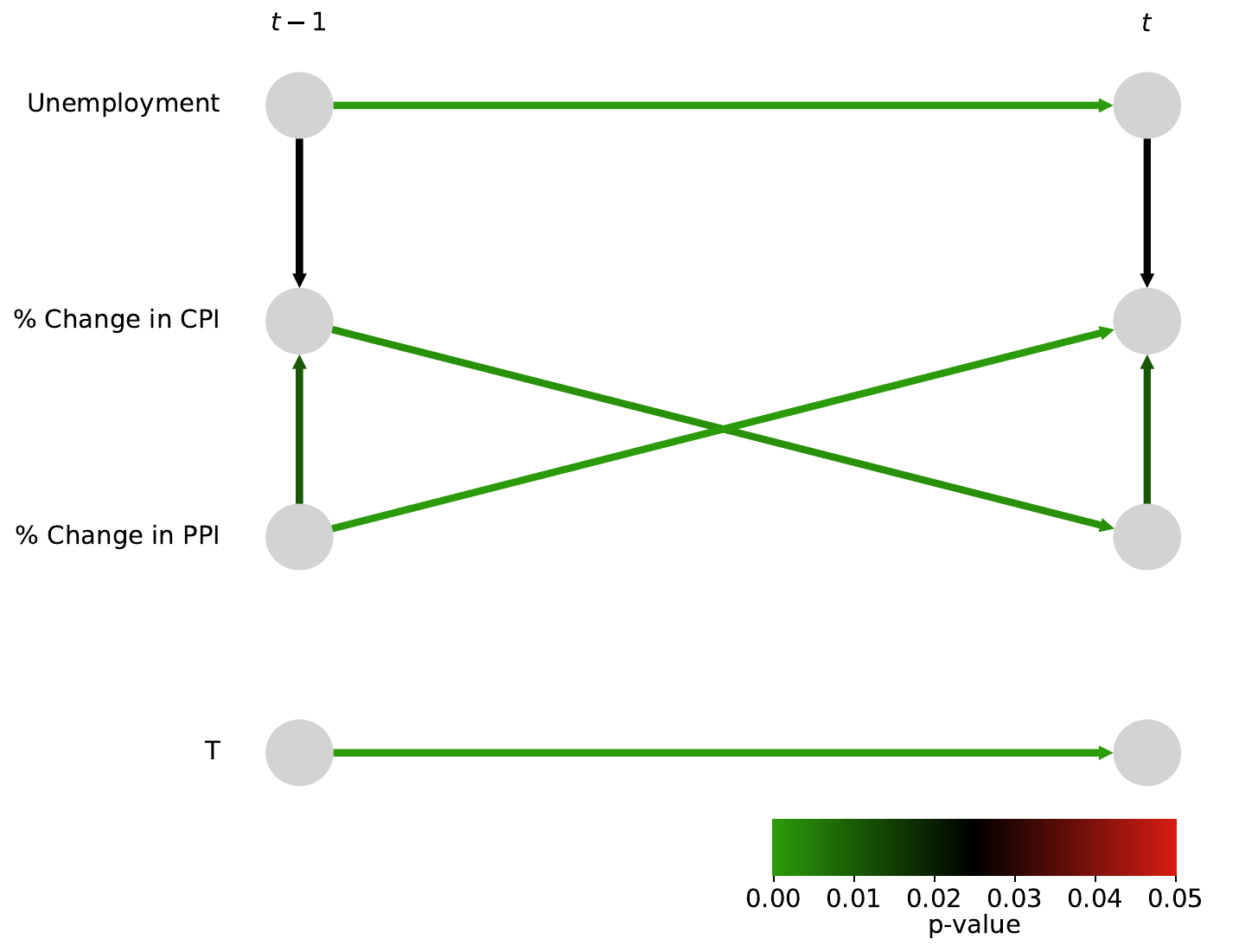}}
    \caption{Italy}
    \end{subfigure}
    \caption{Results of CDNOTS ParCorr ran on six countries' economic variables. }\label{fig:eco_country_wise}
\end{minipage}
\end{figure}
\paragraph{Results and Discussion} In \myref{Figure}{fig:eco_country_wise}, we run CDNOTS with ParCorr on the U.S., Japan, Canada, Italy, France, and the U.K.; we use ParCorr since we have no more than 300 observations per country. We can see that, while the graphs are not identical, they are quite similar, especially if we ignore edges with a higher p-value. Further, we notice the three graphs for France, the U.K., and Italy are very similar, possibly due to their geographic and economic proximity. We repeat the experiment with all countries using KCIT (\myref{Figure}{fig:eco_full}); further, we introduce another static variable (Country) into the graph. Similar to the previous case study in which the graph over all time periods is akin to a superset of the edges of individual time periods, we see a similar behavior here. One difference we see is a new link from $T$ to the percentage change in PPI which is not in any country-specific causal graphs; we interpret this as finding a new relationship through increasing the statistical power by pooling the data.  Note that Country does not point to any other variable suggesting that the behavior across countries is independent of the country; this observation could be used to justify training a single econometric model across countries.  
\begin{figure}[!bt]
\begin{center}
    \begin{subfigure}{0.45\linewidth}
    \centerline{\includegraphics[width=\linewidth]{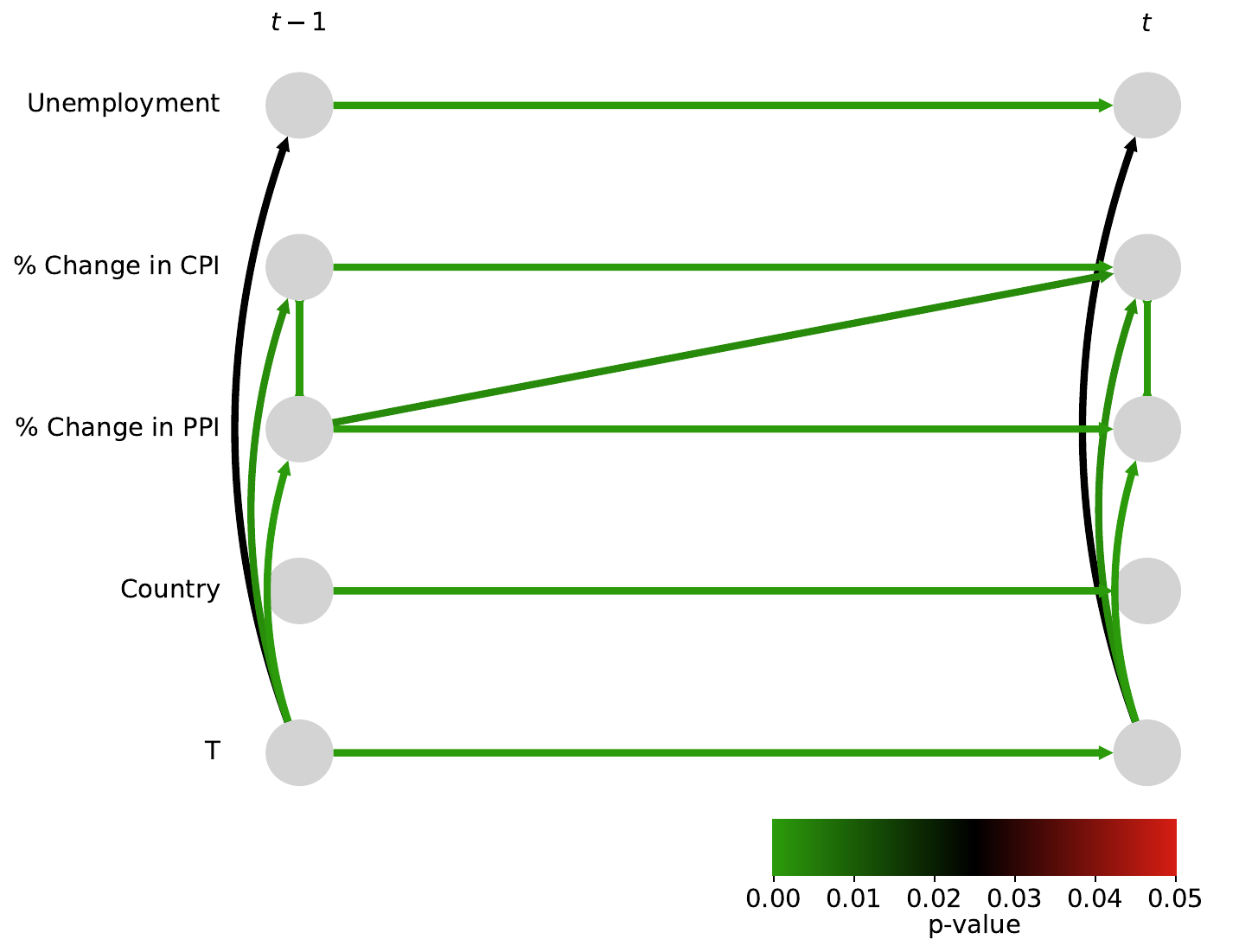}}
    \end{subfigure}
    \caption{Results of CDNOTS KCIT ran on the economic variables of multiple countries. }\label{fig:eco_full}
\end{center}
\end{figure}
\paragraph{Evaluation Assumptions} Similar to the previous case study, we see statistically significant lagged
relationships and statistically significant nonstationarity, implying we would not be able to trust the
results of an algorithm that cannot handle these types of properties. For our linearity test, we test the linearity of the percentage change of CPI and each parent node (both including and excluding the contemporaneous percentage change in PPI since its direction was not inferred). We find that the only time the linearity hypothesis cannot be rejected is when using the contemporaneous percentage change in PPI as a parent and checking the linearity with the one-lag percentage change of CPI. We note that this is inconsistent with our usage of ParCorr in \myref{Figure}{fig:eco_country_wise} since ParCorr is a linear-based CI test. In \myref{Appendix}{sec:addl_eco_plots}, we show the results of using RCoT where we find qualitatively that the graphs are similar.

%%%%%%
\subsection{Company Financials and Returns}

\begin{figure}[!bt]
\begin{minipage}{\linewidth}
    \begin{subfigure}{0.45\linewidth}
    \centerline{\includegraphics[width=\linewidth]{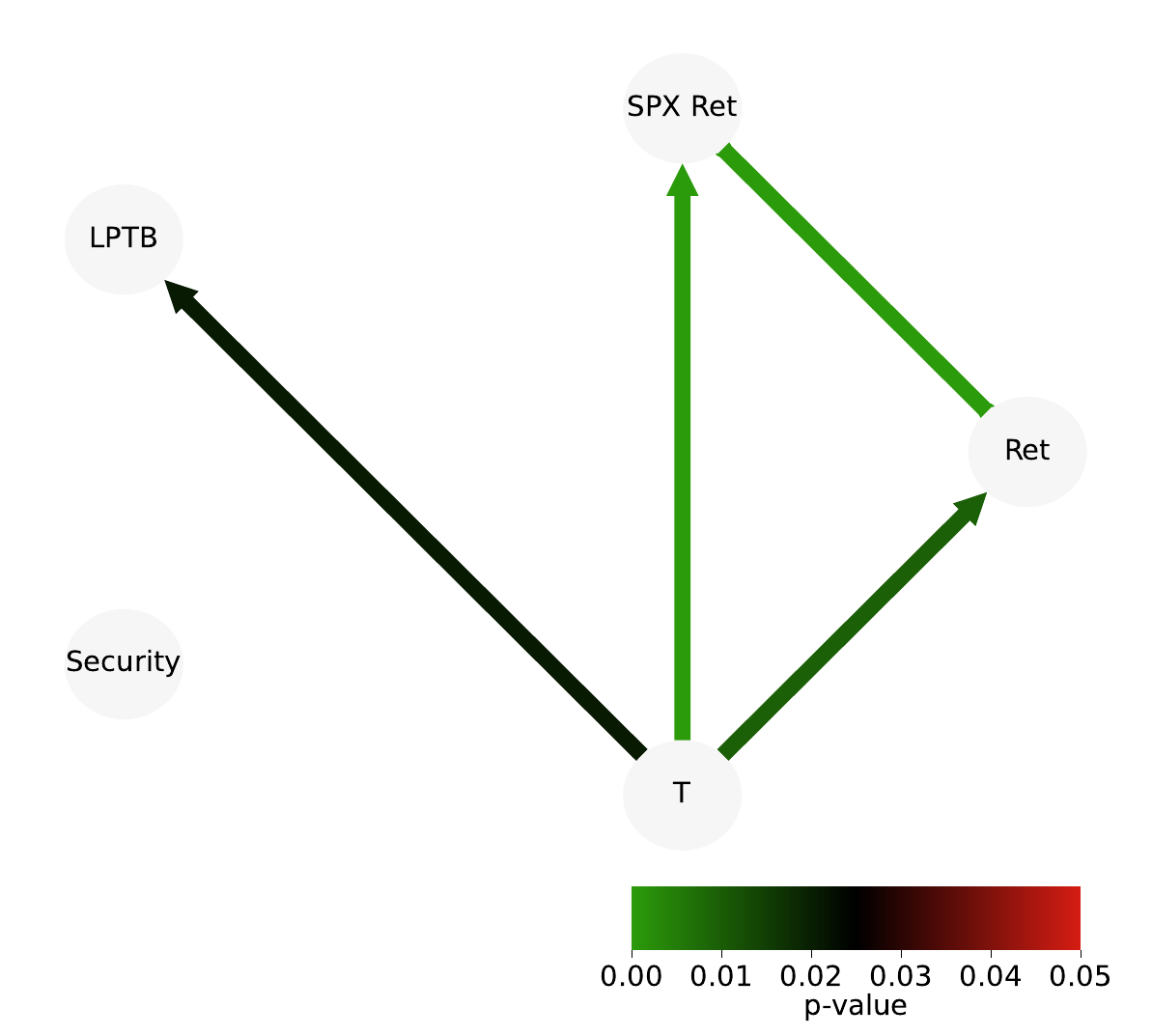}}
    \caption{Contemporaneous Effects}\label{fig:cofi_no_demean_static}
    \end{subfigure}\hfill
    \begin{subfigure}{0.45\linewidth}
    \centerline{\includegraphics[width=\linewidth]{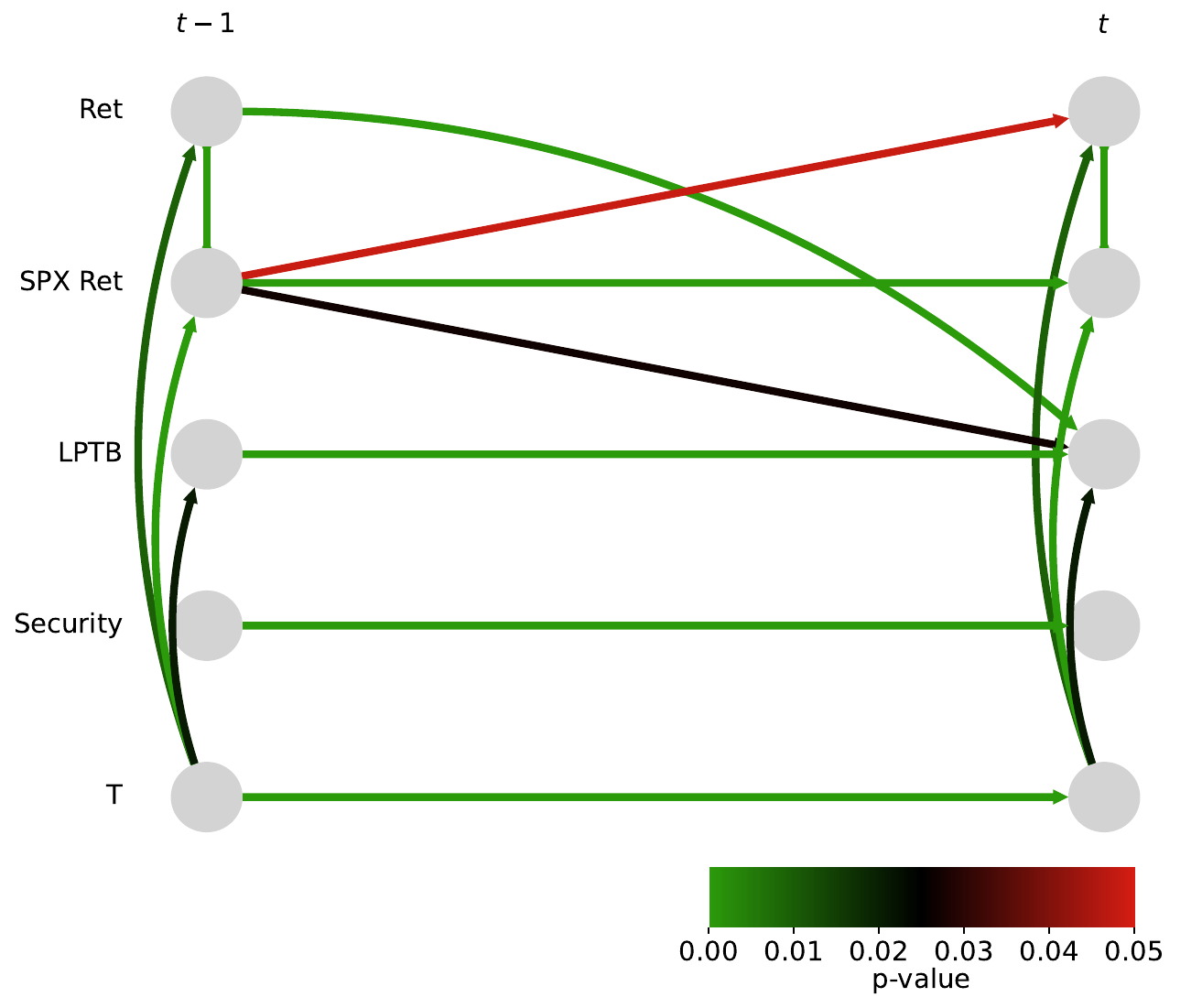}}
    \caption{Lagged Effects}\label{fig:cofi_no_demean_ts}
    \end{subfigure}

    \begin{subfigure}{0.45\linewidth}
    \centerline{\includegraphics[width=\linewidth]{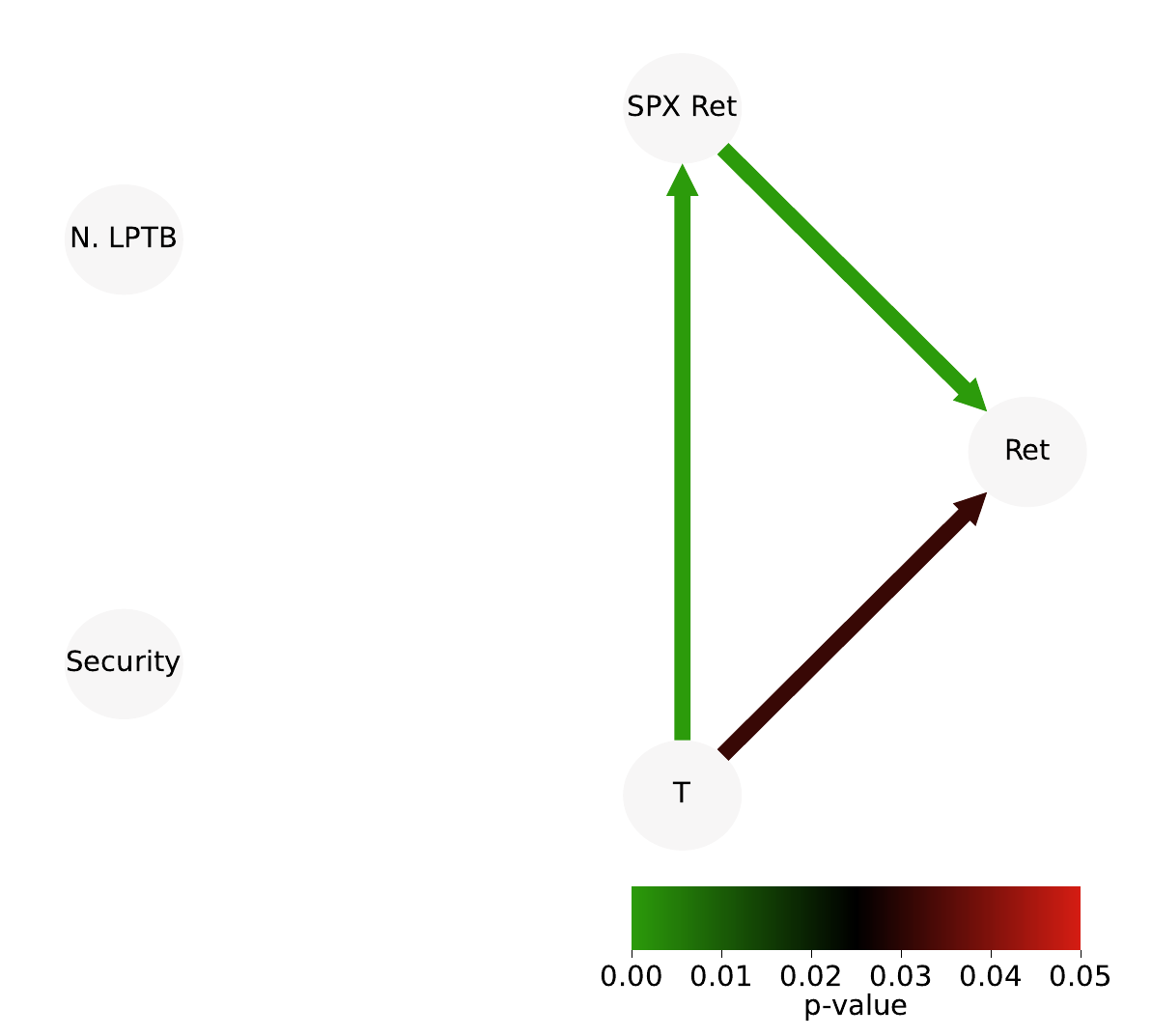}}
    \caption{Contemporaneous Effects}\label{fig:cofi_demean_static}
    \end{subfigure}\hfill
    \begin{subfigure}{0.45\linewidth}
    \centerline{\includegraphics[width=\linewidth]{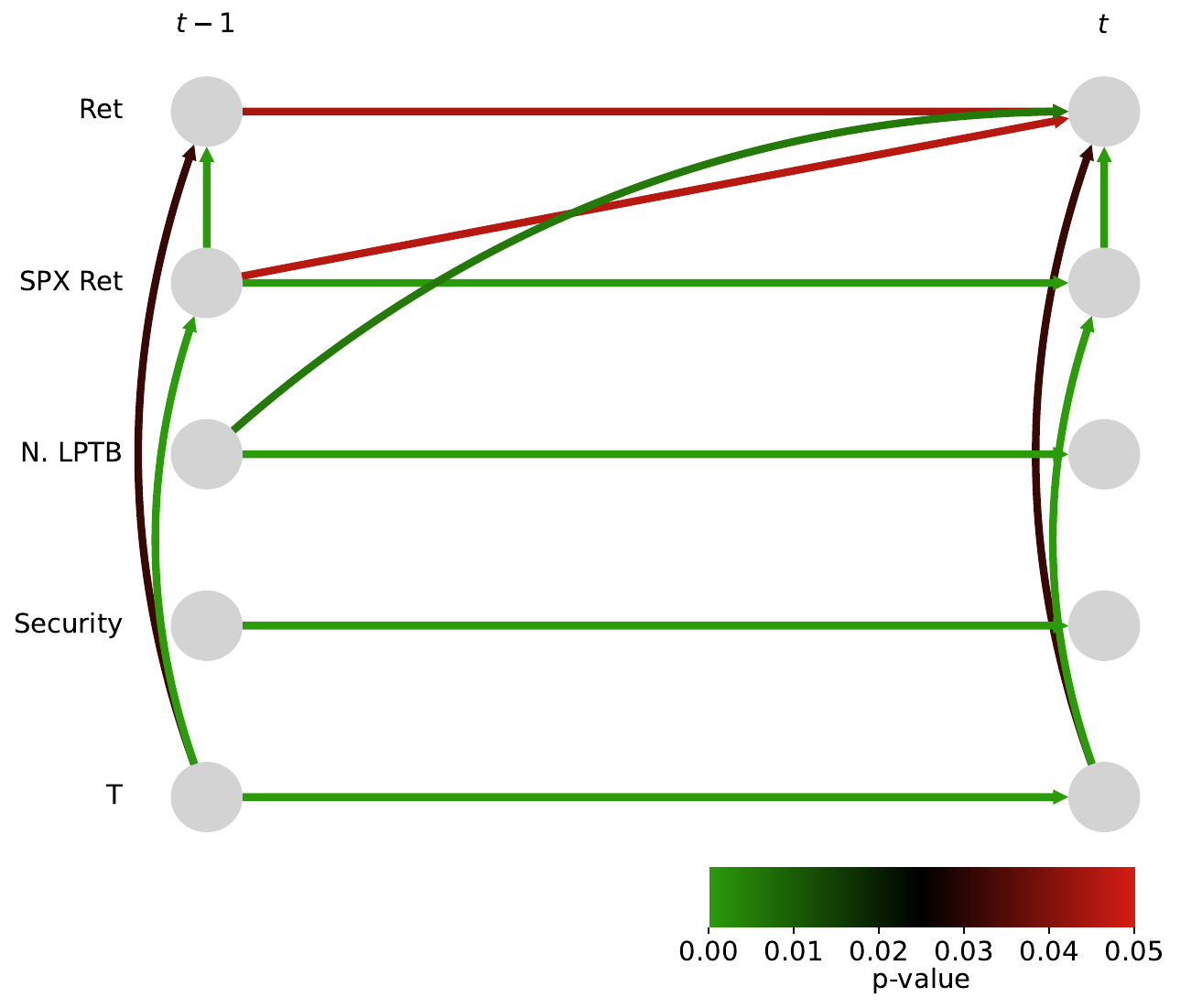}}
    \caption{Lagged Effects}\label{fig:cofi_demean_ts}
    \end{subfigure}

    \caption{Results of CDNOTS KCIT run on financials data for log Price-to-Book ratio (LPTB) and stock returns (Ret). Normalized LPTB (N. LPTB) refers to normalizing the LPTB by the mean and standard deviation of the LPTB of financial companies in that quarter.}

\end{minipage}
\end{figure}
\paragraph{Data} For our third case study, we analyze the relationship between the Price-to-Book ratio and returns of financial companies in the S\&P 500. We use data from the beginning of 2010 to the end of 2023. We utilize reported quarterly numbers, where we set the date to be the end of each quarter. To ensure no look-ahead bias when defining the Price-to-Book ratio, for any given date (i.e., end of quarter), we use the Price-to-Book ratio available that day. We join the next quarter returns with this company's financial information.  For our experiments, we use the log Price-to-Book ratio (LPTB) as well as the normalized LPTB, which we define as the LPTB normalized by the mean and standard deviation of the LPTB of financial companies in that quarter.

\paragraph{Results and Discussion} In \myref{Figure}{fig:cofi_no_demean_ts} and \myref{Figure}{fig:cofi_demean_ts}, we compare the graphs we get running CDNOTS using KCIT using normalized LPTB instead of LPTB. We find that the nonstationarity is removed (T does not point to N. LPTB). Further, we see that the previous quarter's normalized LPTB points to the next quarter's returns with a reasonable p-value suggesting that its inclusion in the graph is not purely a function of the threshold we chose. In \myref{Figure}{fig:lbtp_to_returns}, we compare the next quarter returns and the normalized LPTB and see that there is a small trend in the mean (the black circles). To this end, we create a simple trading strategy where we invest in the subset of the financial securities whose LPTB is greater than some threshold. In \myref{Figure}{fig:cofi_strategy}, we see this simple approach outperforms a market-cap weighted strategy;  prior work \citep{monge23} has found that growth stocks have been outperforming value stocks. Importantly, we note that the goal of this is to not introduce a new signal but to show how causal discovery can be a tool in searching for statistically significant relations which, if the assumptions are true for the data, can be interpreted as causal.
\begin{figure}[!bt]
\begin{minipage}{\linewidth}
    \begin{subfigure}{0.45\linewidth}
    \centerline{\includegraphics[width=\linewidth]{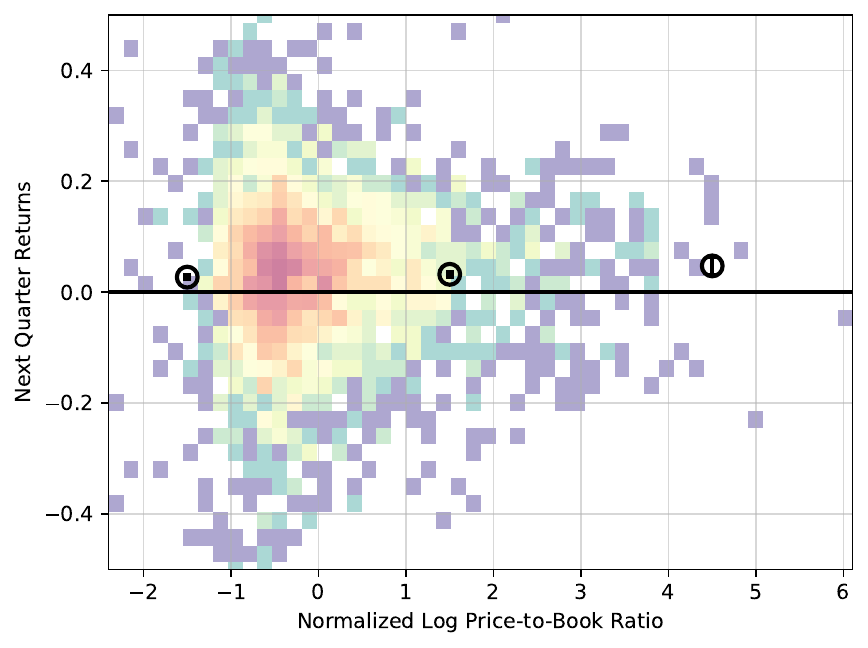}}
    \caption{Comparison of Returns and N. LPTB}\label{fig:lbtp_to_returns}
    \end{subfigure}\hfill
    \begin{subfigure}{0.45\linewidth}
    \centerline{\includegraphics[width=\linewidth]{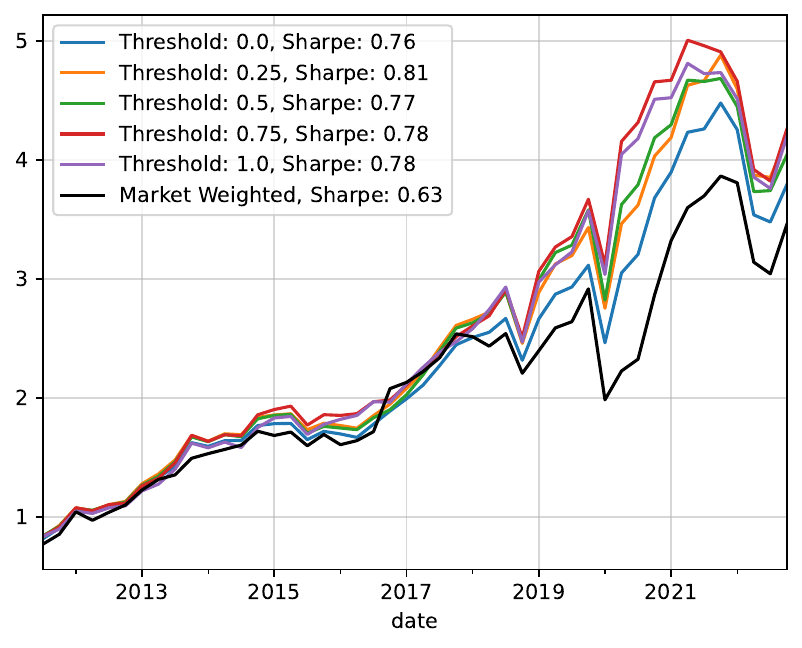}}
    \caption{Cumulative Returns of Strategies}\label{fig:cofi_strategy}
    \end{subfigure}

    \caption{Testing out predictive power of Normalized LPTB for returns of financial companies. For \myref{Figure}{fig:cofi_strategy}, our strategy is long-ing financial companies that have a normalized LPTB greater than the threshold which we compare against the market-cap weighted portfolio.}
\end{minipage}
\end{figure}
\paragraph{Evaluating Assumptions} Similar to previous case studies, we see statistically significant lagged relationships and statistically significant nonstationarity, implying we would not be able to trust the results of an algorithm that cannot handle these types of properties. For our linearity test, we test the linearity of the normalized LPTB and returns conditional on all other parent nodes of returns and find the p-value to be 0.0 (i.e., we can reject the null hypothesis that the relationship is linear).
%%%%%%%%%%%%%%%%%%%%%%%%%%%%%%%%%
\section{Conclusion and Future Work}
In this paper, we developed a novel algorithm, Constraint-based Causal Discovery from Nonstationary Time Series (CDNOTS), which is nonparametric, can handle nonstationarity, and is able to detect both lagged and contemporaneous causalities. We showed using simulated data that, for effective use of CDNOTS, one should use ParCorr for low data regimes (less than 200 data points) and KCIT or RCoT for higher data regimes.

We applied CDNOTS to approximately two decades of Fama-French factors and Apple's returns and found that Apple's returns are nonstationary even after accounting for the Fama-French factors, implying that there is a need to continually update the factor exposures of Apple through time.

Further, we found the causal relationship of economic factors (specifically unemployment, CPI, and PPI) tends to be similar across countries, especially for those geographically and economically similar. This finding implies that we should be able to create stronger models in economics by training a single model across different countries.

Finally, we apply CDNOTS to company financials and stock returns and find a causal relationship between the Price-to-Book ratio and future returns of financial companies, justifying the common usage of Price-to-Book ratio for investing.

Through our case studies, we show the importance of testing the assumptions of causal discovery algorithms. Specifically, we find in all of our case studies nonstationarity, non-linear relationships, and lagged and contemporaneous causal effects. Prior work often assumes some of these properties are not present in the data and thus would lead to spurious relationships and incorrect conclusions. Hence, our contribution improves over prior work by handling all of these scenarios.

To support reproduction and further work, CDNOTS is available as open-source software in the \texttt{causal-ts} Python library (\url{https://github.com/bloomberg/causal-ts}).

% \acks{Part of this research was done while Agathe Sadeghi was an intern at Bloomberg Quant Research team.}

\newpage

\appendix
%%%%%%%%%%%%%%%%%%%%%%%%%%%%%%%%%
\section{Conditional Independence Tests Hyperparameters}\label{sec:hparams}

For all conditional independence tests in our case studies, we use a threshold of 0.05 for the p-value.

\paragraph{KCIT} For KCIT, we use the RBF kernel where the bandwidth is set to the median distance between the first five hundred data points. 

\paragraph{RCoT} For RCoT, similar to KCIT, we use the RBF kernel where the bandwidth is set to the median distance between the first five hundred data points. Further, we use a fixed number of Fourier features, specifically 25 for $Z$ and 5 for $\ddot{X}$, $X$, and $Y$, similar to \citet{strobl17}.

%%%%%%%%%%%%%%%%%%%%%%%%%%%%%%%%%
\section{Fama-French Factors Case Study with RCoT}\label{sec:fama_rcot}

We repeat the plots from our first case study (\myref{Section}{sec:kcit_ff}) using RCoT as opposed to KCIT. One empirical finding is that, due to the randomness of RCoT, many times there could be an inconsistency between the discovered graph and the CI tests, i.e., the graph shows conditional independences that are not found by the CI tests. We leave it to future work to find how to account for these forms of inconsistencies.

\begin{figure}[!h]
\centering
\begin{minipage}{0.8\linewidth}
    \begin{subfigure}{0.45\linewidth}
        \centerline{\includegraphics[width=\linewidth]{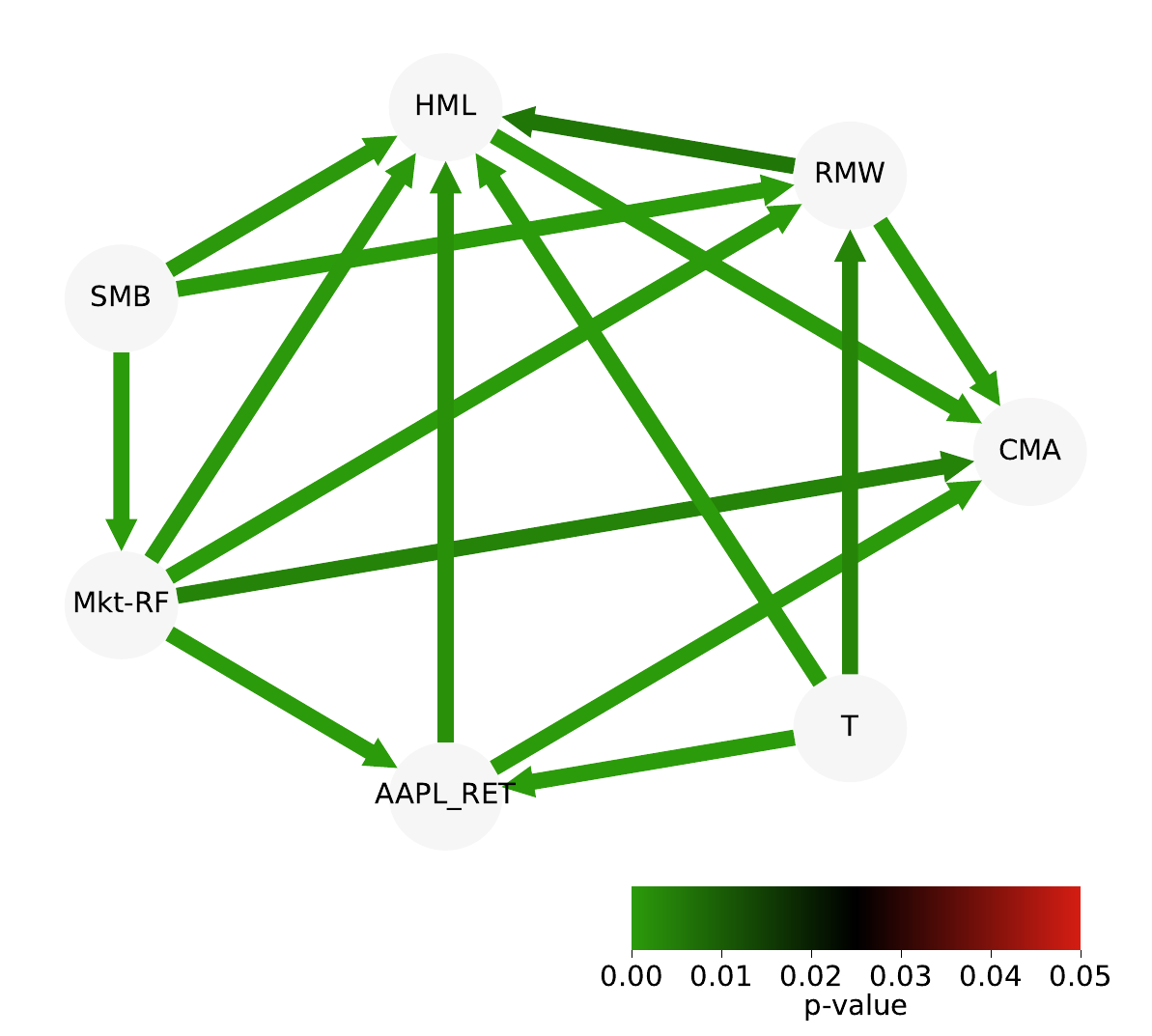}}
    \end{subfigure}\hfill
    \begin{subfigure}{0.45\linewidth}
    \centerline{\includegraphics[width=\linewidth]{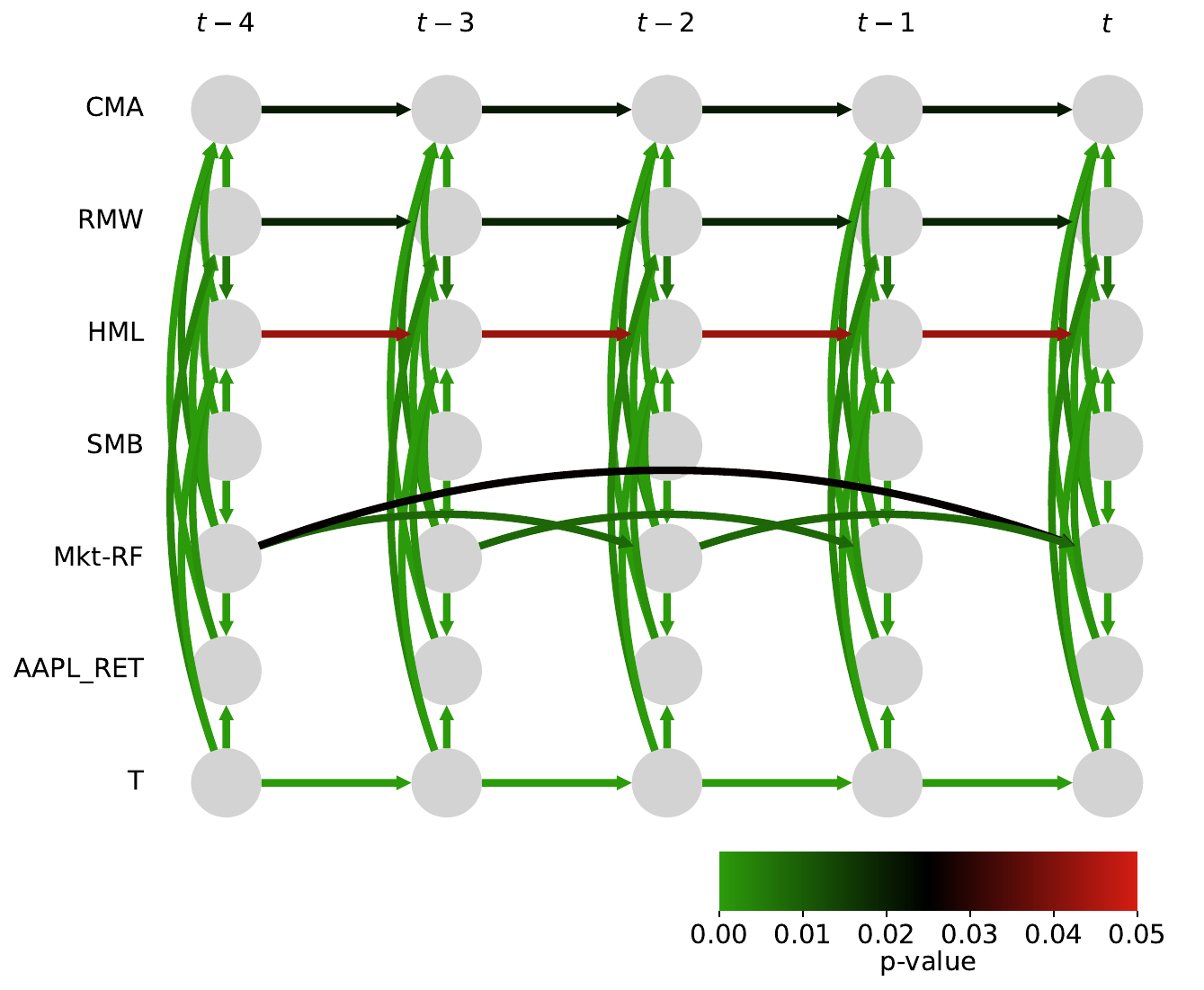}}
    \end{subfigure}
    \caption{Results of CDNOTS RCoT run from beginning of 2000 to end of 2022. }
\end{minipage}
\end{figure}

%%%%%%%%%%%%%%%%%%%%%%%%%%%%%%%%%
\section{Economic Data Case Study with KCIT and RCoT}\label{sec:addl_eco_plots}

Our synthetic data experiments indicate the effectiveness of ParCorr when data is limited, as observed in our country-level experiments. However, in this section, we present results obtained using KCIT and RCoT. Notably, these results appear to lack robustness to specific test used, emphasizing the importance of analyzing countries collectively. Upon comparing KCIT with RCoT, we note several similarities in edges, barring a few exceptions. For instance, RCoT suggests that Unemployment and PPI are stationary, whereas CPI lacks a one-lag relationship with itself. Additionally, it indicates an additional edge from Unemployment in the previous month to the current month's CPI.
\begin{figure}[!hb]
\begin{minipage}{\linewidth}
    \begin{subfigure}{0.45\linewidth}
    \centerline{\includegraphics[width=\linewidth]{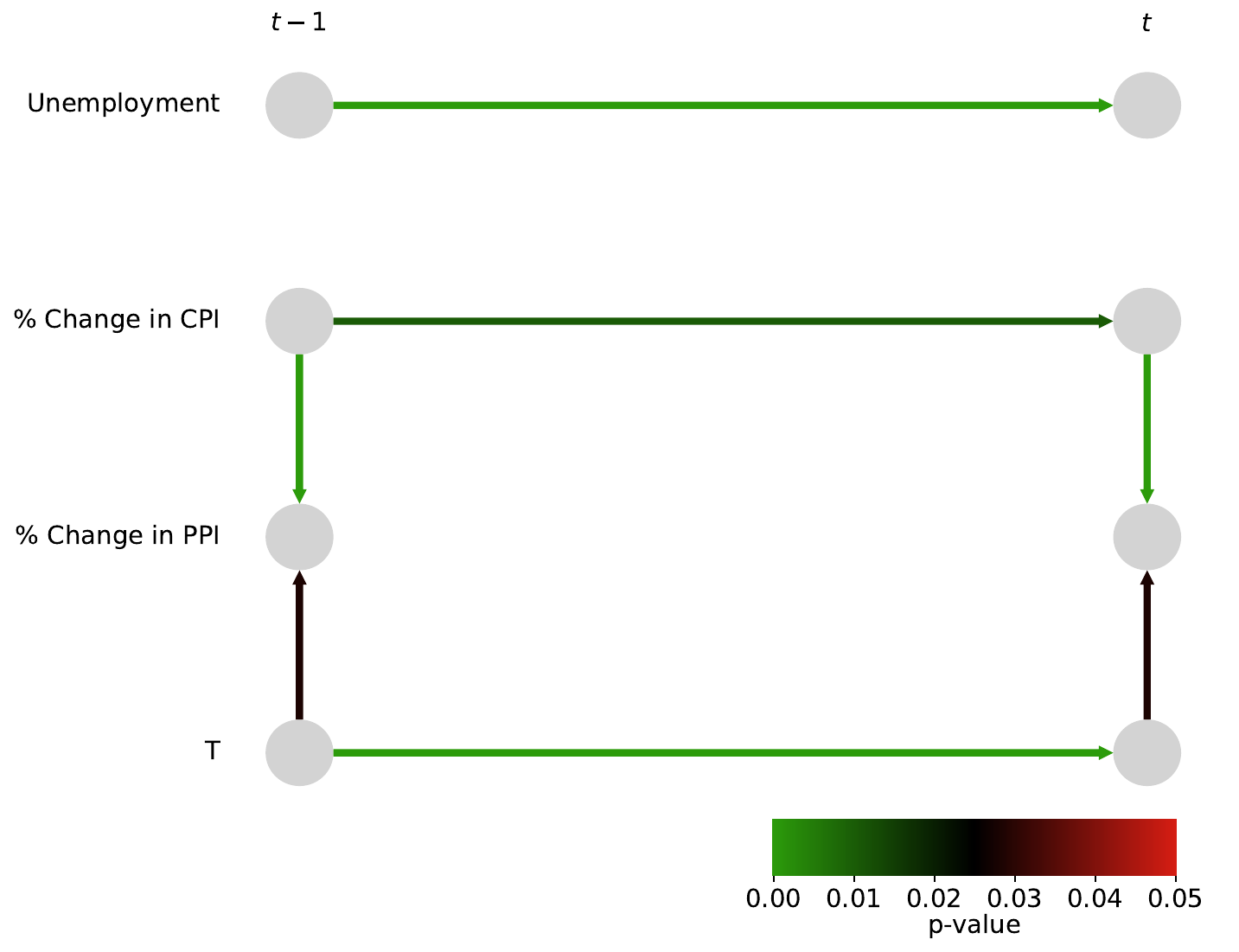}}
    \caption{U.S.}
    \end{subfigure}\hfill
    \begin{subfigure}{0.45\linewidth}
        \centerline{\includegraphics[width=\linewidth]{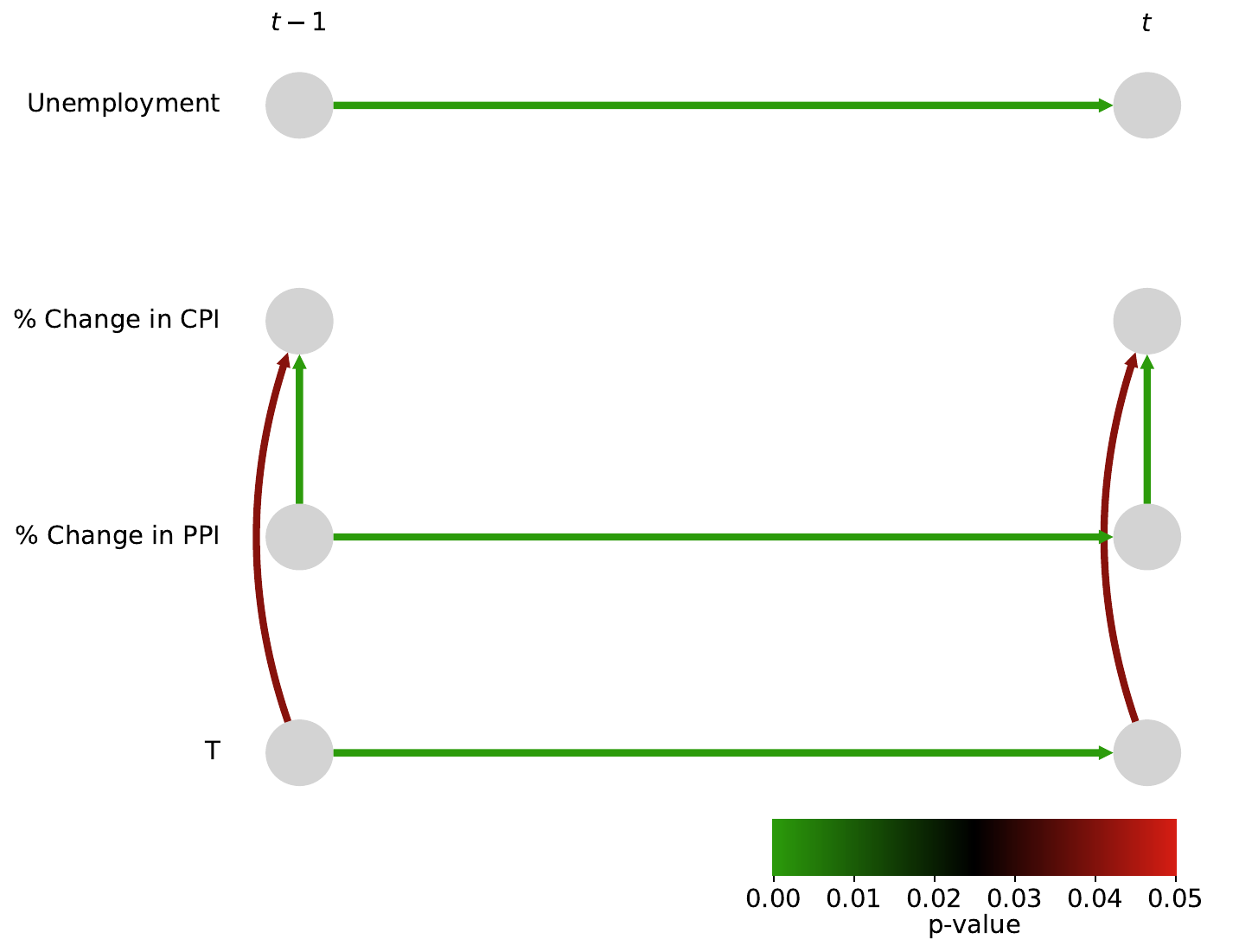}}
    \caption{Canada}
    \end{subfigure}
    \begin{subfigure}{0.45\linewidth}
        \centerline{\includegraphics[width=\linewidth]{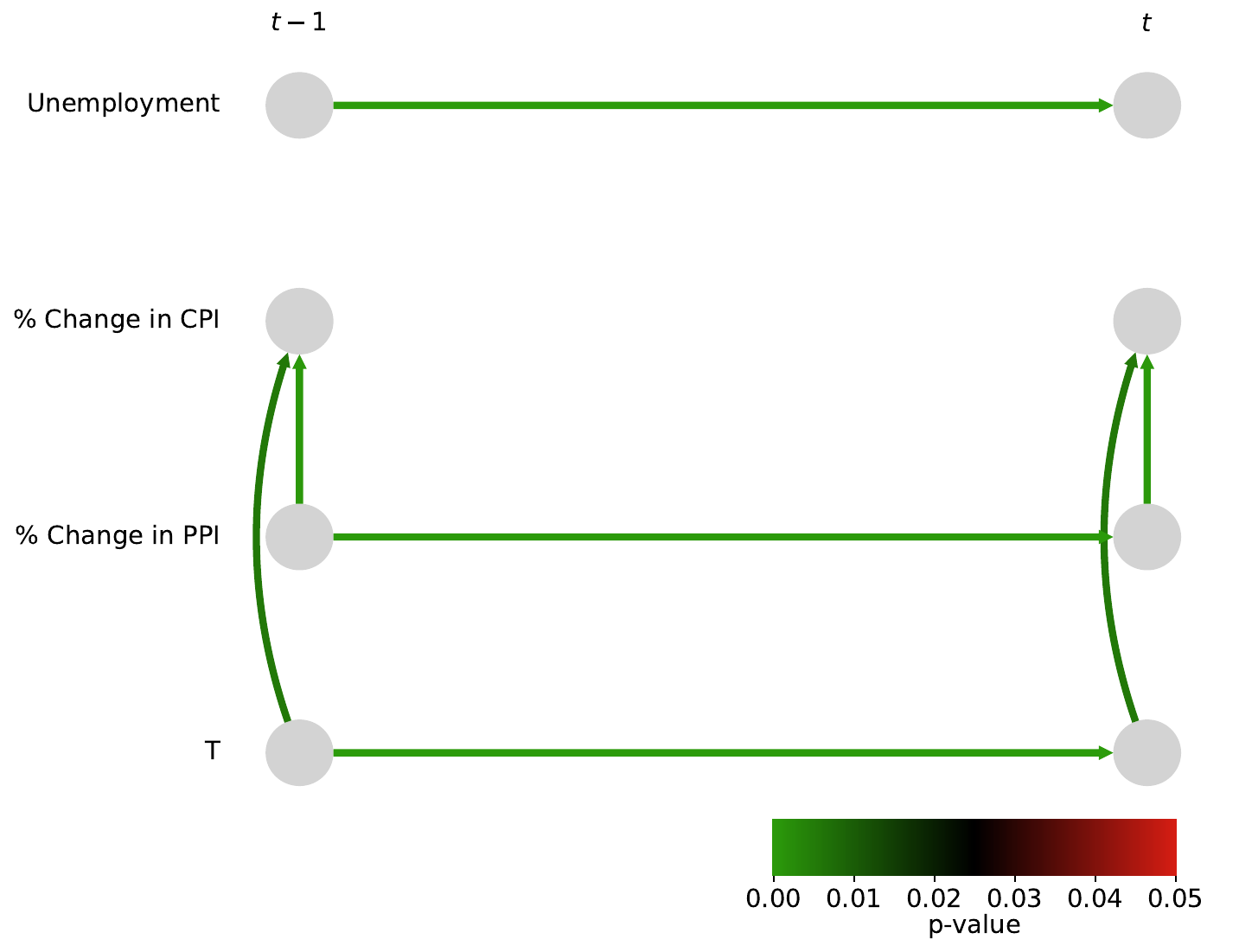}}
    \caption{Japan}
    \end{subfigure}\hfill
    \begin{subfigure}{0.45\linewidth}
    \centerline{\includegraphics[width=\linewidth]{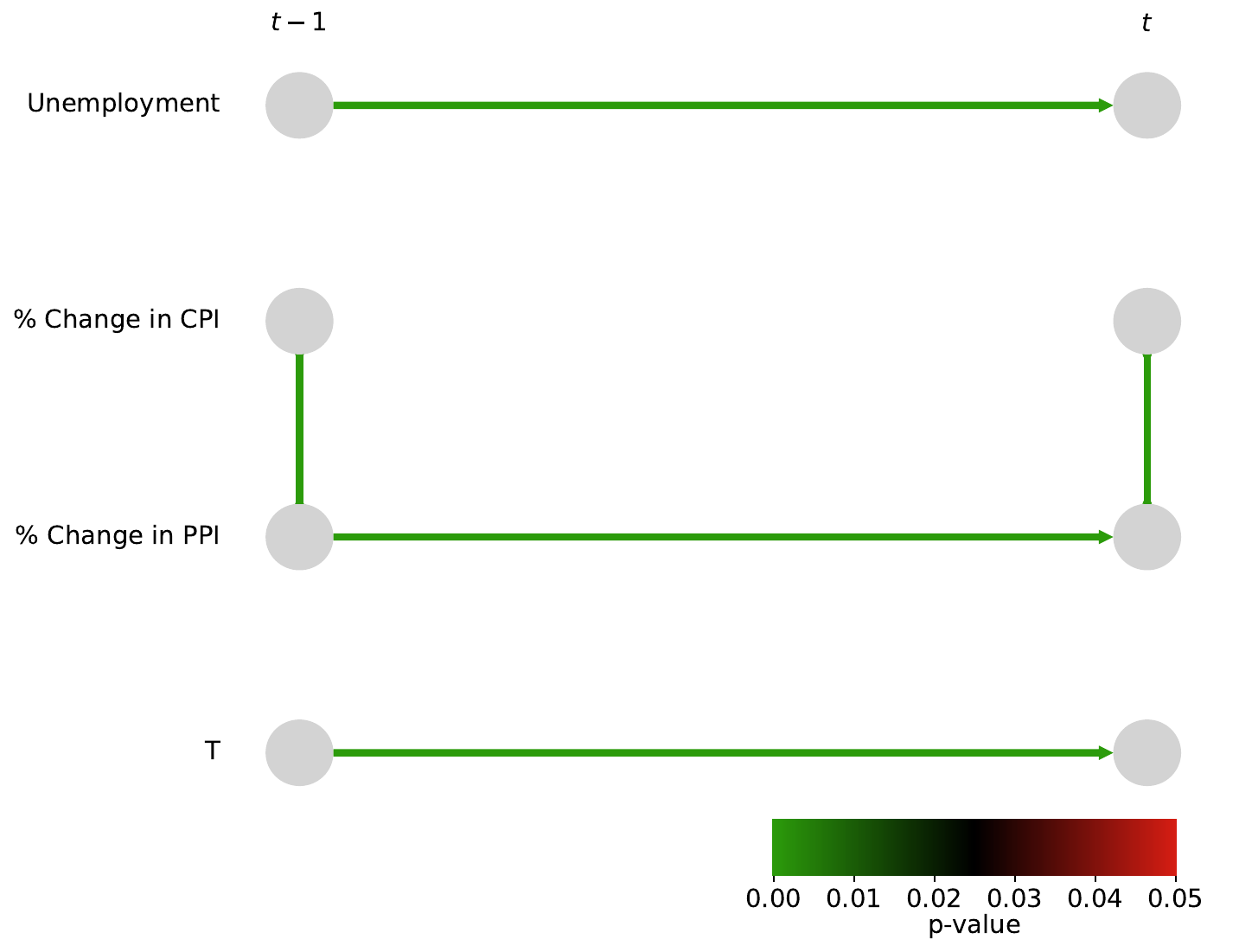}}
    \caption{France}
    \end{subfigure}
    \begin{subfigure}{0.45\linewidth}
        \centerline{\includegraphics[width=\linewidth]{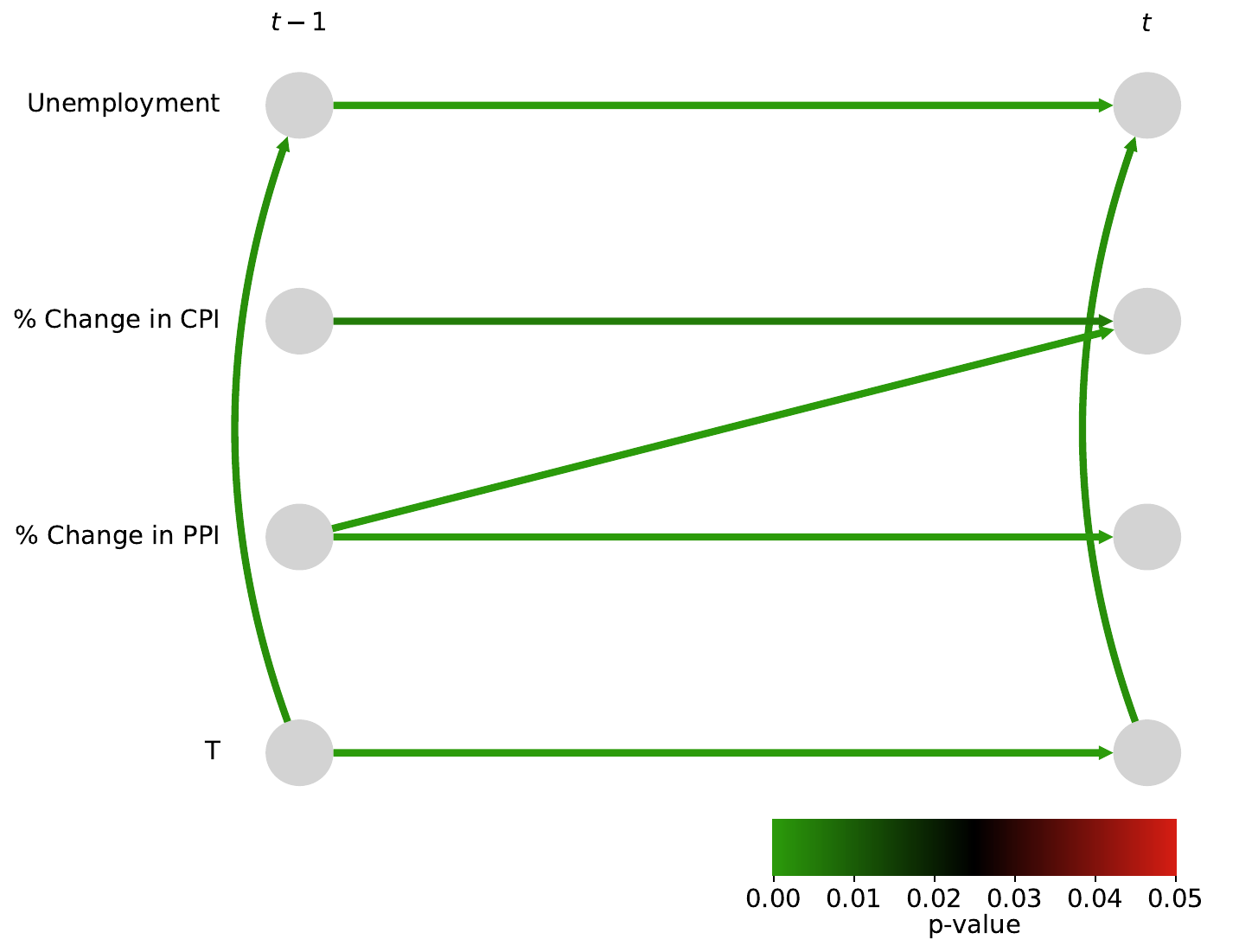}}
    \caption{United Kingdom}
    \end{subfigure}\hfill
    \begin{subfigure}{0.45\linewidth}
        \centerline{\includegraphics[width=\linewidth]{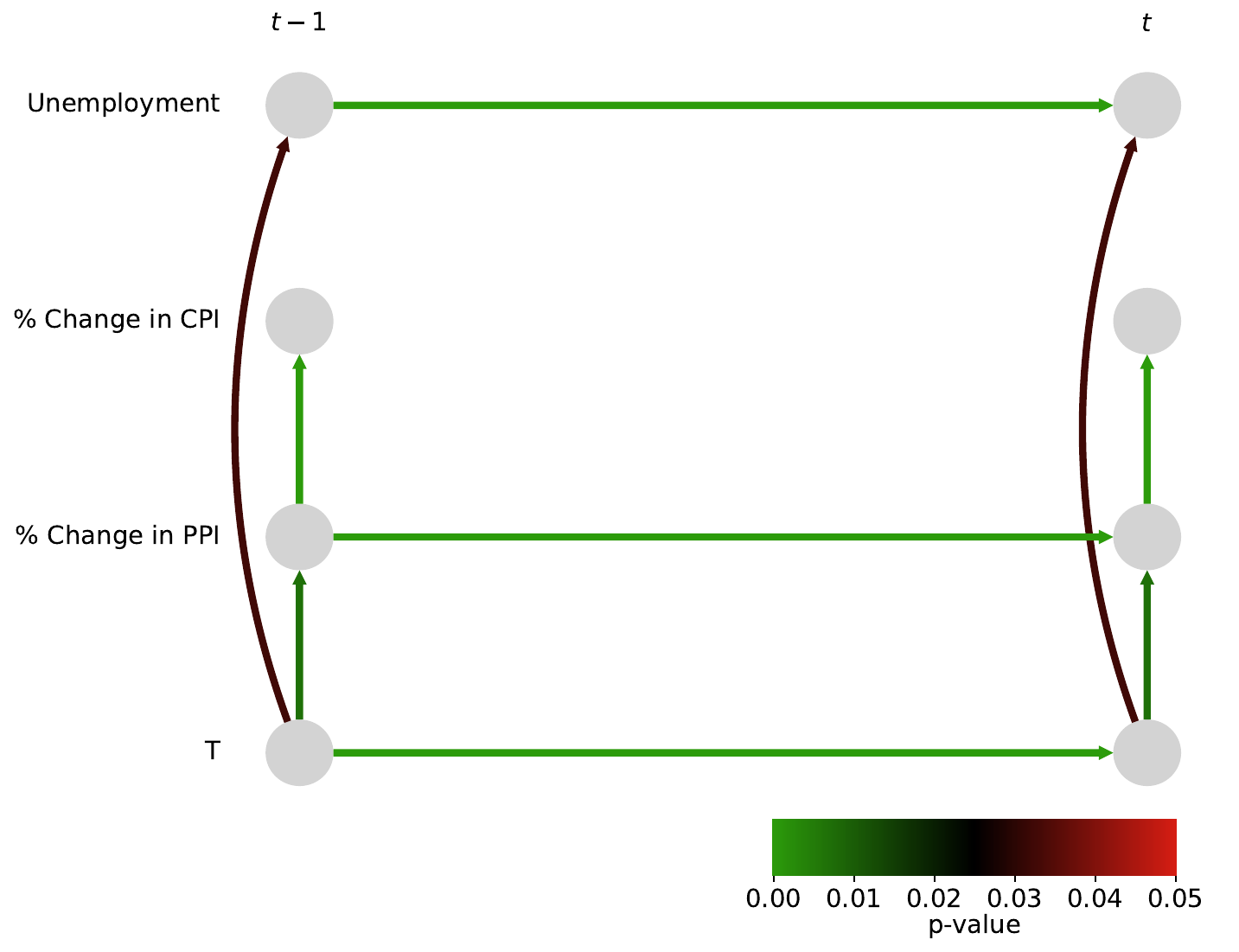}}
    \caption{Italy}
    \end{subfigure}
    \caption{Results of CDNOTS KCIT ran on six countries' economic variables.}
    % \label{fig:eco_country_wise}
\end{minipage}
\end{figure}
\begin{figure}[!bt]
\begin{minipage}{\linewidth}
    \begin{subfigure}{0.45\linewidth}
    \centerline{\includegraphics[width=\linewidth]{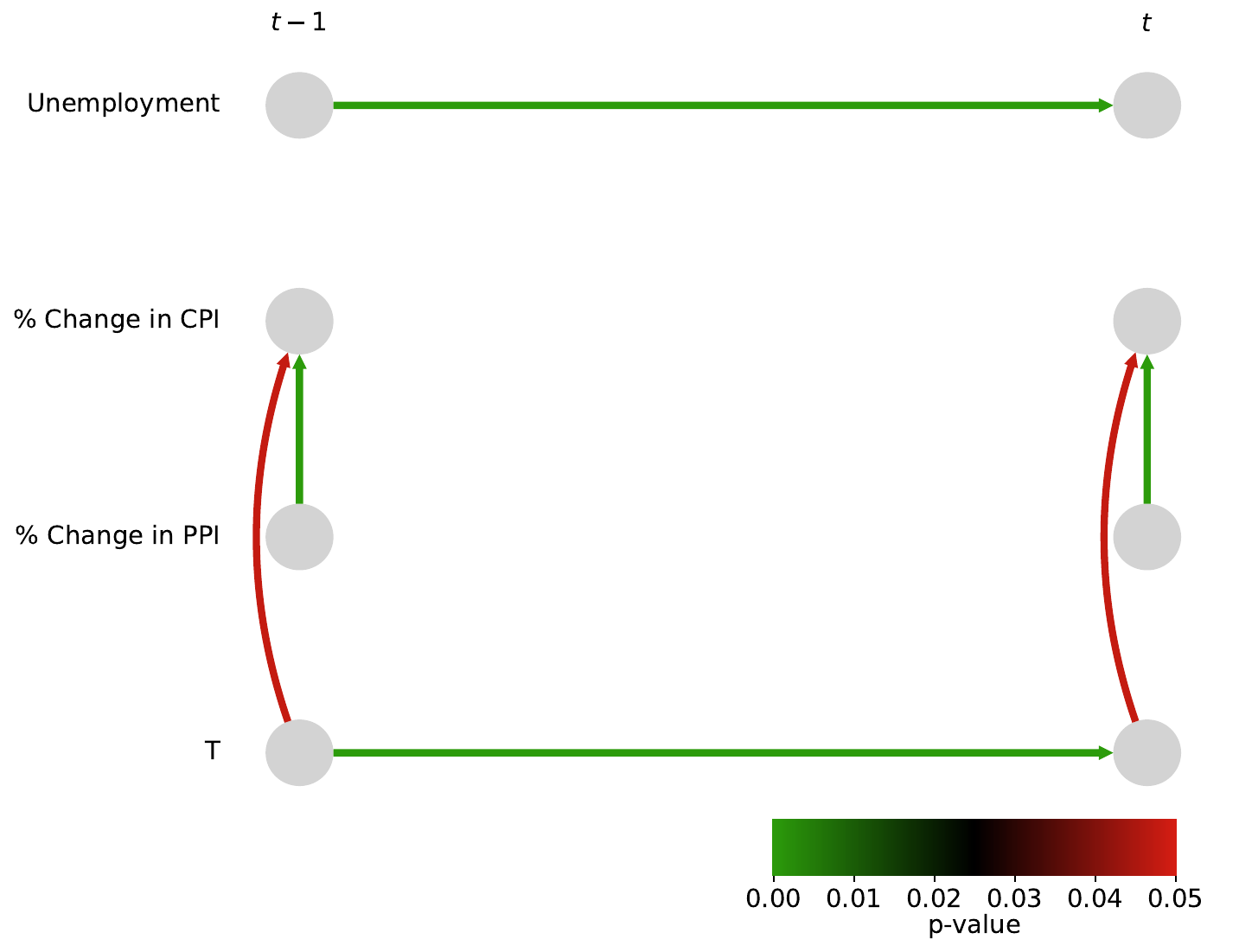}}
    \caption{U.S.}
    \end{subfigure}\hfill
    \begin{subfigure}{0.45\linewidth}
        \centerline{\includegraphics[width=\linewidth]{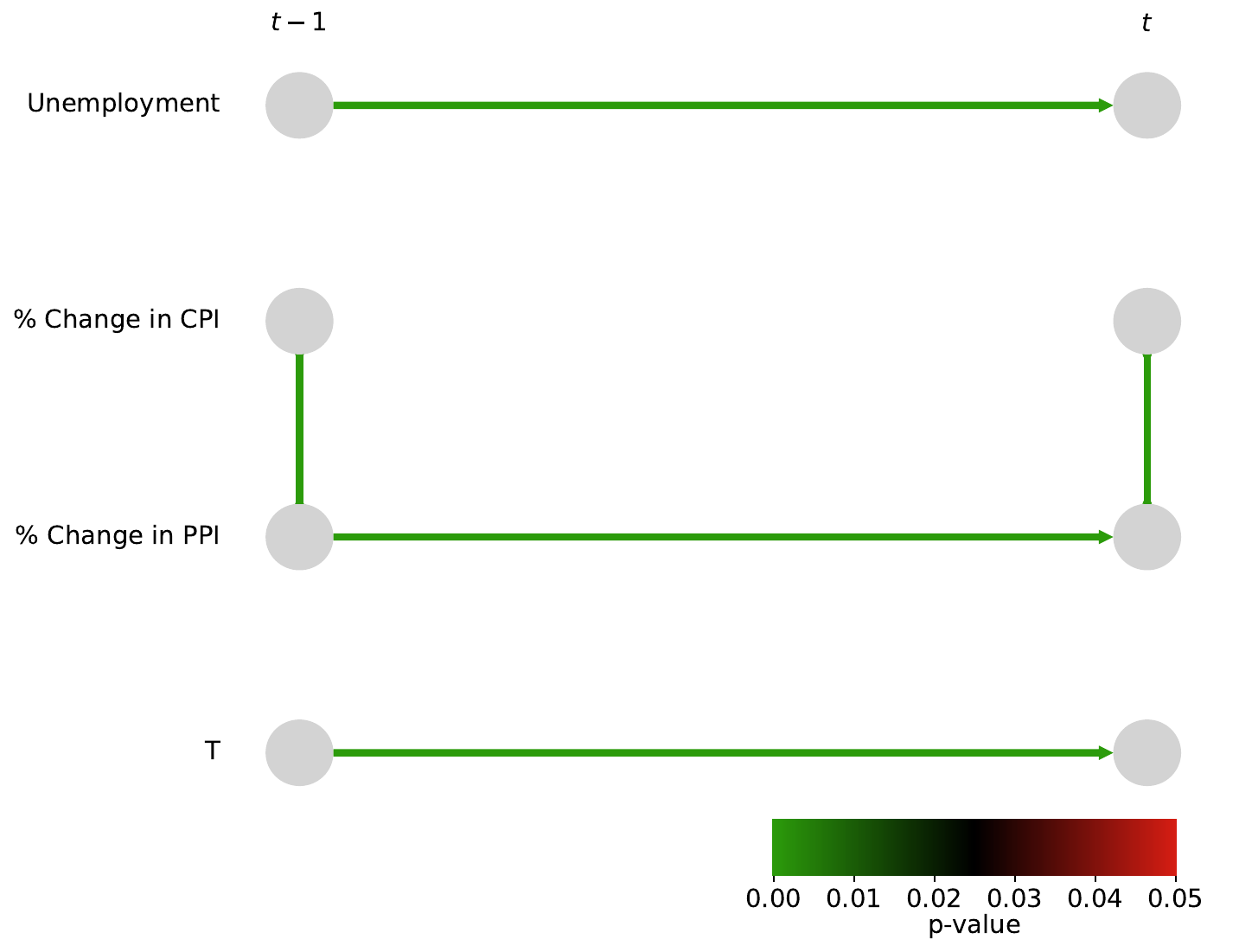}}
    \caption{Canada}
    \end{subfigure}
    \begin{subfigure}{0.45\linewidth}
        \centerline{\includegraphics[width=\linewidth]{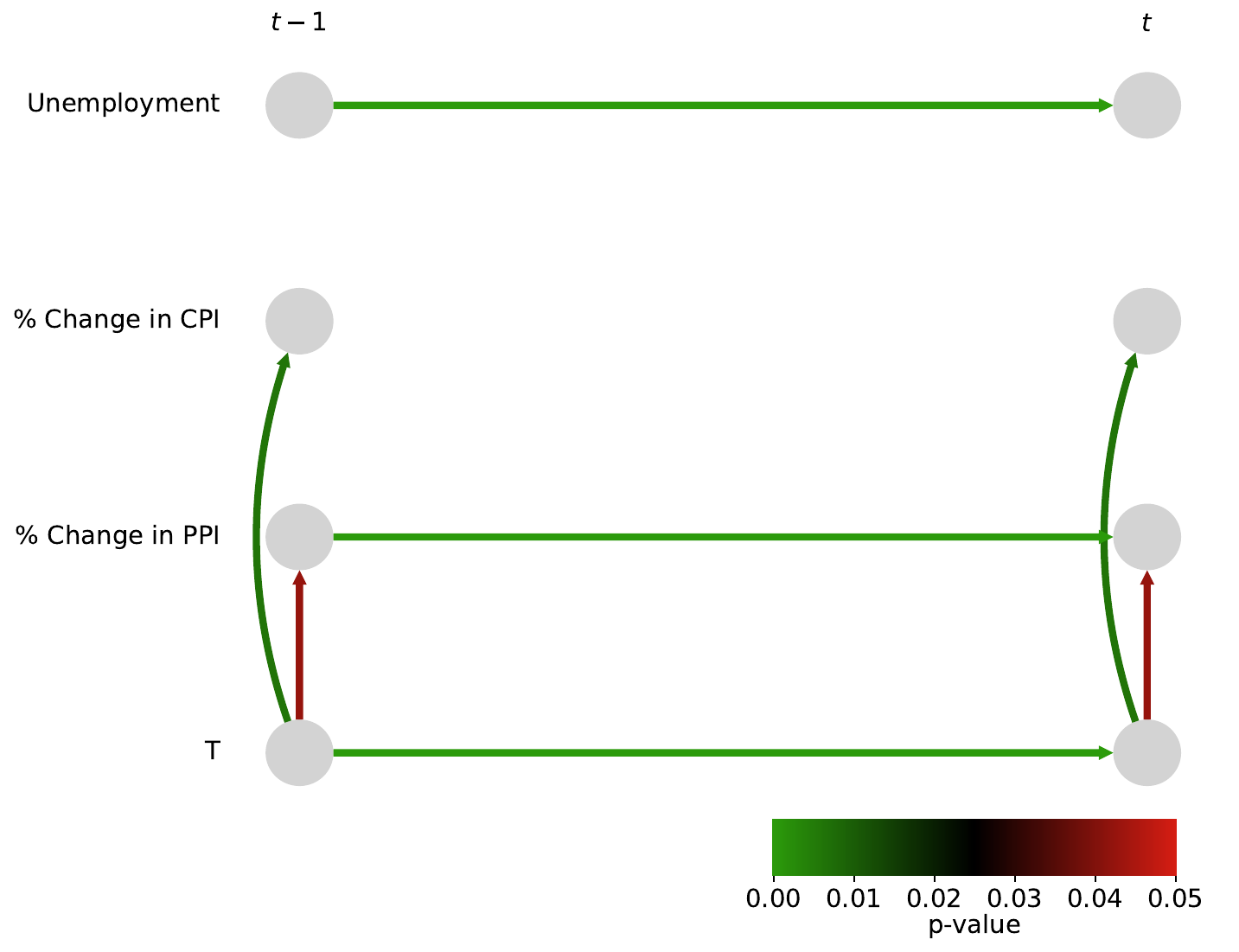}}
    \caption{Japan}
    \end{subfigure}\hfill
    \begin{subfigure}{0.45\linewidth}
    \centerline{\includegraphics[width=\linewidth]{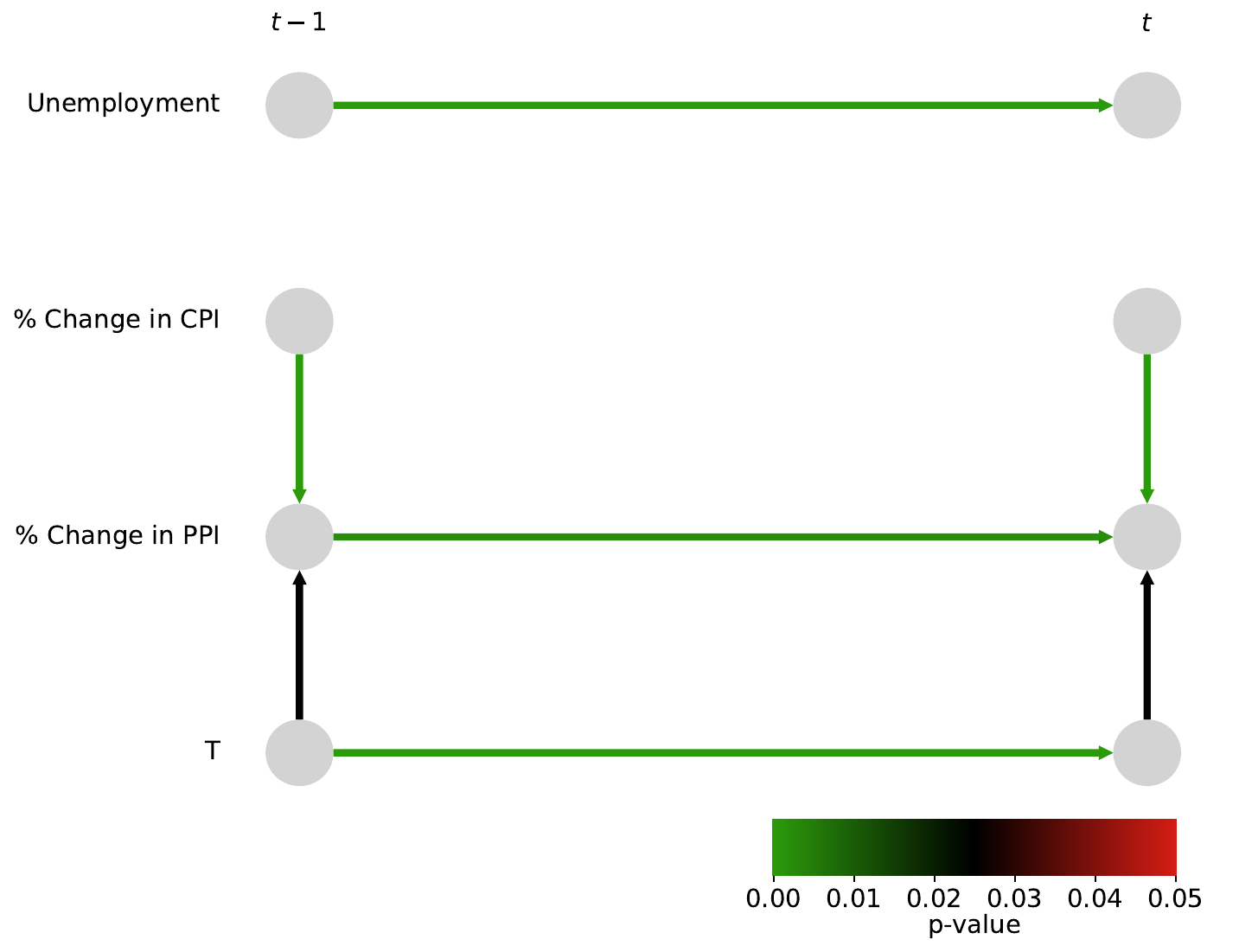}}
    \caption{France}
    \end{subfigure}
    \begin{subfigure}{0.45\linewidth}
        \centerline{\includegraphics[width=\linewidth]{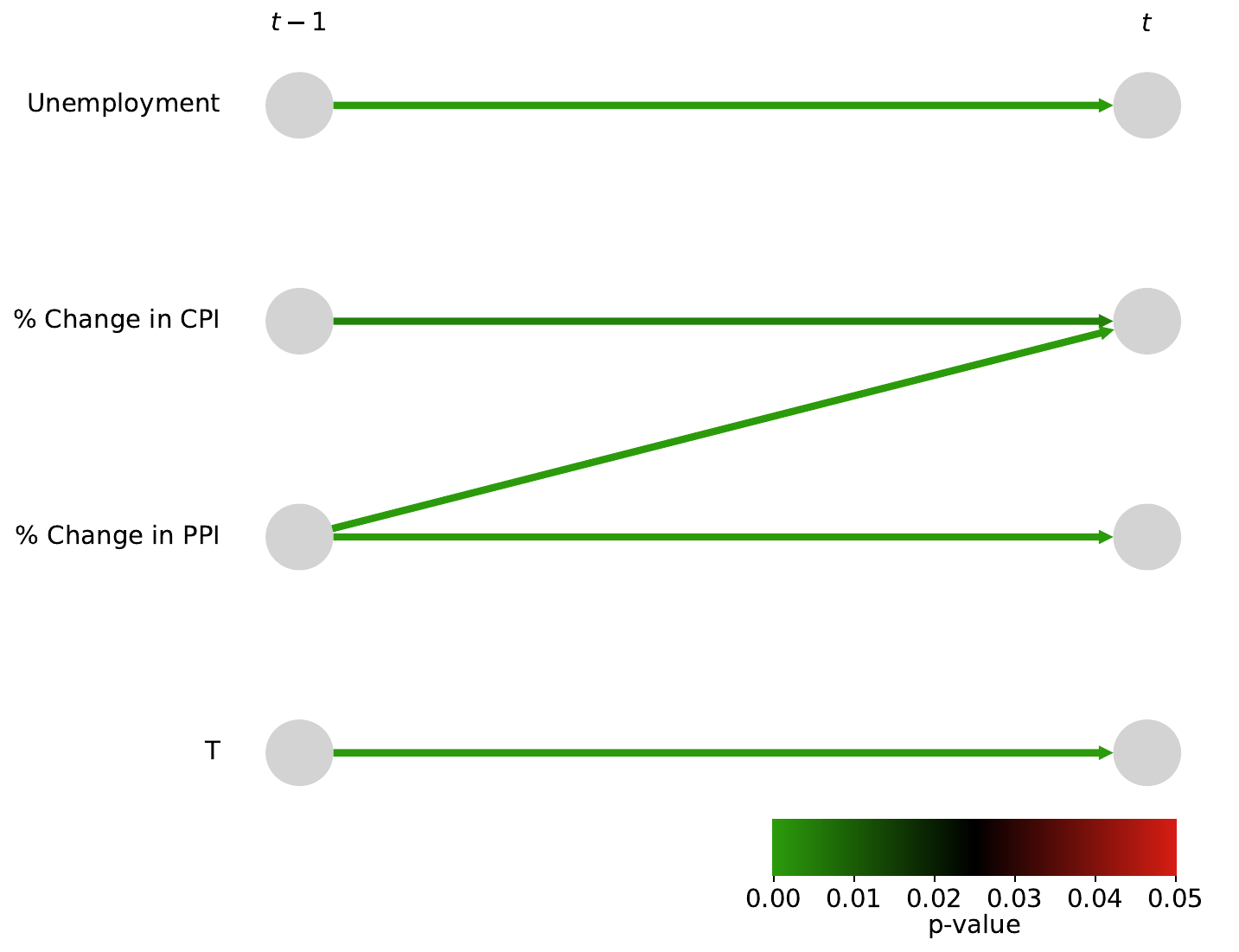}}
    \caption{United Kingdom}
    \end{subfigure}\hfill
    \begin{subfigure}{0.45\linewidth}
        \centerline{\includegraphics[width=\linewidth]{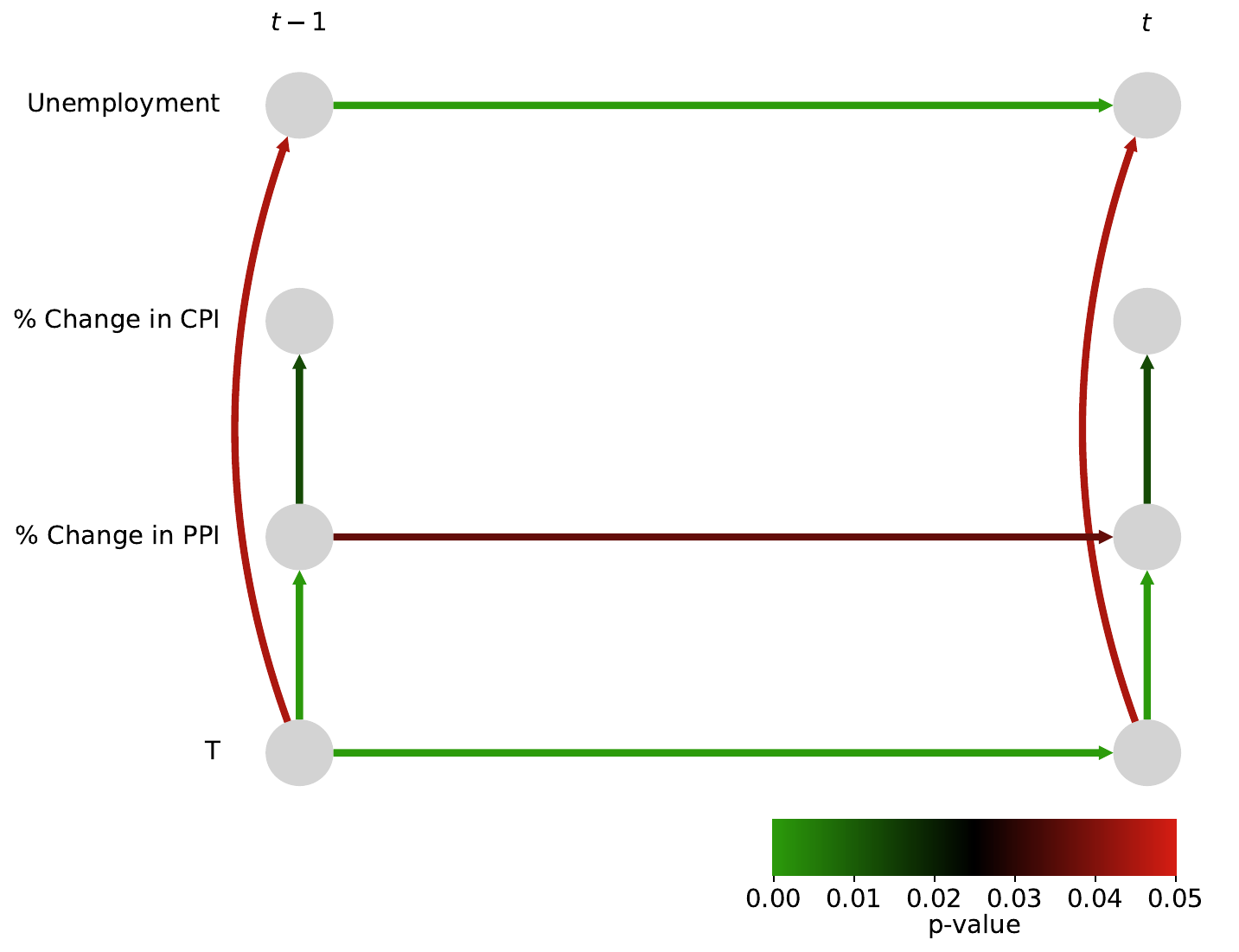}}
    \caption{Italy}
    \end{subfigure}
    \caption{Results of CDNOTS RCoT ran on six countries' economic variables.}
    % \label{fig:eco_country_wise}
\end{minipage}
\end{figure}
%

%\clearpage
%%%%%%%%%%%%%%%%%%%%%%%%%%%%%%%%%
\section{Detailed Simulation Results}
Further findings on simulated data includes an evaluation of precision, recall\footnote{precision$=\frac{tp}{tp+fp}\,$, recall$=\frac{tp}{tp+fn}$ where $tp=$true positive, $fp=$false positive and $fn=$ false negative.} and Structural Hamming Distance (SHD). Note that while precision, recall, and F-score aim for maximization, the objective for SHD is minimization.
\begin{figure}[!t]
\begin{center}
    \begin{subfigure}{0.45\linewidth}
    \centerline{\includegraphics[width=\linewidth]{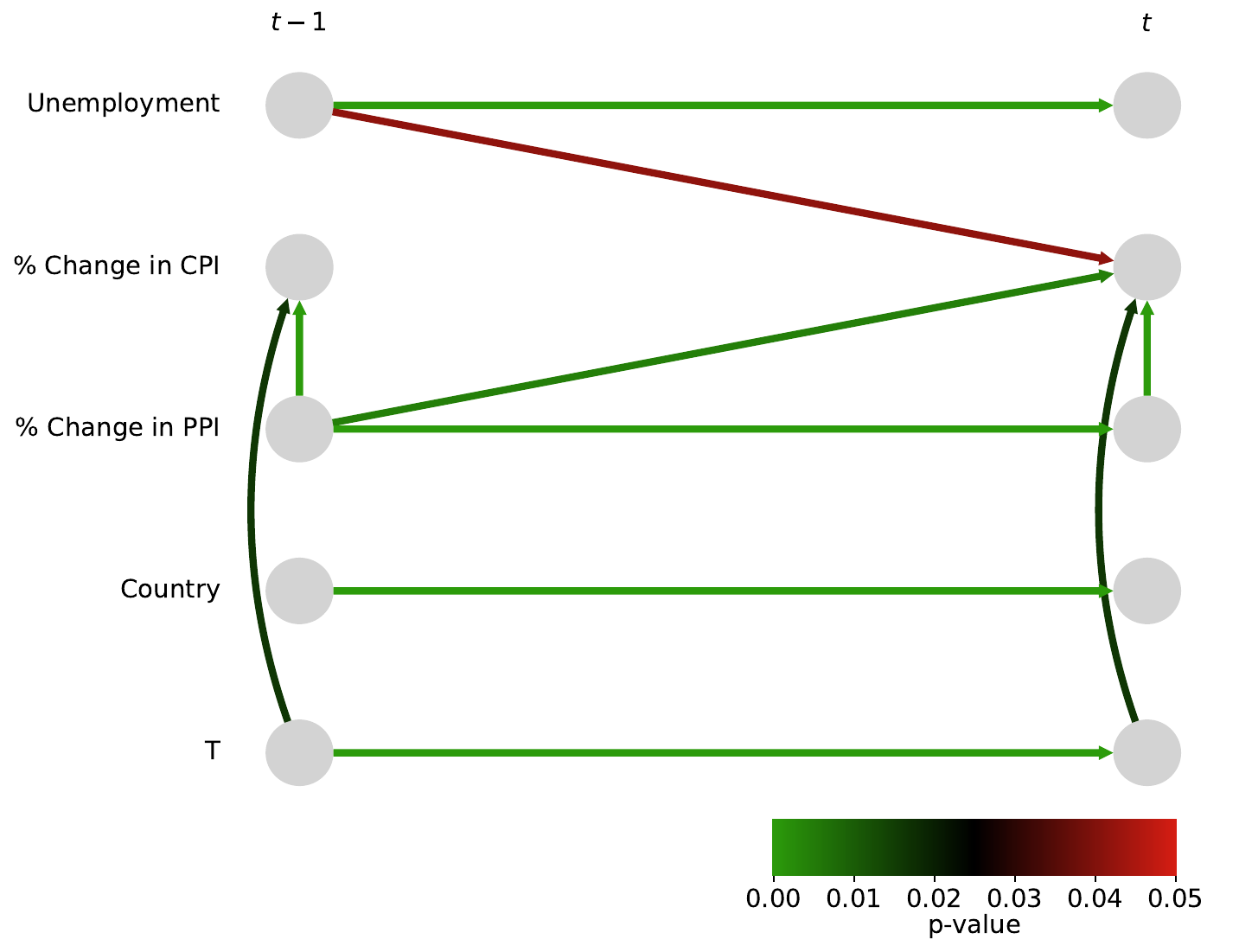}}
    \end{subfigure}
    \caption{Results of CDNOTS RCoT ran on the economic variables of multiple countries.}
\end{center}
\end{figure}
%
%\FloatBarrier
\subsection{Different CI Tests}

At 300 data points, KCIT-SW demonstrates the highest recall but substantially lower precision, resulting in the poorest F-score within the plot. Conversely, in the context of 1000 data points, recall remains consistent across all methods, with discrepancies in F-score attributed to variations in precision. Notably, with 50 data points, the precision curve exhibits an unexpected upward trend, contrary to the anticipated decrease in precision (and recall) as the node count increases. Interestingly, despite these nuances, SHD yields a performance ranking similar to that of F-score across the evaluated scenarios.
\begin{figure}[!bt]
\begin{minipage}{\linewidth}
    \centerline{\includegraphics[width=\linewidth]{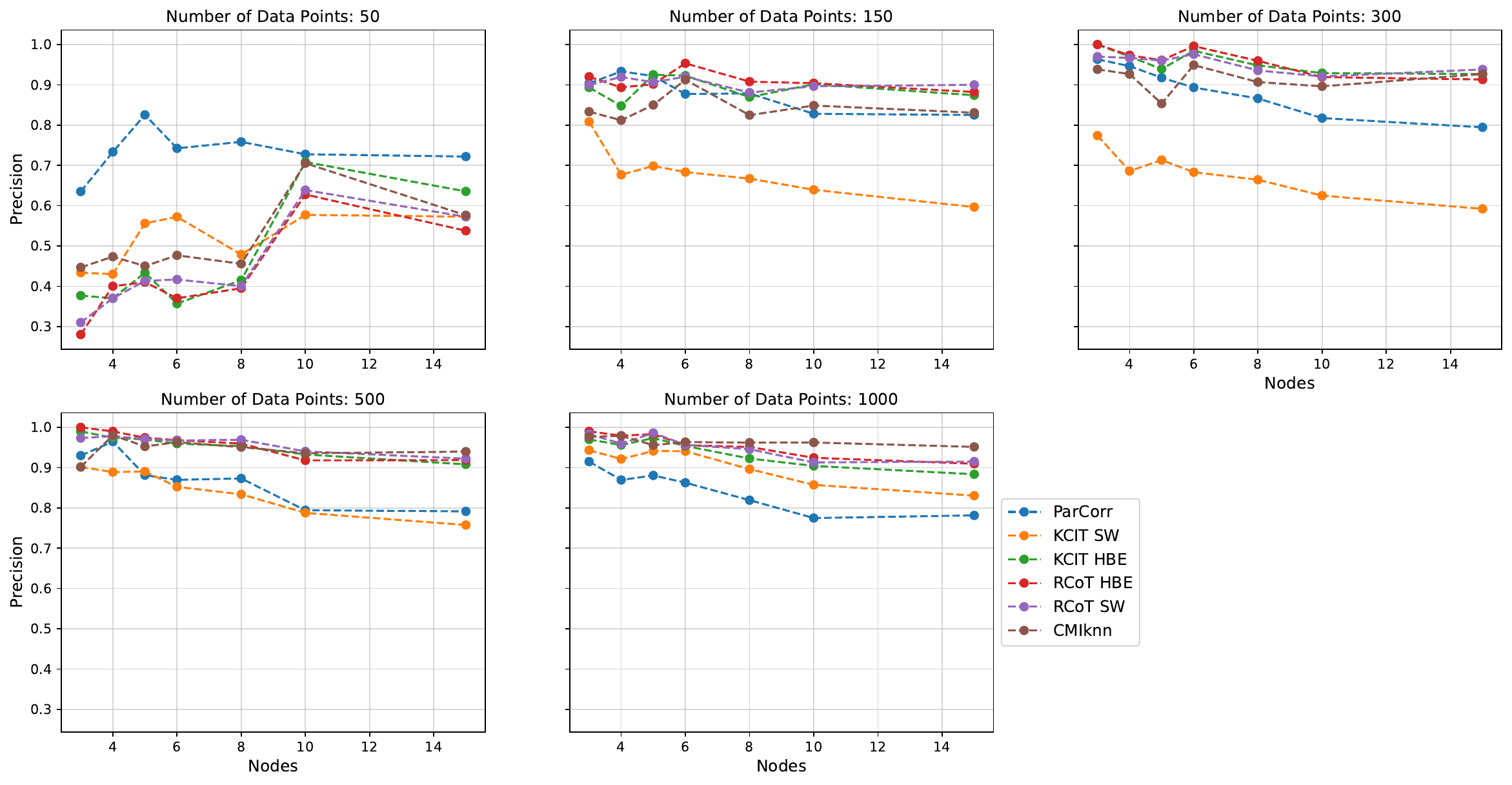}}
    \caption{Precision evaluation for CDNOTS with different CI tests, tested out on different simulated datasets.}
\end{minipage}
\end{figure}
\begin{figure}[!bt]
\begin{minipage}{\linewidth}
    \centerline{\includegraphics[width=\linewidth]{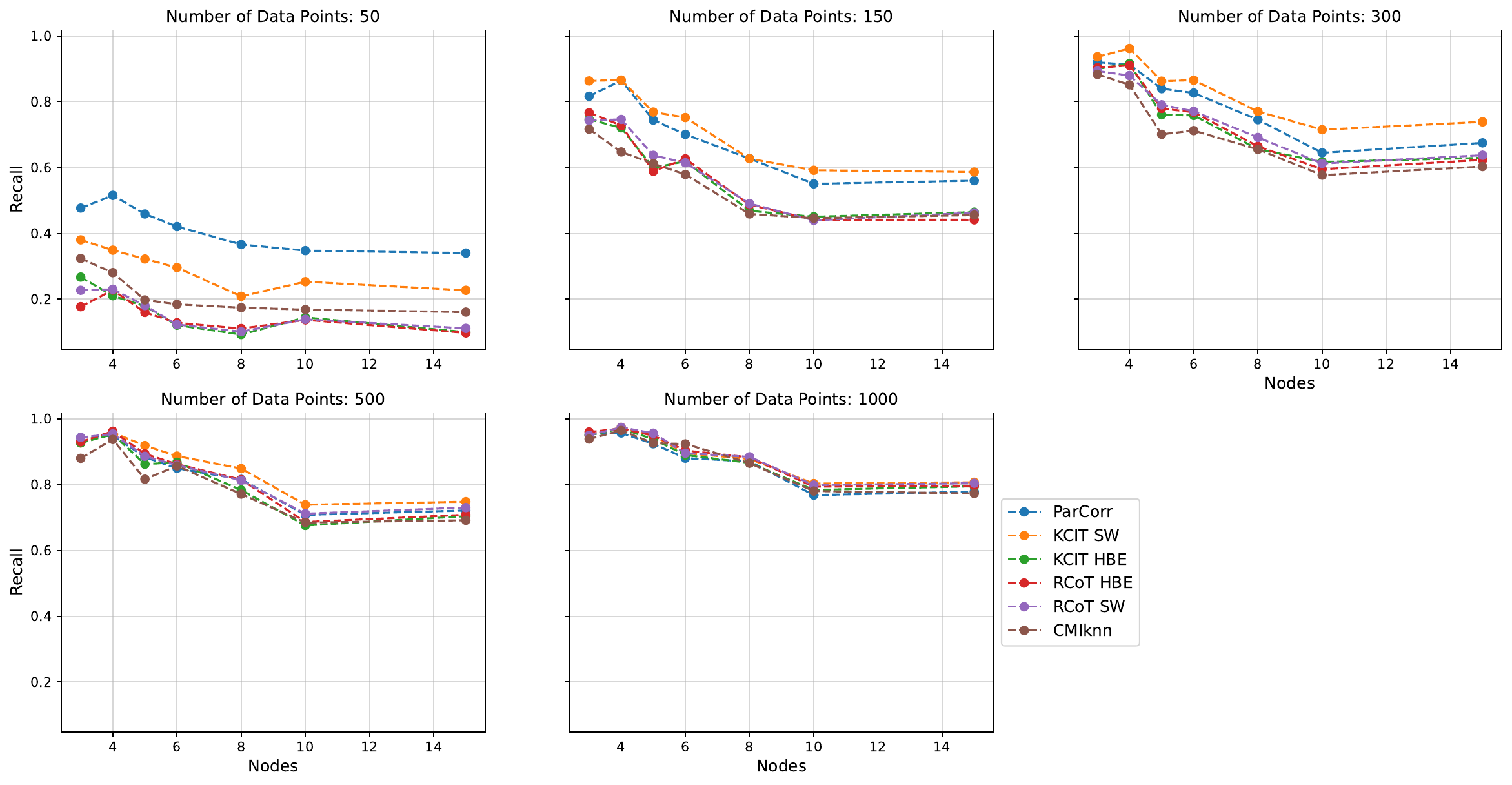}}
    \caption{Recall evaluation for CDNOTS with different CI tests, tested out on different simulated datasets.}
\end{minipage}
\end{figure}
\clearpage
\FloatBarrier
\begin{figure}[!t]
\begin{minipage}{\linewidth}
    \centerline{\includegraphics[width=\linewidth]{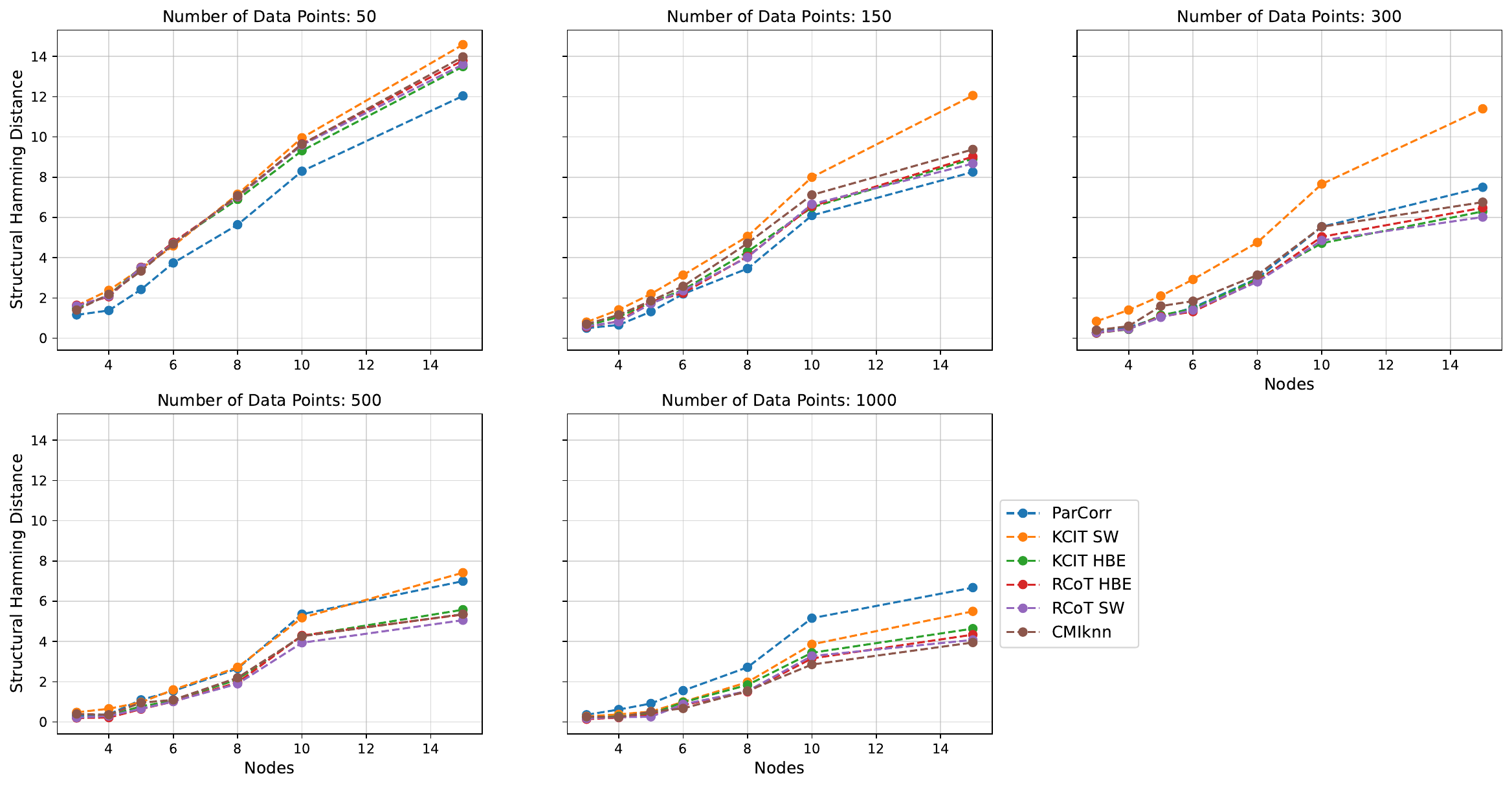}}
    \caption{SHD evaluation for CDNOTS with different CI tests, tested out on different simulated datasets.}
\end{minipage}
\end{figure}
% 
%%%%%%%%%%%%%%%%%%%%%%%%%%%%%%%
%\clearpage
%\FloatBarrier
\subsection{Effect of \(\alpha\)}\label{sec:alpha_cdnod}

Note that the significance level $\alpha$ controls the false positive rate. i.e., when employing a conditional independence test, if the data is truly independent, the test will fail $\alpha$ percent of the time, yet it does not explicitly address false negatives. This phenomenon becomes evident when observing precision plots, where the optimal precision corresponds to the lowest $\alpha$, indicating the lowest false positive rate. Conversely, recall operates inversely, as a decrease in false positives generally coincides with an increase in false negatives. It is this  balance that the F-score metric achieves, although it is worth noting that F-score is not the sole method to combine the two metrics together.

Regarding SHD, it presents another perspective on balancing false positives and false negatives. Unlike F-score, SHD typically advocates for an alpha of 0.01, except for specific scenarios, such as when dealing with 50 and 300 data points in RCoT and CMIknn for 300 data points with larger graphs.

\begin{figure}[!b]
\begin{minipage}{\linewidth}

\begin{subfigure}{\linewidth}
    \centerline{\includegraphics[width=\linewidth]{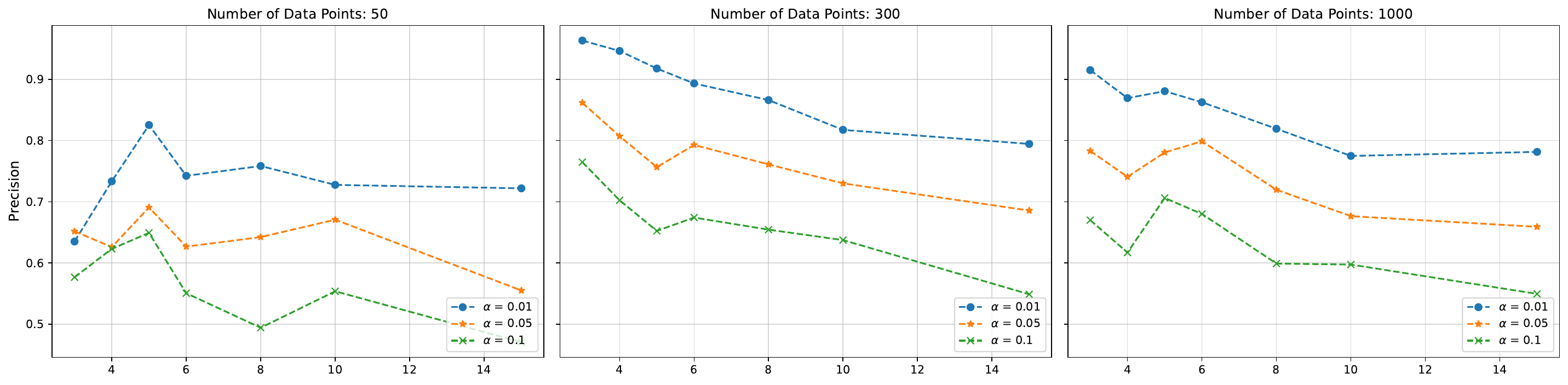}}
    \caption{ParCorr}
    \end{subfigure}
    
    \begin{subfigure}{\linewidth}
    \centerline{\includegraphics[width=\linewidth]{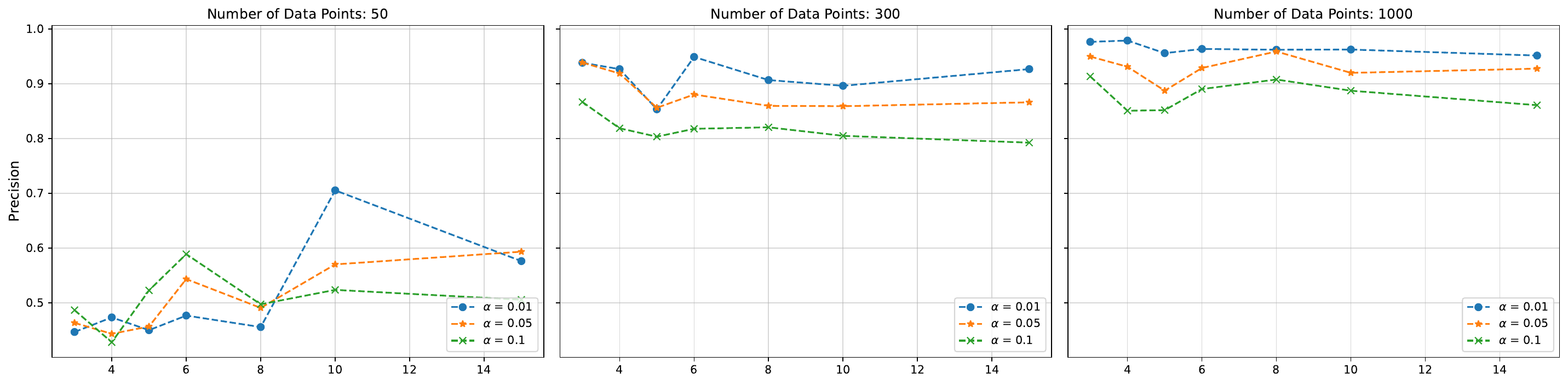}}
    \caption{CMIKnn}
    \end{subfigure}

    \begin{subfigure}{\linewidth}
    \centerline{\includegraphics[width=\linewidth]{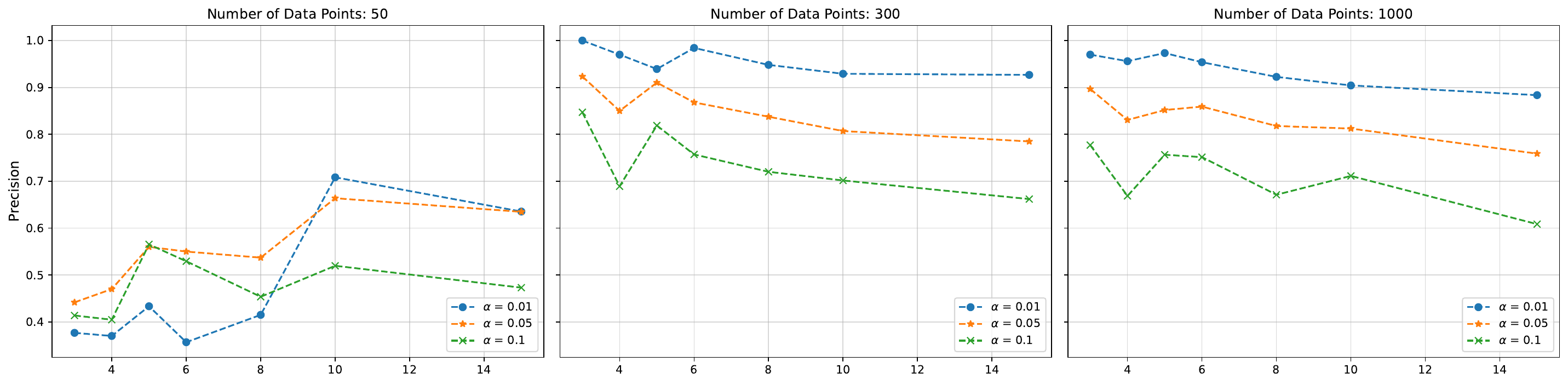}}
    \caption{KCIT}
    \end{subfigure}

    \begin{subfigure}{\linewidth}
    \centerline{\includegraphics[width=\linewidth]{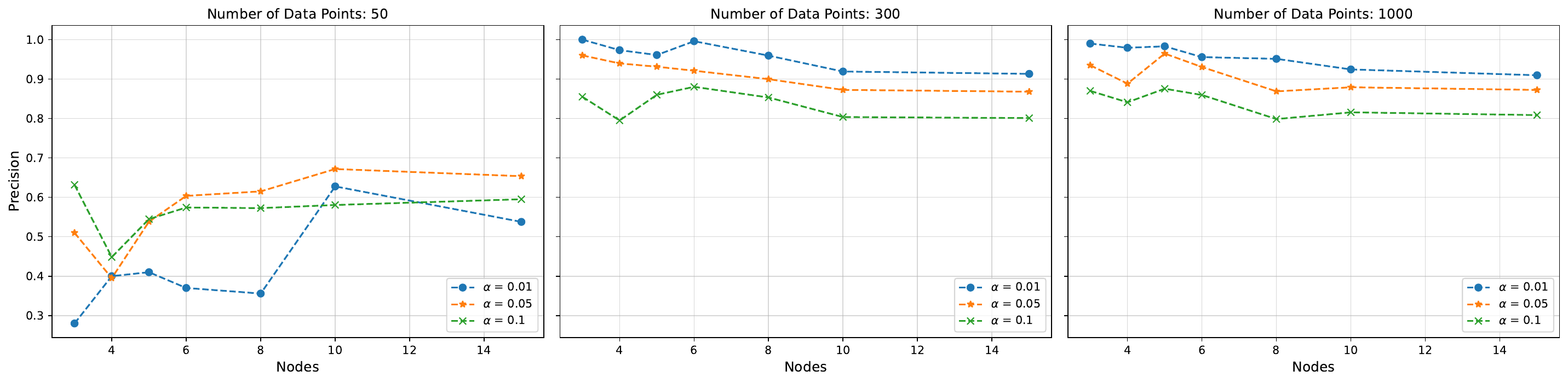}}
    \caption{RCoT}
    \end{subfigure}

    \caption{Effect of \(\alpha\) on Precision}\label{fig:cdnots_alpha_sensitivity_prec3d}

\end{minipage}
\end{figure}

\begin{figure}[!bt]
\begin{minipage}{\linewidth}

\begin{subfigure}{\linewidth}
    \centerline{\includegraphics[width=\linewidth]{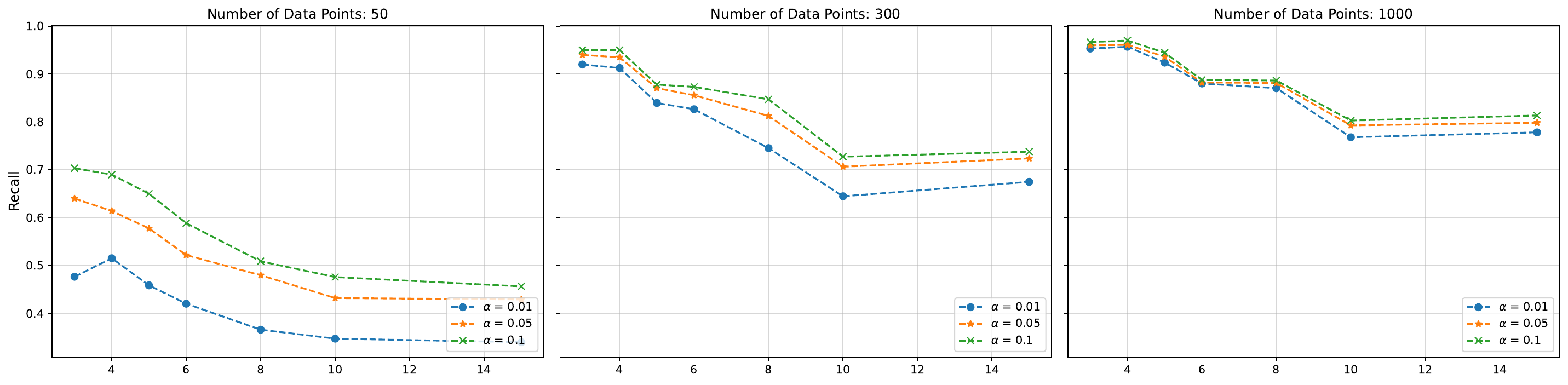}}
    \caption{ParCorr}
    \end{subfigure}
    
    \begin{subfigure}{\linewidth}
    \centerline{\includegraphics[width=\linewidth]{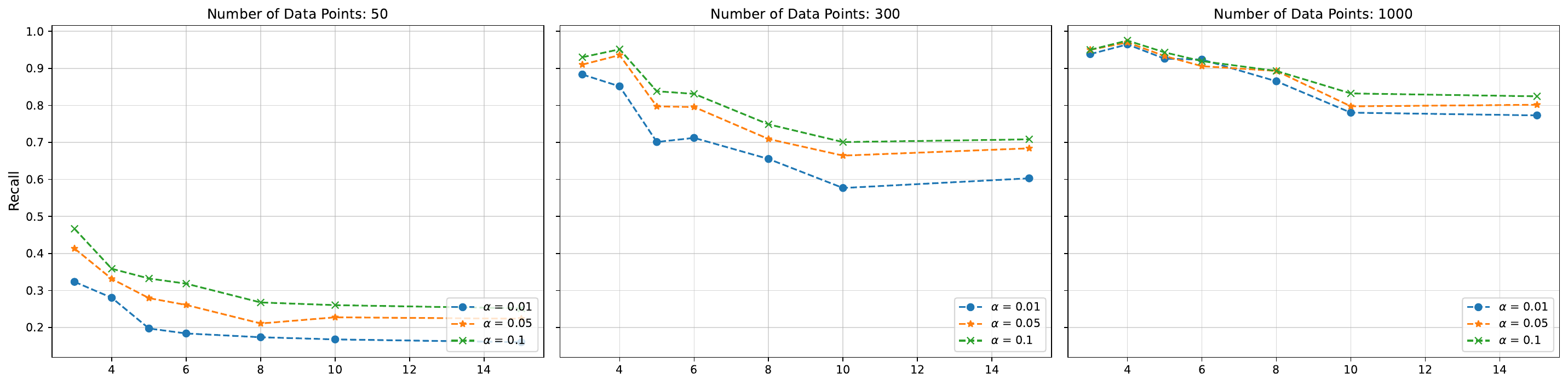}}
    \caption{CMIKnn}
    \end{subfigure}

    \begin{subfigure}{\linewidth}
    \centerline{\includegraphics[width=\linewidth]{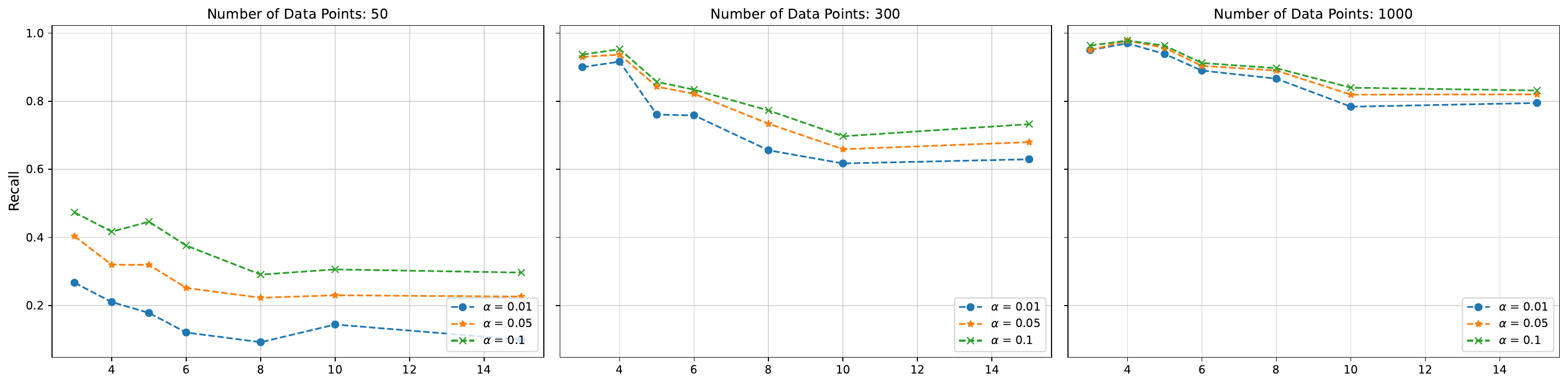}}
    \caption{KCIT}
    \end{subfigure}

    \begin{subfigure}{\linewidth}
    \centerline{\includegraphics[width=\linewidth]{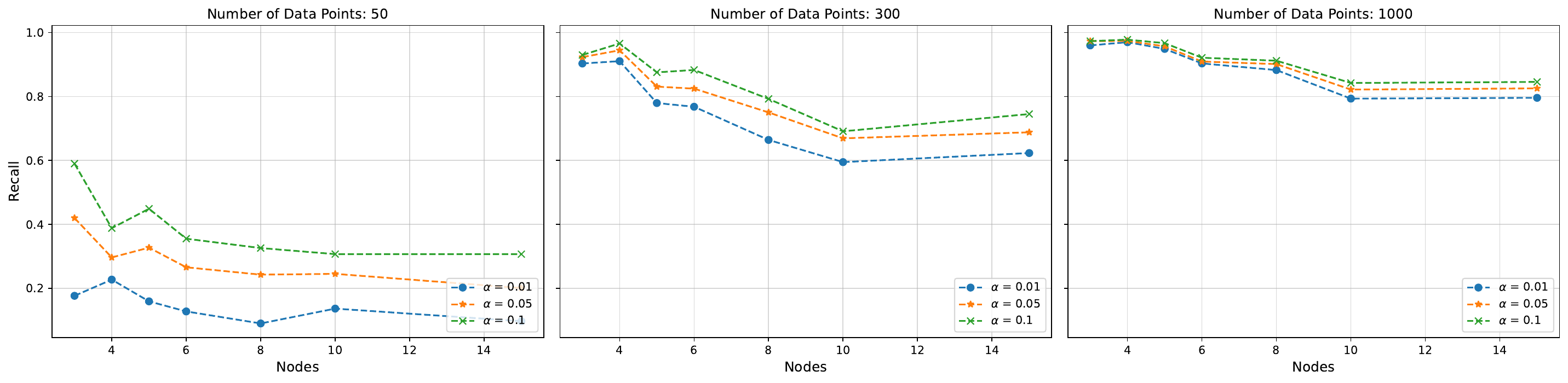}}
    \caption{RCoT}
    \end{subfigure}

    \caption{Effect of \(\alpha\) on Recall}\label{fig:cdnots_alpha_sensitivity_recal3d}

\end{minipage}
\end{figure}

\begin{figure}[!bt]
\begin{minipage}{\linewidth}

\begin{subfigure}{\linewidth}
    \centerline{\includegraphics[width=\linewidth]{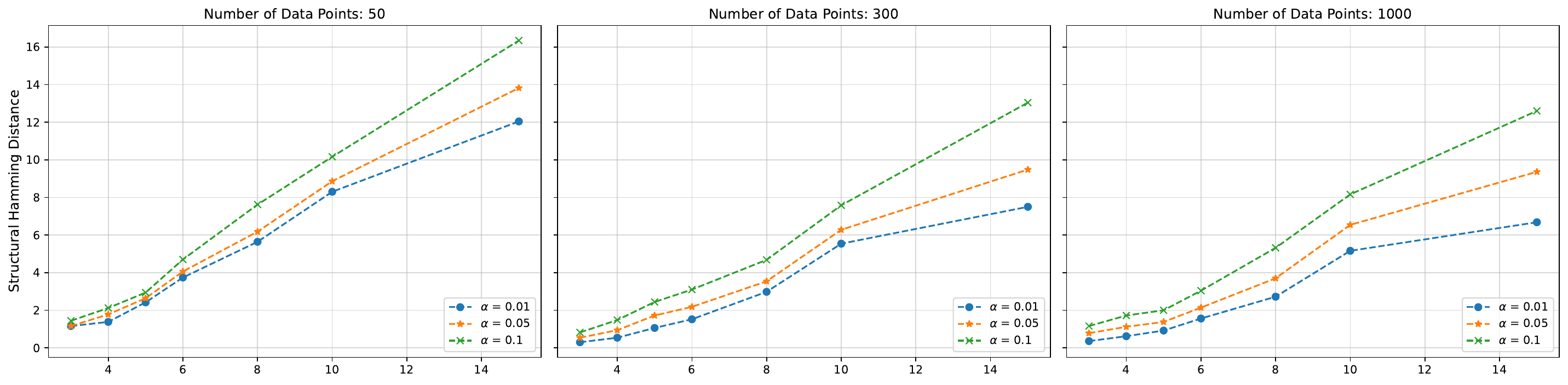}}
    \caption{ParCorr}
    \end{subfigure}
    
    \begin{subfigure}{\linewidth}
    \centerline{\includegraphics[width=\linewidth]{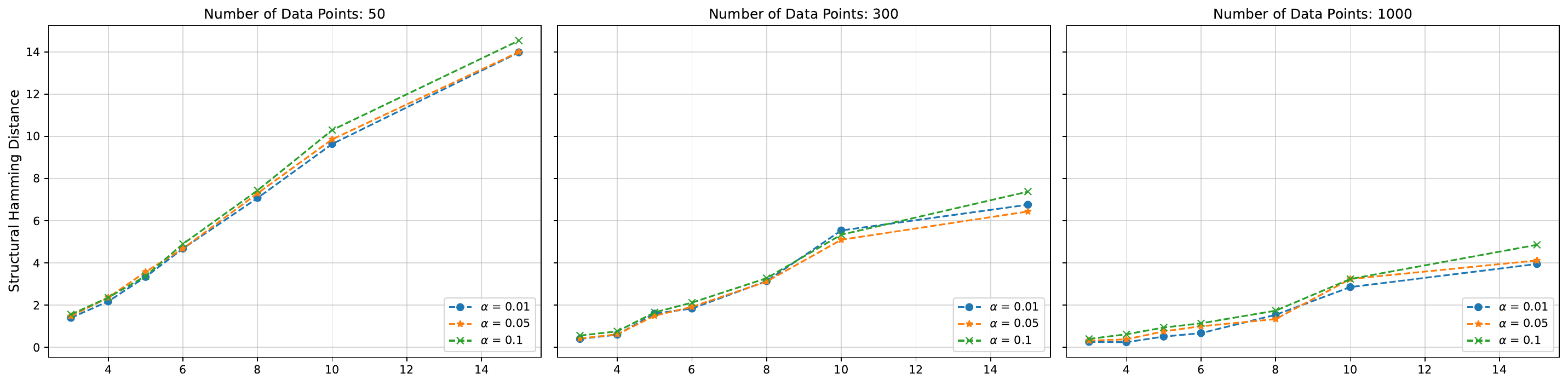}}
    \caption{CMIKnn}
    \end{subfigure}

    \begin{subfigure}{\linewidth}
    \centerline{\includegraphics[width=\linewidth]{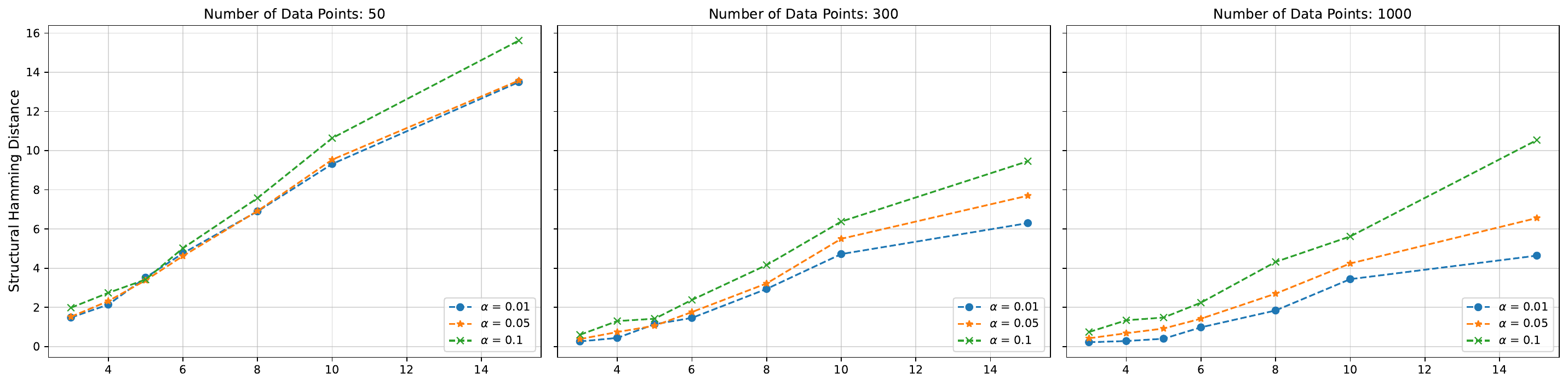}}
    \caption{KCIT}
    \end{subfigure}

    \begin{subfigure}{\linewidth}
    \centerline{\includegraphics[width=\linewidth]{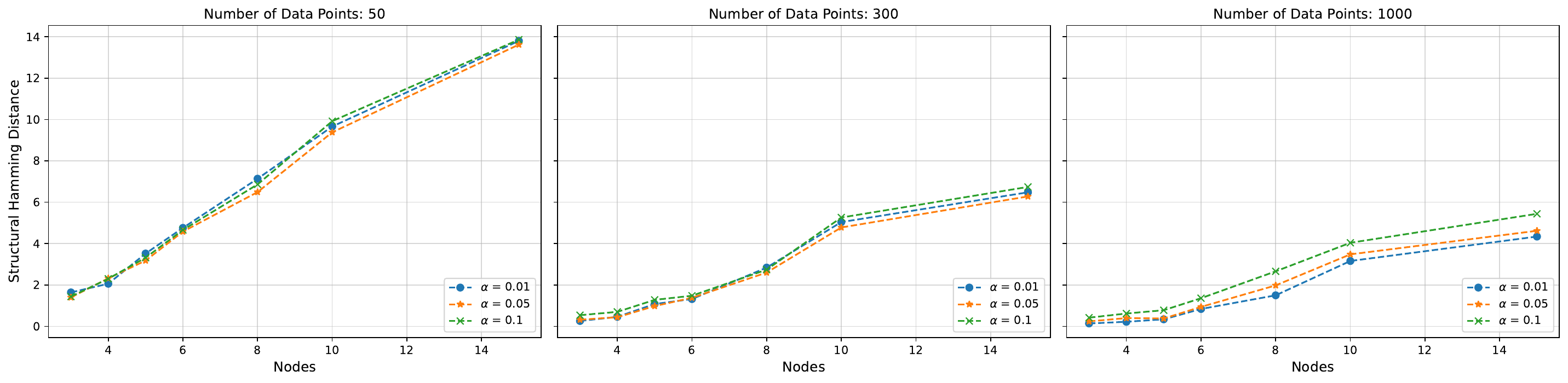}}
    \caption{RCoT}
    \end{subfigure}

    \caption{Effect of \(\alpha\) on SHD}\label{fig:cdnots_alpha_sensitivity_shd3d}

\end{minipage}
\end{figure}

\clearpage
\FloatBarrier
\subsection{Comparison against PCMCI}

CDNOTS consistently outperforms PCMCI in terms of SHD, similar to its dominance in the F-score metric. While PCMCI exhibits better recall than CDNOTS, it does so at the cost of significantly lower precision. This trade-off explains why CDNOTS achieves better performance in both F-score and SHD metrics.

\begin{figure}[!bh]
\begin{minipage}{\linewidth}
    \centerline{\includegraphics[width=\linewidth]{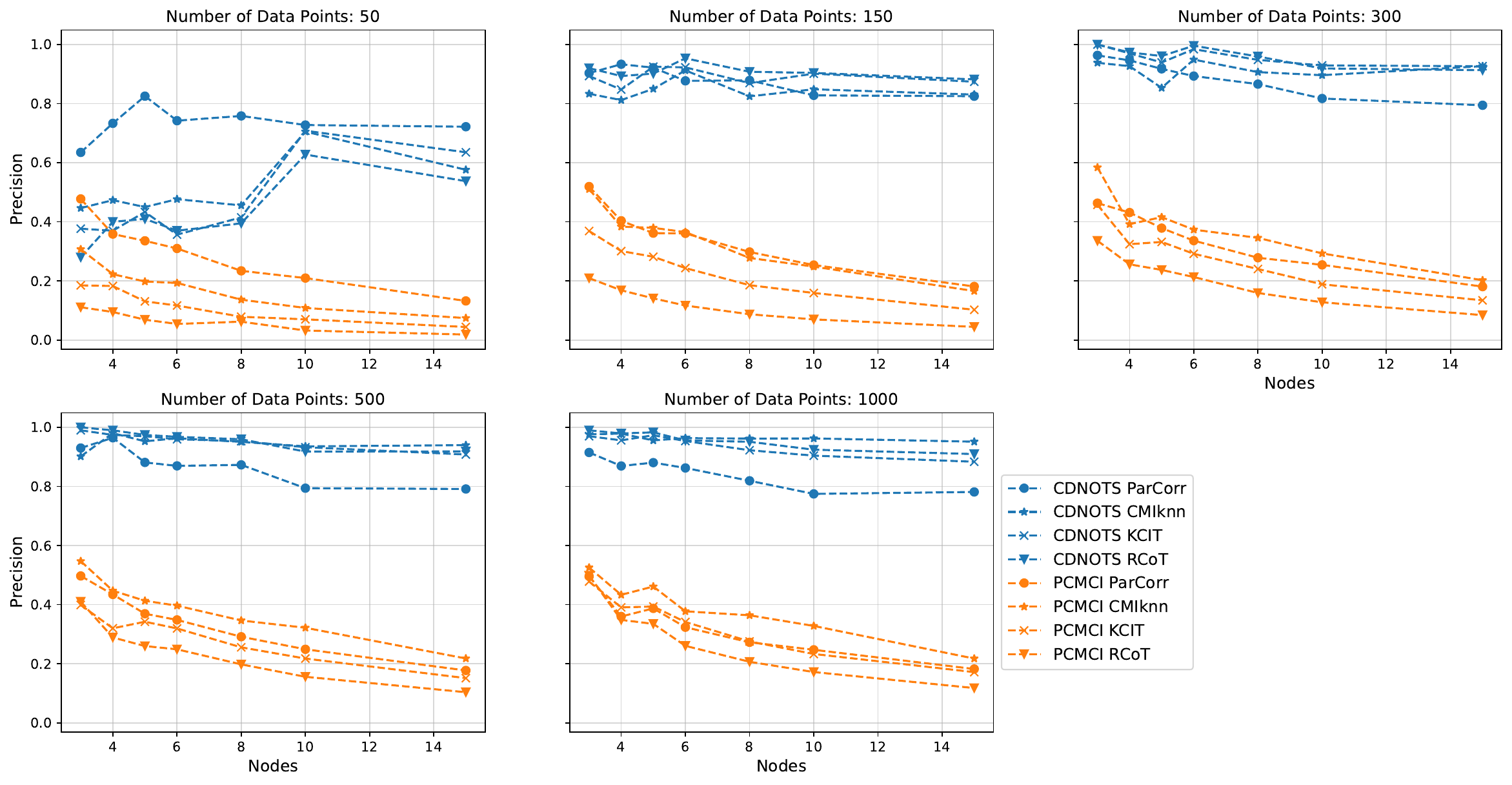}}
    \caption{Precision evaluation, comparing CDNOTS and PCMCI, tested out on different simulated datasets.}
\end{minipage}
\end{figure}

\begin{figure}[!bt]
\begin{minipage}{\linewidth}
    \centerline{\includegraphics[width=\linewidth]{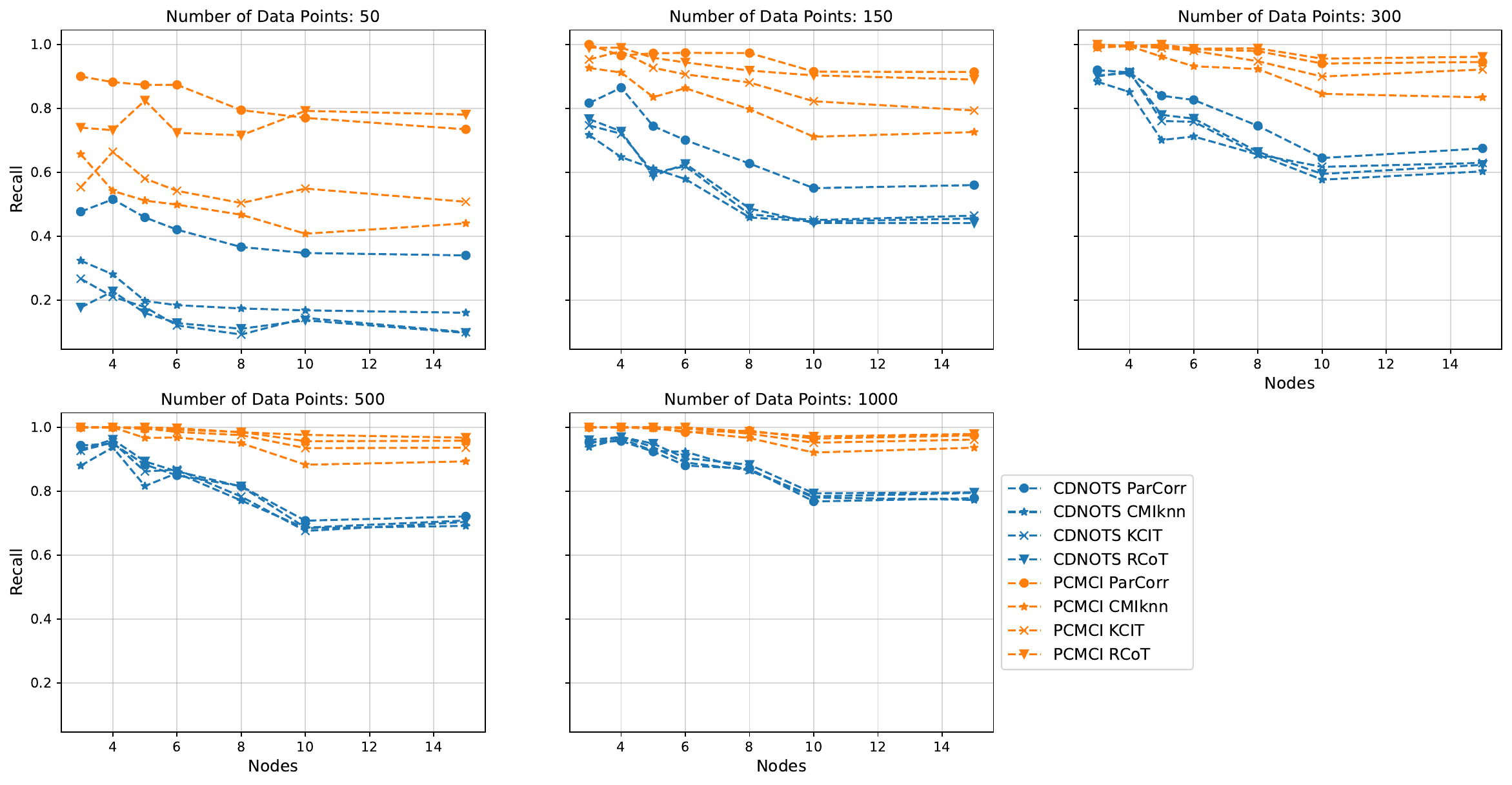}}
    \caption{Recall evaluation, comparing CDNOTS and PCMCI, tested out on different simulated datasets.}
\end{minipage}
\end{figure}

\begin{figure}[!bt]
\begin{minipage}{\linewidth}
    \centerline{\includegraphics[width=\linewidth]{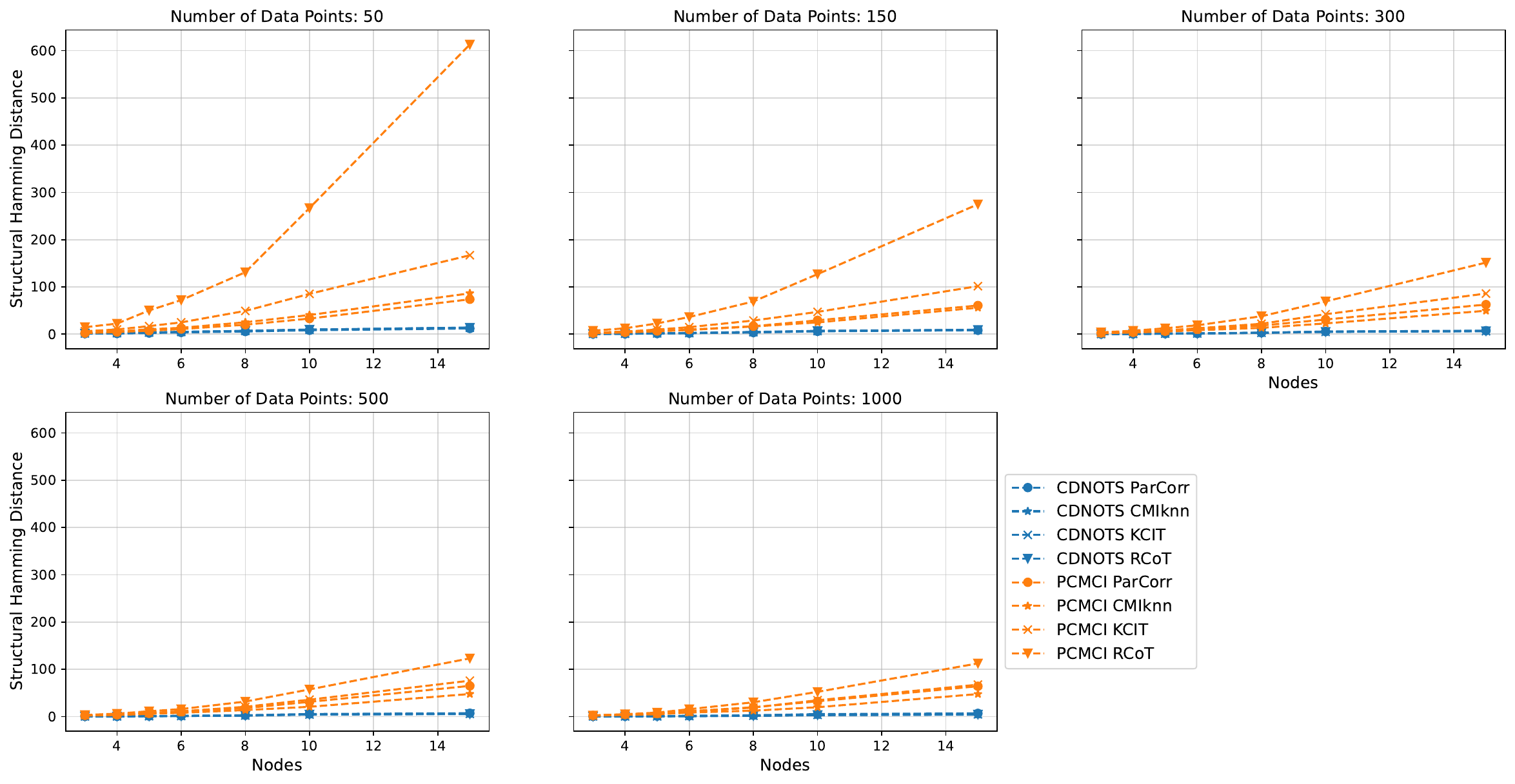}}
    \caption{SHD evaluation, comparing CDNOTS and PCMCI, tested out on different simulated datasets.}
\end{minipage}
\end{figure}

\vskip 0.2in
\bibliography{sample}

\end{document}